\documentclass[reprint,aps,prl,groupedaddress,10pt]{revtex4-2}

\usepackage{packages}
\usepackage{notations}

\newcommand{\qTASEPContinuousLattice}{
\begin{tikzpicture}[
    node distance=0.75cm,
    particle/.style={circle, draw=black, fill=white, thick, minimum size=2mm, inner sep=0pt},
    particle_black/.style={circle, draw=black, fill=black, thick, minimum size=2mm, inner sep=0pt},
    arrow_line/.style={->, >=Stealth, thick, bend left},
    both_arrow_line/.style={<->, >=Stealth, thick}
]

    \foreach \x in {0,...,6} {
        \ifthenelse{\x = 2 \OR \x = 5}{
            \node[particle_black] (P\x) at (\x*0.75,0) {}; % Adjusted spacing for smaller nodes
        }{
            \node[particle] (P\x) at (\x*0.75,0) {}; % Adjusted spacing for smaller nodes
        }
    }

    \draw[both_arrow_line] ([xshift=-1cm]P0.west) -- ([xshift=1cm]P6.east);

    \node[below=0.1cm of P2] {$x_i(t)$};

    \node[below=0.1cm of P5] {$x_{i-1}(t)$};

    \draw[arrow_line] (P2.north) to (P3.north) node[above=2mm] {rate $a_i(1 - q^{\text{gap}_i})$};

\end{tikzpicture}
}

\newcommand{\qTASEPDiscreteLattice}{\begin{tikzpicture}[
    node distance=0.75cm,
    particle/.style={circle, draw=black, fill=white, thick, minimum size=2mm, inner sep=0pt},
    particle_black/.style={circle, draw=black, fill=black, thick, minimum size=2mm, inner sep=0pt},
    arrow_line/.style={->, >=Stealth, thick, bend left=60},
    both_arrow_line/.style={<->, >=Stealth, thick}
]

    \foreach \x in {0,...,6} {
        \ifthenelse{\x = 0 \OR \x = 4 \OR \x = 6}{
            \node[particle_black] (G\x) at (\x*0.75,0) {}; % Nodes 1, 5, and 7 are black
        }{
            \node[particle] (G\x) at (\x*0.75,0) {}; % Other nodes are white
        }
    }

    \draw[both_arrow_line] ([xshift=-1cm]G0.west) -- ([xshift=1cm]G6.east);

    \node[below=0.1cm of G0] {$x_i(t)$};
    \node[below=0.1cm of G4] {$x_{i-1}(t)$};

    \draw[arrow_line] (G0) to (G1);
    \draw[arrow_line] (G0) to (G2);
    \draw[arrow_line] (G0) to (G3);
    \draw[arrow_line] (G4) to (G5);

    \node[above=5mm of G0.north] {$\mathbf{p}_{\text{gap}_i, a_i \alpha_{t+1}}(j)$};
    \node[above=5mm of G5.north] {$\mathbf{p}_{\text{gap}_{i-1}, a_{i-1} \alpha_{t+1}}(j)$};

\end{tikzpicture}}

\begin{document}

\setlength{\abovedisplayskip}{4pt}
\setlength{\belowdisplayskip}{4pt}

\title{Integrable Matrix Probabilistic Diffusions and \\ the Matrix Stochastic Heat Equation }

\author{Alexandre Krajenbrink}
\email{alexandre.krajenbrink@quantinuum.com}
\affiliation{Quantinuum, Partnership House, Carlisle Place, London SW1P 1BX, United Kingdom}
\affiliation{Le Lab Quantique, 58 rue d'Hauteville, 75010, Paris, France}
\author{Pierre Le Doussal}
\email{ledou@lpt.ens.fr}
\affiliation{Laboratoire de Physique de l'\'Ecole Normale Sup\'erieure, CNRS, ENS $\&$ PSL University, Sorbonne Universit\'e, Universit\'e de Paris, 75005 Paris, France}

\date{\today}

\begin{abstract}
We introduce a matrix version of the stochastic heat equation, the MSHE, and obtain its explicit invariant measure in spatial dimension $D=1$.
We show that it is classically integrable in the weak-noise regime, in terms of the matrix extension of the imaginary-time $1D$ nonlinear Schr{\"o}dinger equation which allows us to study its short-time large deviations through inverse scattering. 
The MSHE can be viewed as a continuum limit of the matrix \LGPol on the square lattice introduced recently. We also show classical integrability of that discrete model, as well as of other extensions such as of the semi-discrete matrix O'Connell-Yor polymer and the matrix strict-weak polymer. For all these models, we obtain the Lax pairs of their weak-noise regime, as well as the invariant measure, using a fluctuation--dissipation transformation on the dynamical action. 
\end{abstract}

\maketitle

{\bf Introduction}. Exponential functionals of the Brownian motion $B(t)$, are ubiquitous in both physics \cite{comtet1998exponential,majumdar2007brownian} 
and mathematics \cite{Matsumoto_Yor_2000,yor2001exponential,ChhaibDeformedDufresne}. The simplest example
is the integrated geometric Brownian motion with drift, i.e., the solution $z(t)$ of the
multiplicative noise equation
$dz =(1- \mu z) \rmd t+ z dB(t)$. It arises, for example, in diffusion in 
random media or finance, and leads to stationary inverse gamma distributions
\cite{BOUCHAUD1990285,Dufresne_1990}. Its discrete-time
version is known as Kesten's recursion, which appears in products of random matrices \cite{Kesten_1973}. 
Recently, matrix
generalisations have been studied, such as matrix multiplicative random walks or matrix diffusions.
For geometric drifted matrix Brownian motion, Wishart distributions were shown to
generalize the gamma distributions \cite{rider2016matrix}. In physics, analogous studied have been conducted in the context of Wigner-Smith time-delay matrices in mesoscopic quantum transport in disordered wires \cite{grabsch2020wigner,grabsch2015topological},
matrix Kesten recursions and their relations with interacting fermions  in a Morse potential \cite{GautieMatrixKesten},
and models of weak measurements in chaotic quantum systems
\cite{gerbino2024dyson}. Matrix Kesten recursions were also recently
studied in mathematics \cite{arista2021matsumoto}. Diffusion models have also recently played an important role in artificial intelligence, for instance, in \cite{NEURIPS2020_4c5bcfec, biroli2024dynamical}, sparking additional interest in the study of their matrix generalisation \cite{li2023spd}.

In the presence of an additional spatial dimension, scalar multiplicative noise leads to 
non-trivial space-time correlations. The paradigmatic example in the continuum setting is the stochastic heat equation (SHE),
which is related to the Kardar-Parisi-Zhang equation (KPZ), describing the random growth of interfaces
\cite{kardar1986dynamic,halpin1995kinetic,corwin2012kardar}.
In the discrete setting, multiplicative noise is realized in models of directed polymers in presence 
of random weights on a lattice \cite{seppalainen2012scaling,o2001brownian,O_Connell_2002}.
Some of these models exhibit remarkable
stochastic integrability properties \cite{Corwin_2014,borodin2013log,O_Connell_2012,borodin2014macdonald,corwin2014macdonald},
including the SHE/KPZ equation \cite{calabrese2010free,calabrese2011exact,le2012kpz,dotsenko2010bethe,dotsenko2010replica,amir2011probability,sasamoto2010one,sasamoto2010exact, sasamoto2010crossover}.
These are already manifest in the weak-noise/short-time regime, 
in the form of classical integrable structures, such as Lax pairs,  
which enable the computation of large deviation rate functions
\cite{JanasDynamical,UsWNTDroplet2021,UsWNTFlat2021,tsai2023integrability,krajenbrink2023weak}.
In the large-time limit, the exact solutions led to a very detailed mathematical description
of the universal fixed point, describing the 
1D KPZ class \cite{Matetski_2021}.

Thus, it is natural to ask about possible matrix extensions of these stochastic integrable models. Two such models were
considered very recently. Both the partition functions of the semi-discrete Brownian polymer (also called
O'Connell-Yor (OY) polymer \cite{o2001brownian,O_Connell_2002})
and of the log Gamma polymer on the square lattice \cite{seppalainen2012scaling}
were generalized to matrices in \cite{matrixOYpolymer,MatrixWhittaker2023}. The latter involves matrix-valued interacting random walks with inverse Wishart increments. Certain integrability properties, such as relations to quantum non-abelian Toda lattice, and  matrix Whittaker processes were discovered \cite{matrixOYpolymer,MatrixWhittaker2023}.
However, many properties of the scalar models have not yet been extended to their matrix counterpart, an outstanding problem. 

In this Letter, we introduce a matrix version of the 1D stochastic heat equation (MSHE).
We obtain its invariant measure, and show that its weak-noise theory is classically integrable,
hinting at full integrability beyond weak-noise. The classical integrability is obtained through a mapping to a matrix version of the
imaginary-time nonlinear Schr{\"o}dinger equation, equivalent to a matrix version of the classical Heisenberg spin chain. 
We perform the scattering analysis of the underlying matrix integrable model 
and obtain large deviation rate functions in the short time regime. We then extend our study
to the matrix log Gamma polymer, the matrix OY polymer and the matrix strict-weak polymer for which we obtain invariant measures 
and show weak-noise/classical integrability. The classical integrability of each model is characterised by a Lax pair which we provide explicitly.

\bigskip
{\bf Matrix SHE}. Consider now  a matrix version of the stochastic heat equation (MSHE) with values in ${\cal P}_d$, the set of positive definite $d \times d$ real symmetric matrices. The MSHE describes the time evolution of a space-dependent random matrix field $\mathcal{Z}=\mathcal{Z}(x,t) \in {\cal P}_d$
\be \label{matrixshe} 
     \p_t \mathcal{Z} =\p_x^2  \mathcal{Z} + 
 \mathcal{Z}^{1/2} \mathcal{W} \mathcal{Z}^{1/2} \, .
\ee 
Here, $\mathcal{W} = \mathcal{W}(x,t)$ is a random symmetric matrix 
of i.i.d. space-time Gaussian white noise, drawn from the measure 
$\sim \exp( - \frac{1}{4 g} \iint \rmd t \rmd x \Tr \mathcal{W}^2)$ and we use Ito's prescription.
We focus here on one spatial dimension $x \in \mathbb{R}$ but extensions to higher dimensions $x \in \mathbb{R}^D$ can be studied
(substituting $\rmd x=d^Dx$).
One may use the polar parametrisation of $\mathcal{P}_d$, $\mathcal{Z}= R^{-1} \Lambda R$, where $R \in {\rm SO}(d)$
and $\Lambda={\rm diag}\{\lambda_1,\dots,\lambda_d\}$ is positive diagonal  \cite{Terras_2016}. Writing the eigenvalues
as $\lambda_i=e^{h_i}$, one can view the $h_i=h_i(x,t)$ as interacting fields undergoing a growth process. In that parametrisation
we recall that $\mathcal{Z}^{1/2}= R^{-1} \Lambda^{1/2} R$.

The transition probabilities of the process \eqref{matrixshe}
can be represented by a matrix path integral $\iint {\cal D}\hat Z {\cal D} Z e^{- S[Z,\hat{Z}] }$, where the dynamical action reads \cite{conventions}  
\be
\label{eq:action-MSHE}
S[Z,\hat{Z}]  =  \iint \rmd t \rmd x \,  \Tr\big[ \hat{Z} ( \p_t Z -\p_x^2 Z ) - g
  \hat{Z} Z \hat{Z} Z   \big]
\ee
and $\hat Z$ is the matrix response field, which is also symmetric (but not necessarily positive). 
The path integral measure is proportional to the product of the space-time local Lebesgue measures ${\mathcal D} \mathcal{Z} = 
\prod_{x,t} \rmd \mathcal{Z}(x,t)$ and $\rmd \mathcal{Z} \propto \prod_{ i \leq j} \rmd  \mathcal{Z}_{ij}$. The last term arises from the noise average
$\overline{\exp( - \iint \rmd t \rmd x \Tr[ \hat Z Z^{1/2}\mathcal{W} Z^{1/2}] )}$. 

{\bf Invariant Measure of the MSHE}. A first remarkable property of the process \eqref{matrixshe}, which makes it interesting to study,
is that in $D=1$ its invariant measure can be obtained exactly, much like the KPZ equation.
Indeed, we 
show below that it is given by the following path integral measure over the matrix field $\mathcal{Z}(x)$
\be \label{stat}
{\mathcal D}_\mu \mathcal{Z}  \exp\left( - \frac{1}{2 g} \int \rmd x \Tr[ (\mathcal{Z}^{-1} \partial_x \mathcal{Z})^2 ] \right) 
\ee 
where ${\mathcal D}_\mu \mathcal{Z} = 
\prod_x \mu(d\mathcal{Z}(x))$ and $\mu(d\mathcal{Z}) \propto |\Det \mathcal{Z} |^{- \frac{d+1}{2}} \prod_{ i \leq j} d \mathcal{Z}_{ij}$, which is the natural measure on $\mathcal{P}_d$.
In the above polar parametrisation it reads 
\be
{\mathcal D}_\mu \mathcal{Z} e^{  - \frac{1}{2 g} \int \rmd x   \left( \sum_i (\partial_x \log \lambda_i)^2 + \sum_{i<j} \frac{2 (\lambda_i-\lambda_j)^2}{ \lambda_i \lambda_j} ( M_{ij} )^2
\right) } 
\ee
where $M = \partial_x R R^{-1}$, and the path integral measure is now
written with $\mu(d\mathcal{Z}) \propto  \prod_{i<j} |\lambda_i - \lambda_j| 
\prod_{ i } \lambda_i^{-\frac{d+1}{2}} \rmd \lambda_i  \rmd R$
where $\rmd R$ is the Haar measure of $\mathrm{SO}(d)$. For $d=1$ this reduces to the well known invariant measure of the 1D KPZ equation, 
$dh_1 e^{- \frac{1}{2 g} \int \rmd x (\partial_x h_1)^2}$, see 
Refs.~\cite{forster1977large,parisi1990replica,huse1985huse,bertini1997stochastic,funaki2015kpz,hairer2018strong,gu2024integration}. 
For $d=2$ it is a measure on three fields $\lambda_1,\lambda_2,\theta$
with $M_{12}=\partial_x \theta$ (the associated stochastic equation \eqref{matrixshe} is given in Section~\ref{suppmat:sec:invariant-measureMSHE}).
For $d \geq 2$, different representation of the invariant measure are obtained with different sets of coordinates, e.g. Iwasawa or polar coordinates \cite{SM}. The invariant measure \eqref{stat} has also appeared in the context of Brownian loops \cite{bruned2022geometric}.

It is crucial to point out that, as for the 1D SHE/KPZ equations, the zero mode of the process $\mathcal{Z}(x,t)$ solution
of \eqref{matrixshe} is not fixed, and $h_i(x,t) = \log \lambda_i(x,t)$
may grow unboundedly. Indeed it is easy to see that the above invariant measure \eqref{stat} is preserved by
the transformation $\mathcal{Z}(x) \to X^\intercal \mathcal{Z}(x) X$ where $X$ is any $x$-independent matrix in $GL_d(\R)$. 
This is equivalent to the invariances under $\Lambda(x) \to \Lambda(x) \Lambda_0$
and $R(x) \to R(x) R_0$ with $X=\Lambda_0^{1/2} R_0$.
So at large fixed time we can expect that, as a process in $x$
\be
\begin{split}
& \mathcal{Z}(x,t) \simlaw X(t)^{\intercal} \mathcal{Z}(x) X(t) \\
& \lambda_i(x,t) \simlaw \lambda_i(x) e^{h_i(t)}
,~ R(x,t) \simlaw  R(x) R_0(t)
\end{split}
\ee  
where ${\cal Z}(x)$ is stationary.
So only the analog of the height differences, i.e., the combinations such as $\Tr [\mathcal{Z}(x,t)^{-1}\mathcal{Z}(x',t) ]$
where the zero mode cancels, can be stationary.

To show that \eqref{stat} is an invariant measure, we construct a special time-reversal fluctuation-dissipation (FD) symmetry of the dynamical path integral. Given the time evolution between $t=t_i$ and $t=t_f$, it transforms the pair of matrix fields $\{\hat Z,Z \}$,
into the pair $\{\hat Y, Y \}$ as follows

\be 
\label{fdt00} 
\begin{split}
 Z(x,t) =& Q(x,s)^{-1}  \\
 Z(x,t) \hat Z(x,t) =& \hat Q(x,s) Q(x,s) \\
 &+ \frac{1}{g} \partial_x (Q(x,s)^{-1} \partial_x Q(x,s)) 
 \end{split}
\ee  
where $s=t_i+t_f-t$ is a time reversal transformation. Such a transformation is known
in the scalar case $d=1$. The remarkable fact is that, despite the non-commutativity
of the matrix fields, we show that the more general matrix transformation \eqref{fdt00}
preserves the dynamical action \eqref{eq:action-MSHE}, up to boundary terms. Identifying these
boundary terms, and taking particular care of the Jacobian of the FD transformation (with subtle time regularisation)
leads to the invariance of \eqref{stat}, see \cite{SM} for the derivation. 

{\bf Integrability of the MSHE}.  The second remarkable property of the process \eqref{matrixshe} is that
it is integrable in the weak-noise limit. To consider the weak-noise
limit we consider a noise strength $g \to \varepsilon g$. In the limit $\varepsilon \ll 1$ 
the typical fluctuations of the matrix field $\mathcal{Z}(x,t)$ 
become less interesting, but their large deviations remain
non trivial. Note that, as for the KPZ equation, small noise strength
is equivalent to finite noise but short time, as can be seen
by a simple rescaling \cite{SM}. To probe these large deviations we consider 
the following observable
\be 
G^\varepsilon[J] = \overline{\exp\left(  \frac{1}{\varepsilon} \iint \rmd t  \rmd x \Tr \left[J(x,t) \mathcal{Z}(x,t) \right] \right)}
\ee 
where $J(x,t)$ is a given symmetric matrix source field (assumed to vanish at $t=+\infty$)
and we recall that $\overline{\cdots}$ denotes the average over the noise. Upon rescaling 
the response field $\hat Z \to \hat Z/\varepsilon$ one finds that
this observable can
be expressed as the path integral
\be 
G^\varepsilon[J] = \iint {\cal D}\hat Z {\cal D} Z e^{- \frac{1}{\varepsilon} S_J[Z,\hat{Z}] } 
\ee 
with the modified dynamical action
\be 
S_J[Z,\hat{Z}] = S[Z,\hat{Z}] - \iint \rmd t \rmd x \Tr [ J Z ]
\ee 
In the limit $\varepsilon \ll 1$ the path integral is dominated by
a saddle point configuration for the matrix fields $\hat{Z},Z$. Taking derivatives w.r.t. these fields,
and after some algebraic manipulations \cite{SM}, we obtain
the following matrix saddle point equations 
\begin{equation} 
    \begin{split}
     \label{sp} 
 \p_t Z &=\p_x^2 Z + 2 g 
 Z \hat Z Z \\
 - \partial_t \hat{Z} &= \partial_x^2 \hat{Z} + 2 g \hat{Z} Z \hat Z  + J 
\end{split}
\end{equation} 
The equation for $Z$ must be solved forward in time starting e.g. from
a prescribed initial condition, and the one for $\hat Z$ must be solved
backward with vanishing condition at $t=+\infty$. 

For concreteness we study \eqref{matrixshe} with a deterministic initial condition ${\mathcal Z}(x,0)=Z_0(x)$,
and aim to determine the large deviation form of the probability density function (PDF) ${\cal P}(Z)$
of $Z={\cal Z}(0,t=1)$. To this aim we 
choose the source field $J(x,t) = B \delta(t-1) \delta(x) $ where $B$ is a fixed symmetric matrix. It corresponds to the observable
\bea \label{observableB} 
G^\varepsilon = \overline{\exp\left(  \frac{1}{\varepsilon} \Tr \left[B\mathcal{Z}(0,1) \right] \right)}
\eea 
This choice is equivalent to instead study the system \eqref{sp} for $t \in [0,1]$ setting $J=0$, but 
with the mixed boundary conditions
\be 
\label{eq:init-cond-wnt-kpz}
Z(x,0)= Z_0(x)  \quad , \quad \hat Z(x,1)=  B \delta(x) \, .
\ee 
Below we focus on the delta initial condition $Z_0(x)=Z_0 \delta(x)$ where $Z_0=A$ is a fixed positive definite symmetric matrix,
which is the analog of the point-to-point/droplet problem for the SHE/KPZ equation.
In that case the observable \eqref{observableB} takes the large deviation form \footnote{the KPZ case is recovered setting $g B=-z$} 
\bea \label{largedev} 
G^\varepsilon \underset{\varepsilon\to 0}{\sim} \exp\left( -\frac{1}{\varepsilon} \Psi_A(- g B) \right) 
\eea
The rate function admits a symmetry from the invariance of the equations \eqref{sp} under the transformation 
\be 
Z(x,t) \to X^\intercal Z(x,t) X \quad , \quad \hat Z \to X^{-1} \hat Z(x,t) (X^\intercal)^{-1}
\ee 
for any $X \in GL_d(\mathbb{R})$. In the case $X=A^{-1/2}$, it changes the boundary matrices
as $A\to A'=I_d$ and $B\to B'=A^{1/2} B  A^{1/2}$ and leads to $\Psi_A(-gB) = \Psi_{I_d}(-gA^{1/2}BA^{1/2})$.
Once $\Psi_A$ is known, the large deviation form of the PDF
\be 
{\cal P}(Z) \underset{\varepsilon\to 0}{\sim} e^{- \frac{1}{\varepsilon} \Phi_A(Z)} 
\ee 
can be obtained by Legendre inversion of the variational problem  $\Psi_A(-gB)=\min_{Z \in {\cal P}_d} (\Phi_A(Z) - {\rm Tr} [B Z])$. To compute
$\Psi_A$ we need to solve Eqs.~\eqref{sp} and obtain $Z(0,1)$ as a function of the source field $B$.

The remarkable fact is that the saddle point equations \eqref{sp} (with $J=0$) are integrable,
being an imaginary-time version of the matrix nonlinear Schr{\"o}dinger equation \cite{AblowitzKaup1974,ablowitz2004discrete,ablowitz2004discreteNote}. It is equivalent to the the linear
system $\partial_x \vec v= U_1 \vec v$, $\partial_t \vec v= U_2 \vec v$
where $\Vec{v}=(v_1,v_2)^\intercal$ is a two component pair of $d \times d$-matrices (depending on $x,t,k$) 
and the two Lax matrices are now block matrices of total size $2d \times 2d$ given by
\begin{equation}
\label{eq:LaxPairU}
U_1=
\begin{pmatrix}
-\I \frac{k}{2} I_d  & - g \hat Z(x,t)\\  Z(x,t) & \I  \frac{k}{2} I_d 
\end{pmatrix} \quad , \quad 
U_2= 
\begin{pmatrix}
{\sf A} & {\sf B}\\
{\sf C} & -{\sf A}'
\end{pmatrix}
\end{equation}
where $I_d$ is the $d \times d$ identity matrix and
${\sf A}= \frac{1}{2} k^2 I_d - g \hat Z Z$, ${\sf B}=g (\partial_x - \I k) \hat Z$,
${\sf C}= (\partial_x +  \I k) Z$, ${\sf A}'= \frac{1}{2} k^2 I_d - g Z \hat Z$.
One can now check, taking carefully
into account the ordering of the matrices,
that the compatibility equation $\p_t U_1-\p_x U_2 +[U_1, U_2]=0$ is indeed equivalent to
the system \eqref{sp} (with $J=0$). 

{\bf Scattering of the MSHE and Large Deviations}.
The scattering problem is now a matrix version of the standard one. Let $\Vec{v}=e^{ k^2 t/2} {\phi}$ with $ {\phi}=(\phi_1,\phi_2)^\intercal$ and $\Vec{v}=e^{- k^2 t/2} {\bar{\phi}}$
be two independent solutions of the linear problem which become plane waves 
at $x \to -\infty$, $\phi \simeq (e^{-\I k x/2}I_d,0)^\intercal$ and $\bar \phi \simeq (0,-e^{\I k x/2}I_d)^\intercal$.
Assuming from now on that $\{\hat Z,Z \}$ vanish at infinity, the $x \to +\infty$ behavior of these solutions
defines scattering amplitudes
\be \label{eq:plusinfinity} 
\phi \underset{x \to +\infty}{\simeq}
\begin{pmatrix}
a(k,t)e^{-\frac{\I  kx}{2}}\\b(k,t)e^{\frac{\I  kx}{2}}
\end{pmatrix}  ~,~
\bar{\phi} \underset{x \to +\infty}{\simeq}
\begin{pmatrix}
\tilde{b}(k,t)e^{-\frac{\I  kx}{2}}\\ -\tilde{a}(k,t)e^{\frac{\I  kx}{2}}
\end{pmatrix}
\ee 
where $a,\tilde{a},b,\tilde{b}$ are now $d\times d$ matrices. Plugging this form into the $\partial_t$ equation of the Lax pair at $x \to +\infty$,
one finds a very simple time dependence, $a(k,t)=a(k)$ and $b(k,t)=b(k)e^{-k^2 t}$, 
$\tilde a(k,t)=\tilde a(k)$ and $\tilde b(k,t)=\tilde b(k)e^{k^2 t}$. We can solve the first Lax equation $\partial_x \vec v= U_1 \vec v$ at $t=1$ and $t=0$
using the delta boundary conditions \eqref{eq:init-cond-wnt-kpz}. We obtain \cite{SM} 
\bea  
 \!\!\! \!\!\!  && \tilde b(k)= g e^{-k^2} B  ~,~ b(k)= g A \\
 && \!\!\!\!\!\!   \tilde a(k) =  I_d- g Z_+(k) B  ~,~ a(k) = I_d- g B Z_-(k) \label{solua}
\eea 
where we denote 
$Z_\pm(k)=\int_{\mathbb{R}^\pm} \rmd x \, Z(x',1)e^{- \I k x'}$. We expect $a(k)$ to be analytic in the upper-half plane $\mathbb{H}^+$ and $ \tilde{a}(k)$ in the lower-half plane $\mathbb{H}^-$. The scattering problem implies the normalisation relation
\begin{equation}
    A^{1/2}a(k)A^{-1}\tilde{a}(k) A^{1/2}=I_d-A^{1/2}BA^{1/2}e^{-k^2}
\end{equation}
which forms a matrix Riemann-Hilbert (RH) problem allowing to determine $a(k)$ and $\tilde{a}(k)$
(based on a conjecture, see \cite{SM}). Their
large $k$ expansion then provides the expression of $Z(0,1)$, using \eqref{solua}
and $Z_{\pm}(k)\underset{k\to \infty}{\simeq} \pm\frac{1}{\I k}Z(0,1)$.
 The rate function $\Psi_A$ is then readily obtained from the derivative of the Legendre transform $\Psi_A'(-gB)=g Z(0,1)$. 
We find that the RH problem decouples inside each eigenspace of the matrix $B'=A^{1/2}  B A^{1/2}$.
Denoting $b'_i$ its eigenvalues we obtain \cite{SM}
\be
\Psi_A(-gB) = \sum_{i=1}^d \Psi_{\rm KPZ}(-gb'_i)
\ee 
where $\Psi_{\rm KPZ}(z)$ is the rate function for the KPZ equation with droplet
initial condition \cite{le2016exact,UsWNTDroplet2021}, which has two branches, with main branch (no solition) $\Psi_{\rm KPZ}(z)=-\frac{1}{\sqrt{4\pi}}\mathrm{Li}_{5/2}(-z)$ and a second branch given in \eqref{eq:large-dev-generating-func-with-soliton}.
Upon Legendre inversion we finally obtain that in the large deviation regime the PDF ${\cal P}(Z)$
of $Z={\cal Z}(0,1)$ takes
a product form in terms of the eigenvalues $\lambda_i[Z_0^{-1/2} Z Z_0^{-1/2}]$
of the matrix $Z_0^{-1/2} Z Z_0^{-1/2}$
\bea 
{\cal P}(Z) \sim  \prod_{i=1}^d \, e^{ - \frac{1}{\varepsilon} \hat \Phi_{\rm KPZ}(\lambda_i[Z_0^{-1/2} Z Z_0^{-1/2}])}
\eea 
where $\hat \Phi_{\rm KPZ}(\lambda)=\Phi(H)$ is the rate function of the PDF of the partition sum $\lambda \equiv Z=e^H$ defined and obtained for the KPZ equation
in \cite{UsWNTDroplet2021} (see also 
\cite{le2016exact,tsai2023integrability}, and \cite{SM} for more details). 

To conclude on the MSHE, we expect that an explicit solution to the WNT equations \eqref{sp} can be written using the Fredholm inversion of the Fourier transform of the reflection coefficients as in \cite{UsWNTDroplet2021}, left for future work. The remainder of the paper is devoted to three discrete versions, i.e., matrix
polymer models, which we show are also integrable in the weak-noise limit.
These are useful since they allow to obtain in a more controled way the matrix SHE 
in the continuum limit. Two of these models were introduced and studied very recently by completely
different methods in the mathematical literature \cite{matrixOYpolymer,MatrixWhittaker2023}.

{\bf Matrix \LGPol}
We investigate here the matrix log-Gamma polymer, introduced recently in \cite{MatrixWhittaker2023} as a generalisation
of the scalar case. The scalar log-Gamma polymer is known to be an integrable finite temperature polymer model 
 \cite{seppalainen2012scaling,Corwin_2014,borodin2013log}.
The matrix polymer partition sums $Z_{n,m} \in {\cal P}_d$ are defined by the following stochastic recursion
\cite{conventions}, usually studied on the first quadrant of the square lattice $(n,m)\in \N^2$
\be 
\label{eq:recursion-matrix-LG}
Z_{n,m} \!= \!( Z_{n-1,m}  + Z_{n,m-1} )^{\frac{1}{2}} V_{n,m}  ( Z_{n-1,m}  + Z_{n,m-1} )^{\frac{1}{2}} 
\ee 
The random matrices $V_{n,m} \in {\mathcal P}_d$ (i.e., the noise) are i.i.d distributed with an 
inverse Wishart law from the measure (see~\eqref{eq:def-inverse-wishart-pdf})
\be
\label{eq:def-inverse-wishart-pdf-main}
\rmd P^{iW}_{\alpha,g}[V]= \frac{(2g)^{-\alpha d}}{\Gamma_d(\alpha) } |\det V|^{-\alpha  } e^{- \frac{1}{2g} \Tr \, V^{-1} } \mu(\rmd V)
\ee
We choose $g=1/2$ and use the short-hand notation $\rmd P^{iW}_{\alpha,g=1/2}[V] =\rmd P^{iW}_\alpha[V]$.
Note the normalisability condition $\alpha > \frac{d-1}{2}$. One can check that $\Tr [Z_{n,m}]$ can be written as 
a sum over up-right paths on the lattice, terminating in $(n,m)$, of traces of ordered products 
of the $V_{i,j}$ matrices. For $d=1$ it recovers the standard polymer partition sum with inverse gamma 
distributed random Boltzmann weights.

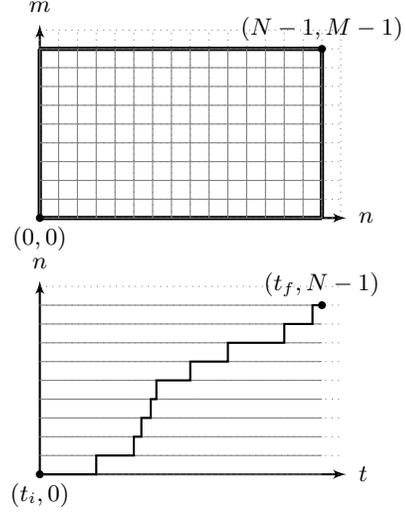
\begin{figure}[t!]
\begin{tikzpicture}[scale=0.25]

%\fill[gray!50] (0,-0.2) rectangle (15.2,0.2);
%\fill[gray!50] (-0.2,-0.2) rectangle (0.2,9);

%\fill[pattern=north west lines, pattern color=black] (0.8,8.8) rectangle (15,9.2);
%\fill[pattern=north west lines, pattern color=black] (14.8,0.8) rectangle (15.2,9.2);
% 
%% axis
\draw[->, thick,>=latex'] (0, 0) -- (0, 10.3);
\draw[->, thick,>=latex'] (0,0)--( 16.3, 0);

%% Light cone for z
\draw[ultra thick] (0,0) -- (0,9) ;
\draw[ultra thick] (0,0) -- (15,0) ;

%% Light cone for ztilde
\draw[ultra thick] (0,9) -- (15,9) node[midway, above, yshift=0.1cm] {};
\draw[ultra thick] (15,0) -- (15,9) node[midway, right, xshift=0.1cm] {};

\foreach \k in {0,1, ..., 16}
	{\draw[gray, dotted] (\k, 0) -- (\k, 10.1);}
\foreach \k in {0,1, ..., 10}
	{\draw[gray, dotted] (0,\k) -- (16.1, \k);}

\clip (-2, -2) rectangle (19.5, 11.5);

\foreach \k in { 0,1, ..., 9}
	{\draw[gray] (0,\k) -- (15, \k);}
\foreach \k in{ 0, 1, ..., 15}
	{\draw[gray] (\k, 0) -- (\k,9);}

 \draw(15,9) node[anchor=south]{$(N-1, M-1)$};
 \fill (15,9) circle(6pt);

 \draw(0,0) node[anchor=north]{$(0, 0)$};
 \fill (0,0) circle(6pt);
% \draw(15,0) node[anchor=north]{$(N-1, 0)$};
% \fill (15,0) circle(6pt);
% \draw(0,9) node[anchor=south west]{$(0, M-1)$};
% \fill (0,9) circle(6pt);

\node[above] at (0,10.5) {$m$};
\node[right] at (16.5,0) {$n$};

\end{tikzpicture}\hfill
\begin{tikzpicture}[scale=0.25]

%\fill[gray!50] (0,-0.2) rectangle (15.2,0.2);
%\fill[gray!50] (-0.2,-0.2) rectangle (0.2,9);

%\fill[pattern=north west lines, pattern color=black] (0.8,8.8) rectangle (15,9.2);
%\fill[pattern=north west lines, pattern color=black] (14.8,0.8) rectangle (15.2,9.2);
% 
%% axis
\draw[->, thick,>=latex'] (0, 0) -- (0, 10.3);
\draw[->, thick,>=latex'] (0,0)--( 16.3, 0);

%% Light cone for z
%\draw[ultra thick] (0,0) -- (0,9) ;
%\draw[ultra thick] (0,0) -- (15,0) ;

%% Light cone for ztilde
%\draw[ultra thick] (0,9) -- (15,9) node[midway, above, yshift=0.1cm] {};
%\draw[ultra thick] (15,0) -- (15,9) node[midway, right, xshift=0.1cm] {};

%\foreach \k in {0,1, ..., 16}
%	{\draw[gray, dotted] (\k, 0) -- (\k, 10.1);}
\foreach \k in {0,1, ..., 10}
	{\draw[gray, dotted] (0,\k) -- (16.1, \k);}

\clip (-2, -2) rectangle (19.5, 11.5);

\foreach \k in { 0,1, ..., 9}
	{\draw[gray] (0,\k) -- (15, \k);}
%\foreach \k in{ 0, 1, ..., 15}
%	{\draw[gray] (\k, 0) -- (\k,9);}

\draw[thick] (0,0) -- (3,0) -- (3,1) -- (5,1) -- (5,2) -- (5.4,2) -- (5.4,3) -- (5.9,3) -- (5.9,4) -- (6.2,4) -- (6.2,5) -- (8,5) -- (8,6) -- (10,6) -- (10,7) -- (13,7) -- (13,8) -- (14,8) -- (14.5,8) -- (14.5,9) -- (15,9) ;

 \draw(15,9) node[anchor=south]{$(t_f, N-1)$};
 \fill (15,9) circle(6pt);

 \draw(0,0) node[anchor=north]{$(t_i, 0)$};
 \fill (0,0) circle(6pt);
% \draw(15,0) node[anchor=north]{$(N-1, 0)$};
% \fill (15,0) circle(6pt);
% \draw(0,9) node[anchor=south west]{$(0, M-1)$};
% \fill (0,9) circle(6pt);

\node[above] at (0,10.5) {$n$};
\node[right] at (16.5,0) {$t$};

\end{tikzpicture}
\caption{Geometry for the \LGPol and the OY polymer, see text.}
\label{fig:loggamma-polymer-lightcone-wnt-main-2}
\end{figure}

{\bf Invariant Measure of the Matrix Log-Gamma Polymer}. For the scalar log-Gamma polymer the stationary setting was obtained in \cite{seppalainen2012scaling}. 
It was shown that when the system is stationary then for any down-right path on the lattice 
the successive partition sum ratios on adjacent sites are distributed as i.i.d inverse gamma random variables. 
Here, we have obtained a family of invariant measures for the matrix process \eqref{eq:recursion-matrix-LG}. 
The ratios are now distributed as i.i.d inverse Wishart random matrices. However one can
define matrix ratios in two different ways so we introduce the notation
\begin{equation}
\label{eq:definition-convention-ratios-main}
   r\left(\frac{A}{B}\right)=B^{-\frac{1}{2}} A B^{-\frac{1}{2}}, \quad  \tilde{r}\left(\frac{A}{B}\right)=A^{\frac{1}{2}} B^{-1} A^{\frac{1}{2}}
\end{equation}
Consider any rectangle $0 \leq n \leq N-1$, $0 \leq m \leq M-1$, see Fig.~\ref{fig:loggamma-polymer-lightcone-wnt-main-2}. We have shown that if the
partition sums on the "initial boundary" (i.e., the down right path $(0,M-1) \to (0,0) \to (N-1,0)$)
are distributed according to the
measure 
\bea  \label{initialM} 
&& \mu(\rmd Z_{0,M-1}) \prod_{m=1}^{M-1}\rmd P^{iW}_{\alpha(1-\kappa)}[\tilde{r}\left(\frac{Z_{0,m}}{Z_{0,m-1}} \right)] \\
&& ~~~~~~~~~~~~~~~~~ \times \prod_{n=1}^{N-1} \rmd P^{iW}_{\alpha \kappa}[r\left(\frac{Z_{n,0}}{Z_{n-1,0}} \right)] \nn 
\eea 
then the partition sums on the "final boundary" (i.e., the down right path $(0,M-1) \to (N-1,M-1) \to (N-1,0)$)
are distributed according to the same measure, i.e., 
\bea  \label{finalM} 
&& \mu(\rmd Z_{0,M-1}) \prod_{m=1}^{M-1}\rmd P^{iW}_{\alpha (1-\kappa)}[\tilde{r}\left(\frac{Z_{N-1,m}}{Z_{N-1,m-1}} \right)] \\
&& ~~~~~~~~~~~~~~~~~ \times \prod_{n=1}^{N-1}\rmd P^{iW}_{\alpha \kappa}[r\left(\frac{Z_{n,M-1}}{Z_{n-1,M-1}} \right)] \nn 
\eea  
The same measure means that in both cases (i) all ratios are sampled independently, (ii) horizontal ratios are defined with the matrix ordering $r$, see \eqref{eq:definition-convention-ratios-main}, 
    and sampled from the inverse Wishart distribution with parameter $\kappa \alpha$, (iii) vertical ratios are defined with the matrix ordering $\tilde{r}$ and sampled from the inverse Wishart distribution with parameter $(1-\kappa)\alpha$, (iv) the zero mode $Z_{0,M-1}$ is common and independently and uniformly distributed over $\mathcal{P}_d$.
Note that this gives a family of invariant measures parametrized by $\kappa$, but that
the measures on the ratios are normalisable only when $\alpha > d-1$ and for 
$\kappa \in ]\frac{d-1}{2\alpha}, 1-\frac{d-1}{2\alpha}[$.

To obtain this invariant measure we have used that the average over the noise of any observable 
${\cal O}(Z)$ of the $Z_{n,m}$ solutions of \eqref{eq:recursion-matrix-LG} 
can be represented by a multiple matrix integral
$\overline{{\cal O}(Z)} = \iiint \mathcal{D}_\mu V\mathcal{D} Z\mathcal{D} \hat{Z} {\cal O}(Z) e^{-S_0[Z,\hat{Z},V]}$ 
where the action $S_0[Z,\hat{Z},V]$ is given in \eqref{eq:app-log-gamma-action-three-fields}
and the integration measures on the fields is
$\mathcal{D}_\mu {V}\mathcal{D} {Z}\mathcal{D} {\hat{Z}}
= \prod_{n,m} \mu(\rmd {V}_{n,m}) \rmd {Z}_{n,m} \rmd {\hat{Z}}_{n,m}$.
Next we constructed a "time reversal" FD symmetry on the fields which leaves
$S_0$ invariant up to boundary terms, a discrete analog of \eqref{fdt00}. 
Identifying these
boundary terms, and taking care of the Jacobian of the FD transformation 
leads to the above invariant measure \eqref{initialM}--\eqref{finalM},
see \cite{SM} for the derivation. Note that for $d=1$
this leads to an independent derivation of the stationarity of the scalar
\LGPol\!, quite different from the probabilistic one
given in \cite{seppalainen2012scaling}.

{\bf Integrability of the Matrix \LGPol \!\!.}
We now show the integrability of the matrix \LGPol
in the weak-noise limit. The weak-noise limit of the scalar \LGPol is treated in more details in a companion paper \cite{us_scalar}. 
It is a further discretization of the weak-noise limit study of the scalar OY polymer in \cite{krajenbrink2023weak}.
It is identified as the limit $\alpha \gg 1$. We will study the 
large deviations in that limit, which retain the full lattice character (while by contrast 
the typical behavior converges upon proper space time rescaling to the continuum model). 
We perform the rescaling $V_{n,m} \to \alpha^{-1} V_{n,m}$ 
and $Z_{n,m} \to  \alpha^{1-n-m} Z_{n,m}$ which leaves the recursion \eqref{eq:recursion-matrix-LG} 
unchanged. We consider the observable (expressed in terms of these rescaled matrix fields) 
\be 
G^\alpha[J] = \overline{\exp\left( \alpha \sum_{n,m} \Tr \left[ J_{n,m} Z_{n,m} \right] \right)}
\ee 
where $J$ is a given source field. Upon rescaling the response field $\hat{Z}_{n,m}\to \alpha^{n+m}\hat{Z}_{n,m}$, in the matrix multiple 
integral mentioned above, this observable can be expressed as a multiple matrix integral
\begin{equation}
    G^\alpha[J] = \iiint \mathcal{D}_\mu {V}\mathcal{D} {Z}\mathcal{D} \hat{Z} e^{-\alpha S_J[Z,\hat{Z},V]}
\end{equation}
where $S_J = S_0 - \sum_{n,m} \Tr [ J_{n,m} Z_{n,m} ]$. For large $\alpha \gg 1$, i.e., a small variance for the noise, the multiple integral is dominated by its saddle point which yields a deterministic nonlinear system
for the fields $Z,\hat Z, V$. Upon eliminating $V$ and introducing a modified response field as $Y_{n,m}= \hat{Z}_{n,m}(I_d-Z_{n,m}\hat{Z}_{n,m})^{-1}$, we find that the saddle point equations simplify into (see \cite{SM} for details)
\begin{equation}
\label{eq:WNT-matrix-LG}
    \begin{split}
           Z_{n,m}(I_d+Y_{n,m}Z_{n,m})^{-1}&=Z_{n-1,m}  + Z_{n,m-1}     \\
      (I_d+Y_{n,m}Z_{n,m})^{-1}Y_{n,m}&=Y_{n+1,m}  + Y_{n,m+1} + J_{n,m} 
    \end{split}
\end{equation}
We now show that this matrix discrete nonlinear system is integrable by exhibiting 
two explicit Lax matrices, which are block matrices of total size $2d \times 2d$ 
\begin{equation}
\label{eq:LaxPair-matrix-LG-1}
    L_{n,m}
    \!=\!
    \begin{pmatrix}
         I_d & 0 \\
    -Y_{n,m} & I_d \\
    \end{pmatrix}\!
    \begin{pmatrix}
         \frac{I_d}{\lambda} & 0 \\
 0 & -\lambda I_d  \\
    \end{pmatrix}\!
    \begin{pmatrix}
         I_d &  Z_{n,m-1} \\
 0 & I_d \\
    \end{pmatrix}
\end{equation}
and
\begin{equation}
\label{eq:LaxPair-matrix-LG-2}
    U_{n,m}
    \!=\!\begin{pmatrix}
         I_d & 0 \\
    Y_{n,m} & I_d \\
    \end{pmatrix} \!\!
    \begin{pmatrix}
         \frac{I_d}{\sqrt{\lambda^2+1} } & 0 \\
 0 & {\scriptstyle\sqrt{\lambda^2+1 }}I_d  \\
    \end{pmatrix}\!\!
    \begin{pmatrix}
         I_d & - Z_{n-1,m} \\
 0 & I_d \\
    \end{pmatrix}
\end{equation}
so that the compatibility equation $L_{n,m+1} U_{n,m} =U_{n+1,m} L_{n,m}$ 
is equivalent to the system \eqref{eq:WNT-matrix-LG} (for $J=0$). This can
be checked explicitly noting that $Z$ and $Y$ matrices alternate in order in
the products. Here $\lambda$ is
the (complex) spectral parameter, and the discrete nonlinear system \eqref{eq:WNT-matrix-LG} is thus equivalent to the linear
system
$\vec v_{n+1,m}=L_{n,m} \vec v_{n,m}$ and $\vec v_{n,m+1}=U_{n,m} \vec v_{n.m}$,
where $\Vec{v}$ is a two component pair of $d \times d$-matrices (depending on $n,m,\lambda$).
Scattering can be performed, as for the MSHE, e.g. to compute 
observables such as $\overline{e^{- \alpha \Tr[B Z_{N-1,M-1}}]}$ for the
point-to-point polymer,  but is left for future studies. 
\\

{\bf Matrix O'Connell-Yor Polymer}. The (scalar) 
OY polymer was introduced and studied in \cite{o2001brownian,O_Connell_2002,O_Connell_2012,borodin2014macdonald}.
A matrix generalisation was introduced and studied recently in \cite{matrixOYpolymer}.  Here, we consider the matrix partition sums $Z_{n}(t) \in {\cal P}_d$ which evolve as a stochastic
process discrete in space and continuous in time (with Ito discretization) as
\be \label{recursionOY-main}
 \p_t Z_{n} = Z_{n-1} - Z_{n} +
 Z_{n}^{1/2} W_{n}(t) Z_{n}^{1/2}
\ee 
where $W_{n,t}$ is $d\times d$ real symmetric Gaussian centered noise matrix
with correlator $\overline{[W_{n}(t)]_{ij}[W_{n'}(t')]_{kl}} = g (\delta_{ik}\delta_{jl}+\delta_{il}\delta_{jk})\delta_{nn'}\delta(t-t')$. For $d=1$ one recovers the scalar OY polymer. 
An example of initial condition is the droplet initial condition 
$Z_n(t_i)=\delta_{n,1} \,  I_d$, $Z_0(t)=0$ for all $t \geq t_i$, for which 
$Z_{N-1}(t_f)$ is the point-to-point OY partition function, 
from $(0,t_i)$ to $(N-1,t_f)$. 

{\bf Invariant Measure of the Matrix O'Connell-Yor Polymer}. To
describe the invariant measure \footnote{In this section we first change $Z_n \to e^{-t} Z_n$ so 
that we actually study the recursion \eqref{recursionOY-main} with the diagonal term $- \mathcal{Z}_{n,t}$ absent.
This amounts to a change in $\kappa$.}, 
consider a more general class of initial condition, which
specifies $Z_1(t),\dots,Z_{N-1}(t)$
at $t=t_i$ together with $Z_0(t)$ for $t \in [t_i,t_f]$, see Fig.~\ref{fig:loggamma-polymer-lightcone-wnt-main-2}. We obtain that
the following measure on the initial condition 
\bea  
\label{eq:invariant-measure-YO-compressed-1-main}
&& \prod_{n=1}^{N-1}\rmd P^{iW}_{\kappa/(2g),g}[ \,  r\left(\frac{Z_{n}(t_i)}{Z_{n-1}(t_i)} \right)]   \\
&& \times \prod_{t=t_i}^{t_f} \mu(\rmd Z_{0}(t)) 
 e^{- \frac{1}{4g} \int_{t_i}^{t_f} \rmd t \, \Tr \big[  (Z_{0}^{-1} \partial_t Z_{0}-\kappa)^2\big]} \nn 
\eea  
is invariant in the sense that it leads to the same measure 
for the final values, i.e., with the replacements $t_i \to t_f$ in the first line of \eqref{eq:invariant-measure-YO-compressed-1-main}
and $Z_0(t) \to Z_{N-1}(t)$ in the second line of \eqref{eq:invariant-measure-YO-compressed-1-main}.
It is a family of invariant measures indexed by $\kappa$ and such that (i) the ratios $\frac{Z_{n}(t)}{Z_{n-1}(t)}$
defined with the matrix ordering $r$, see \eqref{eq:definition-convention-ratios-main}, are i.i.d. with 
Inverse Wishart distribution \eqref{eq:def-inverse-wishart-pdf-main} of parameters $(\kappa /(2g),g)$ (which is normalisable iff $\kappa/(2g)> \frac{d-1}{2} $),
(ii) the processes on the first line $Z_{0}(t)$, and on the last line $Z_{N-1}(t)$, $t \in [t_i,t_f]$ 
are matrix geometric Brownian motions with drift $\kappa \times I_d$
    and diffusion coefficient $2 g$. 
Again this is shown in \cite{SM} by constructing a FD symmetry on the dynamical action associated to the matrix OY polymer. 
For $d=1$ we recover known results by a different method \cite{Imamura_2017,SpohnStationaryYO,YOStationaryScalarValko}. The condition on drift generalizes the $d=1$ case. Note that \cite[Section~3]{matrixOYpolymer} a probabilistic
construction of an invariant measure is also given.

{\bf Integrability of the Matrix O'Connell-Yor Polymer}.
We show the integrability of the matrix OY polymer
in the weak-noise limit. For the scalar OY 
polymer this was done in \cite{krajenbrink2023weak}. Let us replace $g \to \varepsilon g$ 
in \eqref{recursionOY-main} and consider the following observable 
$G^\varepsilon[J]=\overline{e^{  \frac{1}{\varepsilon} \int \rmd t  \sum_n  \, \Tr \left[J_n(t) Z_n(t) \right] }}$.
It admits again a path integral representation and upon rescaling of the response field $\hat{Z} \to \hat{Z}/\varepsilon$
one obtains 
$G^\varepsilon[J]= \iint \mathcal{D}{\hat{Z}}\mathcal{D}Z \,  e^{-\frac{1}{\varepsilon} S_J[Z,\hat{Z}]}$
where the action $S_J[Z,\hat{Z}]$ is given in \eqref{actionOY}. 
In the weak-noise limit $\varepsilon \ll 1$, the path integral is dominated by its saddle point,
i.e., a configuration of the fields 
$Z_n(t),\hat Z_n(t)$
which obeys the nonlinear matrix system 
\begin{equation}
    \begin{split} \label{systemOY-main} 
        \p_t Z_n &= Z_{n-1}-Z_n +2g Z_n \hat{Z}_n Z_n \\
        -\p_t \hat{Z}_n &=\hat{Z}_{n+1}-\hat{Z}_n+2g \hat{Z}_n Z_n \hat{Z}_n  \\
    \end{split}
\end{equation}
where $J_n(t)$ must be added to the r.h.s. of the second equation when a source is present.
This is a discretization of the MSHE saddle point equations \eqref{sp}, which turns out 
again to be integrable, providing an (imaginary-time) integrable discretization of the 
matrix Schr{\"o}dinger equation. We have found Lax matrices such that \eqref{systemOY-main} 
implies the semi-discrete compatibility equation $\p_t L_n=U_{n+1}L_n -L_n U_n$
which ensures equivalence to the linear system $\vec v_{n+1}= L_{n} \vec v_{n}$, $\partial_t \vec v_{n}= U_{n} \vec v_{n}$.
Their explicit forms are (time dependence is implicit) 
    \begin{equation}
    \label{eq:supp-mat-factorisation-lax-Ln}
        L_{n} = \begin{pmatrix}
I_d  &   0 \\ 
-2g\hat{Z}_n &  I_d
\end{pmatrix}
\begin{pmatrix}
\frac{I_d}{\lambda} &   0 \\ 
0 &  \lambda I_d
\end{pmatrix}
\begin{pmatrix}
I_d  &   Z_{n} \\ 
0 &  I_d
\end{pmatrix}
    \end{equation}
and 
    \begin{equation} \label{UOY} 
U_{n}=
\begin{pmatrix}
\frac{\lambda^2-1}{2}I_d  &   -Z_{n-1} \\ 
&\\
2g\hat{Z}_n &  \frac{1-\lambda^2}{2}I_d
\end{pmatrix}
\end{equation}
From there one can study the scattering problem for the linear system, as we did above for the MSHE,
along the lines of what was done for the
scalar case in \cite{krajenbrink2023weak}. This is left for future study.

{\bf Matrix Strict-Weak Polymer}. Finally, for completeness we also briefly discuss a matrix version of the so-called
"strict-weak polymer" (whose scalar version was introduced and studied in 
\cite{o2015tracy,corwin2015strict}) introduced in \cite{MatrixWhittaker2023}. 
We define it by the following recursion for the matrices $Z_{n,t} \in {\cal P}_d$ on the square lattice $(n,t) \in \mathbb{Z}^2$
\begin{equation}  
\label{eq:recursion-matrix-SW}
Z_{n,t+1}=Z_{n,t}^{\frac{1}{2}}Y_{n,t}Z_{n,t}^{\frac{1}{2}} +Z_{n-1,t}
\end{equation}
where the matrices $Y_{n,t} \in {\cal P}_d$ are i.i.d. Wishart distributed, from the measure 
 $\sim |\det Y|^{\alpha} e^{-  \Tr \, Y } \mu(dY)$, with $\alpha>(d-1)/2$.  One can again associate to the random recursion of the strict-weak polymer a dynamical action, given in two equivalent forms in \eqref{eq:app-strict-weak-action-three-fields} 
and \eqref{eq:app-strict-weak-action-two-fields}.

In the limit of large $\alpha$, i.e., a small variance for the noise, this action is dominated by its saddle point which yields the following deterministic nonlinear matrix system
\begin{equation}
\label{eq:WNT-matrix-SW}
    \begin{split}
Z_{n,t+1}&=Z_{n,t}(I_d-\hat{Z}_{n,t}Z_{n,t})^{-1}+Z_{n-1,t}\\
\hat{Z}_{n,t-1}&=\hat{Z}_{n,t}(I_d-Z_{n,t}\hat{Z}_{n,t})^{-1}+\hat{Z}_{n+1,t}
    \end{split}
\end{equation}
We found a Lax pair which obeys the compatibility equation, defined as $L_{n,t+1} U_{n,t} =U_{n+1,t} L_{n,t}$,
with the explicit form for the matrices 
\begin{equation}
\label{eq:LaxPair-matrix-SW-1}
\begin{split}
    L_{n,t}&=\begin{pmatrix}
         I_d & 0 \\
 -  \hat{Z}_{n,t-1} & I_d \\
    \end{pmatrix}
    \begin{pmatrix}
         \frac{I_d}{\lambda } & 0 \\
 0 & \lambda I_d  \\
    \end{pmatrix}
    \begin{pmatrix}
         I_d & Z_{n,t} \\
 0 & I_d \\
    \end{pmatrix}
\end{split}
\end{equation}

and 
\begin{equation}
\label{eq:LaxPair-matrix-SW-2}
    U_{n,t}\!=\!\begin{pmatrix}
         I_d & -Z_{n-1,t} \\
 0 & I_d \\
    \end{pmatrix}\!\!
    \begin{pmatrix}
         \sqrt{1 +\lambda ^2}I_d & 0 \\
 0 & \frac{I_d}{\sqrt{1 +\lambda ^2}}  \\
    \end{pmatrix} \!\!
    \begin{pmatrix}
         I_d & 0 \\
 \hat{Z}_{n,t} & I_d \\
    \end{pmatrix}
\end{equation}

We also found an FD symmetry \cite{SM} pointing towards a stationary measure for the matrix strict-weak polymer consisting in a family of independent Wishart 
and inverse Wishart consecutive ratios parametrized by a continuous parameter. This generalizes the result for the 
$d=1$ case \cite{corwin2015strict}.

{\bf Conclusion and Outlook}. In summary, we have considered several stochastic growth models, either fully discrete (log Gamma and Strict-Weak polymers), semi-discrete (OY polymer), or fully continuous (SHE/KPZ equations), and studied their generalisations to $d \times d$ positive definite symmetric matrices.
For each model we have obtained the invariant measure, using a matrix MSR field theoretical method. This required
to unveil a FD time reversal transformation for each model and to treat carefully the Jacobians of this
transformation. Even for the scalar case $d=1$, where the invariant measures were known from quite
different probabilistic methods, this goes beyond what was done previously using MSR. Next we
identified and studied a weak-noise limit for each of these models, such that the large deviations
are described by the "classical limit" of their associated dynamical action. The resulting
saddle point equations provide deterministic matrix nonlinear difference or differential systems,
which we show are integrable. 
This was achieved by exhibiting a Lax pair in each case. In the case of the MSHE we performed the scattering
analysis and computed explicitly large deviation rate functions, but the same can be
done for each of these models. 
Even for
$d=1$ these weak-noise results (the nonlinear systems, their classical integrability and their Lax pairs)
are new in the case of the fully discrete polymers, and also discussed, together with further extensions, 
in a companion paper \cite{us_scalar}. 

It is important to note that one can take a continuum limit from the fully discrete models to
semi-discrete and then to the MSHE, as was done in the scalar case
(see e.g. \cite{borodin2014macdonald}), 
but we have not worked it out in details here. In that limit the 
Wishart and inverse Wishart ensembles, which play an important role here,
become Gaussian ensembles. The inverse Wishart product measures over ratios of matrix  partition sums become geometric matrix Brownian motions.  It may be possible to generalise the invariant measures obtained here to a periodic geometry (where some boundary terms disappear)
using conditioning, as in the scalar case \footnote{I.~Corwin, private communication. See also \cite{CorwinPitmanPeriodic} }.

Although we have focused here on the "classical" (i.e., weak-noise)
integrability, our results point to a larger full "quantum" (any noise) integrability
for all of these models. For the MSHE the matrix nonlinear Schr{\"o}dinger field theory is the natural candidate to generalize the scalar version, which is equivalent to the delta Bose gas. 
For the matrix \LGPol an integrable structure was unveiled  recently by introducing matrix generalisation of Whittaker processes, as
Markov processes on triangular arrays of matrices. By focusing on one side of the triangle, it allows to obtain a characterization of the fixed-time matrix log Gamma and matrix strict-weak partition sum for analogs of the point-to-point polymer.
However, performing actual explicit calculations of observables using this full integrability remains an open challenge.

It is worth noting that there are known connections between a classical limit of the scalar OY polymer 
and the Toda lattice \cite{o2013geometric,O_Connell_2012}, which were recently 
extended to the matrix version (leading to the non-abelian Toda lattice),
see \cite[Section~8]{matrixOYpolymer}. Interestingly, the Toda lattice 
also appears 
as some degeneration of the weak-noise theory of the scalar OY polymer \cite{krajenbrink2023weak}, and
it would be interesting to understand the connections, and
their matrix extensions.

Although we have focused this work on the $\beta=1$ MSHE,  similar methods can be applied to
construct an integrable $\beta=2$ MSHE, i.e., on the space of Hermitian positive definite matrices
(see \cite{GautieMatrixKesten} for a $D=0$ version which leads to a free fermionic representation for $\beta=2$). 

Application-wise, the matrix polymers and MSHE studied in this work could be used to investigate further applications of matrix data in $\mathcal{P}_d$. We refer to \cite{SDP-application,SDP-application2} and reference therein for machine learning algorithms applications in computer vision related e.g. to diffusion tensor imaging and functional MRI. One could additionally apply the polymer models studied in this work in the context of the diffusion of covariance matrices for generative processes.

We finally note an upcoming work on stationary measures for the matrix log Gamma polymer
using different methods \cite{BarraquandInPreparation}.

\bigskip

\begin{acknowledgments}
\paragraph{Acknowledgments.}  
We thank G.~Barraquand, A.~Borodin, I.~Corwin, H.~Desiraju, M.~Hairer and N.~O'Connell for discussions, as well as J.~P.~Bouchaud and T.~Gauti\'e for earlier collaborations on related topics. 
PLD acknowledges support from  ANR grant ANR-23-CE30-0020-01 EDIPS. We acknowledge support from MIT-France MISTI Global Seed Funds project “Exact Solutions in Field Theories via Integrable Probability” and the MIT Mathematics department for hospitality.
\end{acknowledgments}

\bibliography{discrete-wnt.bib}

\newpage

%%%%%%%%%%% Merge with supplemental materials %%%%%%%%%%
%\pagebreak
%%\widetext
%\begin{widetext} 
%\begin{center}
%\textbf{\large Supplemental Materials: Title for main text}
%\end{center}
%%%%%%%%%%% Merge with supplemental materials %%%%%%%%%%
%%%%%%%%%%% Prefix a "S" to all equations, figures, tables and reset the counter %%%%%%%%%%
%\setcounter{equation}{0}
%\setcounter{figure}{0}
%\setcounter{table}{0}
%\setcounter{page}{1}
\makeatletter
\renewcommand{\theequation}{S\arabic{equation}}
\renewcommand{\thefigure}{S\arabic{figure}}
%\renewcommand{\bibnumfmt}[1]{[S#1]}
%\renewcommand{\citenumfont}[1]{S#1}
%%%%%%%%%%% Prefix a "S" to all equations, figures, tables and reset the counter %%%%%%%%%%

\setcounter{section}{0}

\setcounter{secnumdepth}{2}

\begin{widetext} 

\begin{large}
\begin{center}

Supplementary Material for\\  {\it Integrable Matrix Probabilistic  Diffusions and the Matrix Stochastic Heat Equation  }

\end{center}
\end{large}

We give the principal details of the calculations described in the main text of the Letter. 
We also give additional information about the results displayed in the text.

{\hypersetup{linkcolor=black}
\setcounter{tocdepth}{1}
\tableofcontents
}

\section{Compendium of useful properties of random positive symmetric matrices}

Let $\mathcal{P}_d$ be the set of $d \times d$ real symmetric matrices with positive eigenvalues. 
For a detailed reference on the properties of these matrices, see \cite{Terras_2016,dolcetti2018differential}.
It is a non-compact symmetric space (Riemannian manifold with an inversion isometry). 
The Lebesgue measure for symmetric matrices is defined as
\begin{equation}
\label{eq:measure-symmetric-matrices}
    dY=\prod_{i \leq j} dY_{ij}    
\end{equation}
and the natural measure for $Y\in {\cal P}_d$ is
\begin{equation}
\label{eq:measure-psd}
\mu(dY)= |\det Y|^{- \frac{d+1}{2} } dY
\end{equation}

Such a measure is invariant under the action of the group of $d\times d$ invertible real matrices $GL_d(\R)$, i.e., the transformation $Y\to X^\intercal Y X$ for $X\in GL_d(\R)$ preserves the measure, i.e., $\mu(dY)=\mu(d  X^\intercal Y X)$. This can be seen using the identity
\begin{equation} \label{detg}
    \Det \left| \frac{\delta (g Yg^\intercal)}{\delta Y}\right|=|\det \, g|^{d+1}
\end{equation}
Furthermore the measure $\mu$ is invariant the inversion of $Y \to Y^{-1}$, i.e., $\mu(dY)=\mu(d Y^{-1})$. 
The inversion invariance is a consequence of the formula
\begin{equation} \label{detinverse} 
    Z=Y^{-1}, \quad \det \left|\frac{\delta Z}{\delta Y}\right|=|\det Y|^{-(d+1)}
\end{equation}
The first order perturbation of the square root of a positive matrix is
\be
\label{eq:taylor-sqrt-matrix}
(Z + \rmd Z)^{1/2} - Z^{1/2} = \int_0^{+\infty}\rmd t \,  e^{- t Z^{1/2} } \rmd Z e^{- t Z^{1/2} } + \mathcal{O}(\rmd Z^2) 
\ee

The following identity will be used throughout the manipulation of various matrix models. At the first order in $\mathcal{O}(\rmd Z)$ we have for any matrices $Z,\hat{Z},V$
\be
\label{eq:saddle-point-sqrt}
\begin{split}
  \Tr [\hat{Z} \sqrt{Z + \rmd Z} V \sqrt{Z + \rmd Z} ]&=  \int_0^{+\infty} \rmd t \, \Tr [\hat{Z} e^{- t \sqrt{Z} } \rmd Z e^{- t \sqrt{Z} } V \sqrt{Z}]
+ \int_0^{+\infty} \rmd t \, \Tr [\hat{Z} \sqrt{Z} V e^{- t \sqrt{Z} } \rmd Z e^{- t \sqrt{Z} }]  \\
& =  \int_0^{+\infty} \rmd t \, \Tr [e^{- t \sqrt{Z} } V \sqrt{Z} \hat{Z} e^{- t \sqrt{Z} } \rmd Z]  + \int_0^{+\infty} \rmd t \, \Tr [e^{- t \sqrt{Z} } \hat{Z} \sqrt{Z} V e^{- t \sqrt{Z} } \rmd Z ]
\end{split}
\ee  
To take derivatives with respect to the fields we set the variation of the action to zero and use that for any
symmetric matrix $A$
\be 
\Tr[ A \, \rmd Z ] = 0  \quad \text{for any real symmetric matrix} \, \rmd Z \Rightarrow \quad A=0
\ee 
We will use the following identity from \cite{matrixOYpolymer} valid for $X,Y,A \in {\mathcal P}_d$
\begin{equation}
\label{eq:inversion-identity}
    Y=AXA \quad \Longleftrightarrow  \quad A=X^{-1/2}(X^{1/2}YX^{1/2})^{1/2} X^{-1/2}
\end{equation}
We generally refer to Ref.~\cite{matrixcookbook} for a number of algebraic operations on matrix equations.
\subsection{Parametrisation of positive symmetric matrices}

Any positive symmetric matrix $Z \in \mathcal{P}_d$ admits a partial Iwasawa decomposition
\begin{equation}
\label{eq:iwasawa-coordinates}
    Z = 
    \begin{pmatrix}
        I_{d-1} & 0 \\
        y^\intercal & 1
    \end{pmatrix}
    \begin{pmatrix}
        V & 0 \\
        0 & w
    \end{pmatrix}
    \begin{pmatrix}
        I_{d-1} & y \\
        0 & 1
    \end{pmatrix}
\end{equation}
where $V\in \mathcal{P}_{d-1}$, $w>0$, $y \in \R^{d-1}$. In the Iwasawa coordinates, the natural measure \eqref{eq:measure-psd} reads
\begin{equation}
    \mu(dZ)=w^{\frac{1-d}{2}}|\Det V|^{1/2}  \mu(dV)\, \frac{\rmd w}{w}\, \rmd y
\end{equation}
see Ref.~\cite[Eq.~(1.36)]{Terras_2016}. Any positive symmetric matrix $Z \in \mathcal{P}_d$ has a unique Cholesky decomposition
$Z = L^\intercal  L$ where $L$ is upper triangular with positive elements on the diagonal. This is seen as a full Iwasawa decomposition, see \cite{Terras_2016}.\\

Any positive symmetric matrix $Z \in \mathcal{P}_d$ admits a polar decomposition
\begin{equation}
\label{eq:polar-coordinates}
    Z=R^\intercal \Lambda R
\end{equation}
where $R\in {\rm SO}(d)$ and $\Lambda= \mathrm{diag}\{\lambda_1, \dots, \lambda_n \}$ is diagonal with positive entries. In the polar coordinates, the natural measure \eqref{eq:measure-psd} reads
\begin{equation}
    \mu (dZ)= c_d \prod_{\ell=1}^{d} \lambda_\ell^{-(d-1)/2} \prod_{i<j} |\lambda_i-\lambda_j| \prod_{\ell=1}^d \frac{\rmd \lambda_\ell}{\lambda_\ell} \, \rmd R
\end{equation}
with $\rmd R$ the Haar measure of ${\rm SO}(d)$, see Ref.~\cite[Eq.~(1.37)]{Terras_2016}.\\

Finally, one can also consider another parametrisation as $Z= e^H$ where $H$ belongs to the space of symmetric matrices.

\subsection{Product measures and path integral measures}
\label{subsec:definition-measures}
In this paper, we will use two kind of product measures (for discrete models) or path integral measures (for continuum models). We will use an interpolation of the two for semi-discrete models.  The first are Lebesgue product measures
\be \label{path1} 
{\cal D} W = \prod_{n,m} dW_{n,m}, \quad  {\cal D} W = \prod_{n,t} dW_{n}(t)   , \quad {\cal D} W = \prod_{x,t} dW(x,t)   \, .
\ee 
and the second is a product measure of 
$GL_d(\R)$ invariant type for symmetric positive definite matrices $Z_n$ and $Z(x)$ 
\begin{equation} \label{pathmu} 
    {\cal D}_\mu Z = \prod_{n} \mu(dZ_{n}), \quad {\cal D}_\mu Z = \prod_{x} \mu(dZ(x))
\end{equation}

\section{Zero dimensional version $D=0$ -- previous work}

The zero dimensional version of the MSHE \eqref{matrixshe} (i.e., without spatial diffusion) for Dyson index $\beta=1,2$ 
is the positive real symmetric/hermitian matrix process in time ${\cal Z}(t)$,
solution of the Ito stochastic equation
\be
     d Z = \sigma  Z^{1/2} dB(t) Z^{1/2} 
\ee 
with $B(t) = ({H_t + H_t^{\intercal}})/{\sqrt{2\beta}} + \I  \delta_{\beta=2} ({\tilde{H}_t- \tilde{H}_t^{\intercal}})/2  $ where the entries of $H_t$ and $\tilde{H}_t$ are $2 N^2$ independent standard Brownian motions (in the notations of the present paper $\sigma^2=g$). Some of its properties can be obtained from 
the study in Ref.~\cite{GautieMatrixKesten} by keeping only the terms involving $\sigma$ there (see equation (27) there, equivalently, 
by taking the large $\sigma$ limit). It amounts to the following SDE for the eigenvalues
\begin{equation}
\label{eq:EvolutionLambdai}
{\rm d}\lambda_{i}= \sigma^2\sum\limits_{\substack{1 \leqslant j \leqslant N \\j\neq i}} \frac{\lambda_{i} \lambda_{j}}{\lambda_{i}-\lambda_{j}} {\rm d}t+\sqrt{\frac{2}{\beta}} \sigma \lambda_{i} {\rm d}W_{i}(t)
  \quad  \quad \quad \quad \text{(It\^o)}
\end{equation} 
where the $(W_i(t))_{1 \leqslant i \leqslant N}$ are $N$ independent standard Brownian motions. 
Note that the evolution of the eigenvalues is obtained with no a priori knowledge of the evolution of
the eigenvectors. Under the change of variable $\lambda_i = e^{x_i}$, we obtain the process 
\be 
{\rm d}x_i =  - \frac{\sigma^2}{\beta} +  \sigma^2 \sum\limits_{\substack{1\leqslant j\leqslant N\\ j\neq i}}  \frac{1}{e^{ x_i - x_j} -1} + \sqrt{\frac{2}{\beta}} \sigma {\rm d}W_i(t) \, ,
\ee 
which can be mapped via a Doob transform onto fermions with mutual interactions via a Sutherland interaction potential 
$V_\mathrm{int}(x_i,x_j ) = 
   \frac{\beta (\beta -2) }{16} 
  \frac{1 }{\sinh\big(\frac{x_i-x_j}{2}\big)^2}$
  and without external potential \cite{GautieMatrixKesten}.

\section{Matrix SHE}

\subsection{Definition of MSHE}

In this Section we define the matrix SHE and construct its dynamical action. We do it for both $\beta=1$ and
$\beta=2$ although in the remainder of the paper we focus only on $\beta=1$.  Consider ${\cal H}_{ij}(x,t)$ and ${\cal \tilde H}_{ij}(x,t)$ two independent sets, each of $d^2$ i.i.d real standard (centered) space time white noises, i.e., 
with correlator
\be 
\overline{ {\cal H}_{ij}(x,t) {\cal H}_{kl}(x',t') } = \delta_{ik} \delta_{jl} \delta(x-x') \delta(t-t') 
\ee 
The noise matrix ${\cal W}(x,t)$ is $d \times d$ real symmetric for $\beta=1$, and $d \times d$ complex hermitian for $\beta=2$, and defined as
\be 
{\cal W}(x,t) = \sqrt{g} \left( \frac{{\cal H}(x,t) + {\cal H}(x,t)^{\intercal}}{\sqrt{2 \beta} } + \I \delta_{\beta,2} 
\frac{{\cal \tilde H}(x,t) - {\cal \tilde H}(x,t)^{\intercal}}{\sqrt{2 \beta} } 
\right) 
\ee 
and has thus correlator  
\bea 
&& \overline{ {\cal W}_{ij}(x,t) {\cal W}_{kl}(x,t) } = g ( \delta_{ik} \delta_{jl} + \delta_{il} \delta_{jk} ) \delta(x-x') \delta(t-t') \quad , \quad \beta=1 \\
&& \overline{ {\cal W}_{ij}(x,t) {\cal W}^*_{kl}(x,t) } = g  \delta_{ik} \delta_{jl}  \delta(x-x') \delta(t-t') \quad , \quad \beta=2 
\eea  
and its probability measure is the path integral measure (in agreement with \eqref{path1}) 
\be \label{path2}
{\cal D} {\mathcal W} \exp\left( - \frac{\beta}{4 g} \iint \rmd t\rmd x  \Tr {\cal W}^2 \right)  ~ , ~ {\cal D} {\mathcal W} 
= 
\prod_{x,t}  d {\cal W}(x,t)  ~,~ 
 d {\cal W}  = \begin{cases} \prod_{i \leq j} d {\cal W}_{ij} & \beta = 1 \\
\prod_i d {\cal W}_{ii} \prod_{i < j} d {\rm Re} {\cal W}_{ij} d {\rm Im}  {\cal W}_{ij} & \beta =2 
\end{cases} 
\ee 
Note that in the zero-dimensional case $D=0$ (no space) $\int_0^t dt' {\cal W}(t')$ is a Dyson Brownian motion
and the spectrum of $\int_0^t dt' {\cal W}(t')/\sqrt{d}$ converges in density at large $d$ to a semi-circle of support 
$[- 2 \sqrt{g t},2 \sqrt{g t}]$ for any $\beta$. \\

Consider now the evolution equation for the matrix field $\mathcal{Z}= \mathcal{Z}(x,t)$ with Ito prescription
\be \label{matrixshe-app} 
     \p_t \mathcal{Z} =\p_x^2  \mathcal{Z} + 
 \mathcal{Z}^{1/2} \mathcal{W} \mathcal{Z}^{1/2} 
\ee 
where $\mathcal{Z}$ is a $d \times d$ real symmetric matrix for $\beta=1$, and a $d \times d$ complex hermitian matrix for $\beta=2$.
the matrix $\mathcal{Z}$ is positively defined in both cases, a condition which is preserved by the flow. 
Indeed the evolution under the first term in the r.h.s. of \eqref{matrixshe-app} is equivalent to a
convolution by the heat kernel. This convolution preserves positivity as can be seen by taking the scalar product
left and right by any vector. The evolution under the second term also preserves positivity, indeed one can check on \eqref{eq:EvolutionLambdai} that the resulting flow for eigenvalues cannot cross zero \cite{GautieMatrixKesten}. One can finally consider the sum of the two evolutions through Trotterization where each individual term preserves the positivity. This property can also be seen from the limit of the matrix polymer models \cite{MatrixWhittaker2023}, although we do not
study that limit in detail here. Before going to the field theory, let us give the MSHE in the case $d=2$. 

\subsection{Polar representation of the MSHE in $d=2$}

Let us consider the MSHE for $Z(x,t)=\mathcal{Z}(x,t)$ (we drop the calligraphic notation for $\mathcal{Z}$ in this Section)
\be \label{matrixshe-app-2} 
     d Z =\p_x^2  Z \rmd t+  Z^{1/2} dW Z^{1/2} 
\ee 
where the white noise of the previous subsection is written as ${\cal W}(x,t)= dW(x,t)/dt$.
In this subsection we focus on $d=2$ and $\beta=1$. We use the polar coordinates 
\be 
Z= R^{\intercal} \Lambda R \quad , \quad \Lambda= \begin{pmatrix} \lambda_1 & 0 \\  0 & \lambda_2
\end{pmatrix}   \quad , \quad R = \begin{pmatrix} c & s \\ - s & c
\end{pmatrix}  
\ee 
where we denote $c = \cos \theta$ and $s= \sin \theta$. The polar coordinates thus involve
the three variables $(y_1,y_2,y_3)=(\lambda_1,\lambda_2,\theta)$. The original variables
are the three independent entries of the matrix ${\cal Z}$, and are denoted $x_1,x_2,x_3$, hence
they are given by
\be  \label{paramZ} 
Z = \begin{pmatrix}
    x_1 & x_2\\
    x_2 & x_3
\end{pmatrix}  = R^{\intercal} \Lambda R  =
\begin{pmatrix} c^2 \lambda_1 + s^2 \lambda_2 & (\lambda_1-\lambda_2) c s  \\ (\lambda_1-\lambda_2) c s   & 
s^2 \lambda_1 + c^2 \lambda_2
\end{pmatrix}
\ee

To consider the evolution of the polar coordinates, we split the evolution onto several parts.
\begin{enumerate}
    \item  Let us first consider the deterministic part of the evolution $d Z =\p_x^2 Z dt$. 
We evaluate $\partial_t Z$ on one side, using the expression of the matrix $Z$ in terms of $(\lambda_1,\lambda_2,\theta)$ in \eqref{paramZ},
and do the same for $\partial_x^2 Z$. We then equate the two equations which we solve as a linear system in terms of the variables  $(\partial_t \lambda_1, \partial_t \lambda_2,\partial_t \theta)$.
The result simplifies into

\begin{equation}
\label{eq:mshe-2d-deterministic-part}
    \begin{split}
\begin{pmatrix}
    \rmd \lambda_1 \\ \rmd \lambda_2 \\ \rmd \theta
\end{pmatrix}_{\rm deterministic}
=
\begin{pmatrix}
    (\partial_x^2 \lambda_1 + 2 (\lambda_2-\lambda_1) (\partial_x \theta)^2)\rmd t\\
    (\partial_x^2 \lambda_2 + 2 (\lambda_1-\lambda_2) (\partial_x \theta)^2)\rmd t\\
    (\partial_x^2 \theta + 2 (\partial_x \theta) \partial_x( \log|\lambda_1-\lambda_2|) )\rmd t
\end{pmatrix}
\end{split}
\end{equation}

\item Let us now consider the stochastic part of the evolution $d Z = 
 Z^{1/2} dW Z^{1/2}$, we will treat subsequently the drift term
 resulting from the Ito prescription. The stochastic equation reads, in components
\bea  
&& d Z = \begin{pmatrix}
    \rmd x_1 & \rmd x_2\\
    \rmd x_2 & \rmd x_3
\end{pmatrix}  = d \begin{pmatrix} c^2 \lambda_1 + s^2 \lambda_2 & (\lambda_1-\lambda_2) c s  \\ (\lambda_1-\lambda_2) c s   & 
s^2 \lambda_1 + c^2 \lambda_2 \end{pmatrix}
\\
&& = 
\begin{pmatrix} c & - s \\ s & c  \end{pmatrix}  
\begin{pmatrix} \lambda_1^{1/2} & 0 \\  0 & \lambda_2^{1/2} \end{pmatrix}  
\begin{pmatrix} c &  s \\ - s & c  \end{pmatrix} 
\begin{pmatrix} dW_{11} & dW_{12} \\ dW_{12}  & dW_{22}  \end{pmatrix}  
\begin{pmatrix} c & - s \\ s & c  \end{pmatrix}  
\begin{pmatrix} \lambda_1^{1/2} & 0 \\  0 & \lambda_2^{1/2} \end{pmatrix}  
\begin{pmatrix} c &  s \\ - s & c  \end{pmatrix} \\
&& = \begin{pmatrix} c^2 \lambda_1^{1/2} + s^2 \lambda_2^{1/2} & (\lambda_1^{1/2} -\lambda_2^{1/2} ) c s  \\ (\lambda_1^{1/2} -\lambda_2^{1/2}) c s   & 
s^2 \lambda_1^{1/2} + c^2 \lambda_2^{1/2}
\end{pmatrix} 
\begin{pmatrix} dW_{11} & dW_{12} \\ dW_{12}  & dW_{22}  \end{pmatrix} 
\begin{pmatrix} c^2 \lambda_1^{1/2} + s^2 \lambda_2^{1/2} & (\lambda_1^{1/2} -\lambda_2^{1/2} ) c s  \\ (\lambda_1^{1/2} -\lambda_2^{1/2}) c s   & 
s^2 \lambda_1^{1/2} + c^2 \lambda_2^{1/2}
\end{pmatrix} 
\eea  
Inverting these equations as a linear system in the differentials $(\rmd \lambda_1,\rmd \lambda_2, \rmd \theta)$ one finds
\begin{equation}
    \begin{split}
\begin{pmatrix}
    \rmd \lambda_1 \\ \rmd \lambda_2 \\ \rmd \theta
\end{pmatrix}_{\rm noise}   =
\begin{pmatrix}
            \lambda _1 \left( \rmd W_1+\rmd W_2 \cos (2 \theta )+\rmd W_3 \sin (2 \theta )\right)  \\
  \lambda _2 \left(\rmd W_1-\rmd W_2 \cos (2 \theta )-\rmd W_3 \sin (2
   \theta )\right) \\
  \frac{\sqrt{\lambda _1} \sqrt{\lambda _2}}{\lambda_1-\lambda_2} \left(-\rmd W_2 \sin (2 \theta
   )+\rmd W_3 \cos (2 \theta )\right) \\
        \end{pmatrix}
    \end{split}
\end{equation}
where we have defined the three independent noises with $dW_j(x,t) dW_k(x',t) = g \delta_{jk} \delta(x-x') dt$
\be 
W_1 = \frac{W_{11} + W_{22}}{2} \quad , \quad W_2 = \frac{W_{11} - W_{22}}{2} \quad , \quad W_3 =  W_{12}
\ee 

\item   Finally, since we have defined the MSHE using Ito's prescription, when we perform the change of variable from $\{x_1,x_2,x_3\}$ to 
$\{y_1,y_2,y_3\}$ it results in a non-trivial Ito drift term. To compute it 
let us define the matrix $A_{ij}=\frac{\p x_i}{\p y_j}$, which is such that 
\be 
\begin{pmatrix}
    \rmd x_1 \\ \rmd x_2 \\ \rmd x_3
\end{pmatrix}  = A 
\begin{pmatrix}
    \rmd \lambda_1 \\ \rmd \lambda_2 \\ \rmd \theta
\end{pmatrix}  
\quad , \quad 
A =   \begin{pmatrix}
 c^2 & s^2 & 2 s c  (\lambda_2-\lambda_1)  \\
 c s & - c s & (c^2-s^2) (\lambda_1 - \lambda_2)  \\
 s^2  & c^2  & 2 s c (\lambda_1-\lambda_2) \\
\end{pmatrix} 
\ee 

Then, using Ito's rule we obtain the evolution in the variables $\{y_1,y_2,y_3\}= (\lambda_1,\lambda_2,\theta)$
as 
\bea 
&&   \rmd y_j = \frac{\partial y_j}{\partial x_i} \rmd x_i + \frac{1}{2} \frac{\p^2 y_j }{\p x_i \p x_k} \rmd x_i \rmd x_k \\
&& = A^{-1}_{ji}\rmd x_i -\frac{1}{2}A^{-1}_{ji} \frac{\p^2 x_i }{\p y_\ell \p y_m}A^{-1}_{\ell r}A^{-1}_{ms}\rmd x_r \rmd x_s
\eea 
with 
\begin{equation}
\begin{pmatrix}
    \rmd \lambda_1 \\ \rmd \lambda_2 \\ \rmd \theta
\end{pmatrix}  = A^{-1} 
\begin{pmatrix}
    \rmd x_1 \\ \rmd x_2 \\ \rmd x_3 
\end{pmatrix} 
\quad , \quad  
    A^{-1}=
    \begin{pmatrix}
 \cos ^2(\theta ) & \sin (2 \theta ) & \sin ^2(\theta ) \\
 \sin ^2(\theta ) & -\sin (2 \theta ) & \cos ^2(\theta ) \\
 -\frac{\sin (\theta ) \cos (\theta )}{\lambda _1-\lambda _2} & \frac{\cos (2 \theta )}{\lambda _1-\lambda _2} & \frac{\sin (\theta ) \cos (\theta )}{\lambda
   _1-\lambda _2} \\
\end{pmatrix}
\end{equation}
The first term linear in $\rmd x_i$ will provide back the term \eqref{eq:mshe-2d-deterministic-part} and the Ito drift will come from the quadratic variation. Performing the calculation and using $dW_i dW_j= g \delta_{ij}$ one finds the Ito term as
\be 
\begin{pmatrix}
    \rmd \lambda_1 \\ \rmd \lambda_2 \\ \rmd \theta
\end{pmatrix}_{\rm Ito \,  drift}  =
\begin{pmatrix}
    \frac{g \lambda_1 \lambda_2}{\lambda_1-\lambda_2} \rmd t  \\ 
    g \frac{\lambda_1 \lambda_2}{\lambda_2-\lambda_1} \rmd t\\ 
    0 
\end{pmatrix}
\ee 
\end{enumerate}
Putting all together one finally finds the MSHE in polar coordinates
\begin{equation}
    \begin{split}
\begin{pmatrix}
    \rmd \lambda_1 \\ \rmd \lambda_2 \\ \rmd \theta
\end{pmatrix}   =
\begin{pmatrix}
             \left(2 (\lambda_2-\lambda_1)
   (\p_x\theta)^2 +\p_x^2 \lambda_1+\frac{g \lambda _1 \lambda _2}{\lambda _1-\lambda _2}\right)\rmd t+\lambda _1 \left(\rmd W_1+\rmd W_2 \cos (2 \theta )+ \rmd W_3 \sin (2 \theta )\right)  \\
  \left(2 (\lambda_1-\lambda_2) (\p_x\theta)^2+\p_x^2 \lambda_2-\frac{g \lambda _1 \lambda _2}{\lambda _1-\lambda _2} \right)\rmd t+\lambda _2 \left(\rmd W_1-\rmd W_2 \cos (2 \theta )-\rmd W_3 \sin (2
   \theta )\right) \\
  \left(2 \p_x \theta \p_x \log |\lambda_1-\lambda_2|  +\p_x^2\theta
   \right)\rmd t+\frac{\sqrt{\lambda_1 \lambda_2}}{\lambda_1-\lambda_2} \left(-\rmd W_2 \sin (2 \theta)+\rmd W_3 \cos (2 \theta )\right) \\
        \end{pmatrix}
    \end{split}
\end{equation}

Note that these equations would be the same in Stratonovich by setting the Ito term to zero (formally $g \rmd t\to 0$ in the
above equation). However if we work in Ito there is a further simplification. Indeed one sees that the 
three noises are orthogonal. Hence we can rewrite the stochastic variation as
\bea \label{stoch1}
&& d \lambda_1 =  \lambda_1 \rmd B_1  \\
&& d \lambda_2 = \lambda_2 \rmd B_2  \\
&& d \theta = \frac{\sqrt{\lambda_1 \lambda_2}}{\lambda_1-\lambda_2}  \frac{\rmd B_3}{\sqrt{2}} 
\eea 
where $\rmd B_i(x,t) \rmd B_j(x',t)=2 g \delta(x-x') dt$ are independent (space-time) Brownian motions. This leads to our
final result for the MSHE in polar coordinates 

\begin{equation}
    \begin{split}
\begin{pmatrix}
    \rmd \lambda_1 \\ \rmd \lambda_2 \\ \rmd \theta
\end{pmatrix}   =
\begin{pmatrix}
             \left(2 (\lambda_2-\lambda_1)
   (\p_x\theta)^2 +\p_x^2 \lambda_1+\frac{g \lambda _1 \lambda _2}{\lambda _1-\lambda _2}\right)\rmd t+\lambda _1 \rmd B_1  \\
  \left(2 (\lambda_1-\lambda_2) (\p_x\theta)^2+\p_x^2 \lambda_2-\frac{g \lambda _1 \lambda _2}{\lambda _1-\lambda _2} \right)\rmd t+\lambda _2 \rmd B_2\\
  \left(2 \p_x \theta \p_x \log |\lambda_1-\lambda_2|  +\p_x^2\theta
   \right)\rmd t+\frac{\sqrt{\lambda_1 \lambda_2}}{\lambda_1-\lambda_2} \frac{\rmd B_3}{\sqrt{2}} \\
        \end{pmatrix}
    \end{split}
\end{equation}

\section{Dynamical action of the MSHE}
The expectation value over the noise of any functional ${\cal O}[\mathcal{Z}]$ of the space time matrix field $\mathcal{Z}= \mathcal{Z}(x,t)$
can be written as a MSR path integral (see \cite{martin1973statistical,janssen1976lagrangean,dominicis1976techniques}, \cite{Cugliandolo2022path} and references therein for recent progress)
\be 
\overline{  {\cal O}[\mathcal{Z}] } = \iiint {\cal D} {\mathcal W}  {\cal D}\hat Z  {\cal D} Z \, {\cal O}[Z] \, 
e^{-  \iint \rmd t  \rmd x \,  \Tr [ \hat{Z} ( \p_t Z -\p_x^2 Z - Z^{1/2} {\cal W} Z^{1/2}) + \frac{1}{4 g}  {\mathcal W}^2
    ] }
\ee 
Here, the path integral over $Z(x,t)$ is by definition over symmetric/hermitian matrices (restricted to positive definite). 
For the matrix response field $\hat Z(x,t)$ one 
also uses symmetric/hermitian matrices (not restricted to positive definite). 
The role of the response field is to enforce the evolution equation \eqref{matrixshe-app}. 
The definitions of the measures in the MSR path integrals have been given in 
\eqref{path1} and \eqref{path2} and involve Lebesgue measures on matrix elements. 
The term in the exponential has then the form
\be 
\sum_i \hat Z_{ii} ( \partial_t Z_{ii} - A_{ii}) +  \sum_{i<j} ( \hat Z_{ij} ( \partial_t Z_{ij} - A_{ij}) + 
\hat Z_{ij}^* ( \partial_t Z_{ij}^* - A_{ij}^*) ) 
\ee 
and thus, up to some constants absorbed in the normalisation of the path integral (which we will not specify
in detail) enforces correctly the evolution equation for all $i \leq j$. Integrating over the noise
one obtains 
\be \label{obs}
\overline{  {\cal O}[\mathcal{Z}] } = \iint {\cal D}\hat Z  {\cal D} Z \, {\cal O}[Z] \, e^{- S[Z,\hat{Z}] } 
\ee 
in terms of the dynamical action which reads  
\be \label{dynS-app}
S[Z,\hat{Z}]  =  \iint \rmd t \rmd x \,  \Tr\bigg[ \hat{Z} ( \p_t Z -\p_x^2 Z ) - g
  \hat{Z} Z \hat{Z} Z   \bigg]
\ee
where we have used the cyclicity of the trace and that for any real symmetric/complex hermitian matrix $A$ one has
\be 
\label{eq:matrix-Gaussian-Integral}
\int {\cal D} {\mathcal W} e^{- \iint \rmd t\rmd x  \Tr [A {\mathcal W} + \frac{\beta}{4 g}  {\mathcal W}^2 ] } = e^{ \iint \rmd t\rmd x g \Tr [A^2] } 
\ee 
with $A= Z^{1/2} \hat Z Z^{1/2}$.\\

Note that at this stage we have not specified any initial condition. One can now consider a fixed initial data
$Z_i = \{ \mathcal{Z}_i(x,t_i) \}_{x \in \mathbb{R}}$ at time $t_i$. The average over the noise of the observable $\mathcal{O}$ of the final value 
of the process $Z_f = \{ \mathcal{Z}_f(x,t_f) \}_{x \in \mathbb{R}}$ at time $t_f>t_i$, conditioned to that initial data, can
obtained as the MSR path integral
\begin{equation}
\label{pathP}
    \E [\mathcal{O}(Z_f)| Z_i] = 
\iint \prod_{t=t_i^+}^{t_f }\prod_x \rmd \hat Z(x,t)  \rmd  Z(x,t) \mathcal{O}(Z_f)e^{-  \int_{t_i^+}^{t_f} \rmd t \int \rmd x \, 
 \Tr\big[ \hat{Z} ( \p_t Z -\p_x^2 Z ) - g
  \hat{Z} Z \hat{Z} Z   \big]}
\end{equation}

The two fields $Z,\hat{Z}$ are integrated only for $t>t_i$ where the MSHE dynamics is enforced since here the initial condition is fixed (and thus the expectation is conditional). Since the observable involves only the value of the process at time $t=t_f$, the integration over $\hat{Z}$ can be carried for all $t>t_f$, giving unity from the normalisation, finally leading to \eqref{pathP}.  Here, $t_i^+ = t + 0^+$, is a time regularisation. As we will see below, this regularisation will not be needed in the dynamical action but will be very important when dealing with the MSR measure.\\

From now on we restrict to the case $\beta=1$ of symmetric matrices.

\section{Weak-noise limit and saddle point of the matrix SHE}

We now replace $g \to \varepsilon g$.
As discussed in the text we consider 
the observable 
\be 
\mathcal{O}[\mathcal{Z}] = \exp\left(  \frac{1}{\varepsilon} \iint \rmd t  \rmd x \, \Tr \left[J(x,t) \mathcal{Z}(x,t) \right] \right)
\ee 
where $J(x,t)$ is a given symmetric matrix source field (assumed to vanish at $t=+\infty$). We then 
insert it into Eqs.~\eqref{obs}--\eqref{dynS-app} and obtain
\be 
G_\epsilon[J] :=\overline{  {\cal O}[\mathcal{Z}] }= \iint {\cal D}\hat Z {\cal D} Z e^{ - \iint \rmd t \rmd x \,  \Tr\big[ \hat{Z} ( \p_t Z -\p_x^2 Z ) - \varepsilon g
  \hat{Z} Z \hat{Z} Z  - \frac{1}{\varepsilon} J Z \big]    } 
\ee 
Upon rescaling 
the response field $\hat Z \to \hat Z/\varepsilon$ one obtains (up to an unimportant scaling factor) 
\be 
G_\epsilon[J] = \iint {\cal D}\hat Z {\cal D} Z e^{- \frac{1}{\varepsilon} S_J[Z,\hat{Z}] } 
\ee 
with the modified dynamical action 
\be 
S_J[Z,\hat{Z}] = S[Z,\hat{Z}] - \iint \rmd t  \rmd x\,  \Tr [ J Z ]
\ee 
In the limit $\varepsilon \ll 1$ the path integral is dominated by
a saddle point configuration for the matrix fields $\{\hat{Z},Z\}$. This is called the weak-noise limit.

\subsection{Method I - saddle point on two fields}

Taking derivatives w.r.t. these fields,
and after some algebraic manipulations, we obtain
the following matrix saddle point equations:
\begin{equation}
    \begin{split}
      \label{eq:sp-app} 
 \p_t Z &=\p_x^2 Z + 2 g 
 Z \hat Z Z \\
 - \partial_t \hat{Z} &= \partial_x^2 \hat{Z} + 2 g \hat{Z} Z \hat Z  + J 
\end{split}
\end{equation}

\subsection{Method II - saddle point on three fields}

Another method consists in keeping the noise.

\be 
G_\epsilon[J] = \iiint {\cal D} {\mathcal W} {\cal D}\hat Z  {\cal D} Z \,  
e^{- \frac{1}{\varepsilon} \iint \rmd t  \rmd x \,  \Tr [ \hat{Z} ( \p_t Z -\p_x^2 Z - Z^{1/2} {\cal W} Z^{1/2}) + \frac{1}{4 g} {\mathcal W}^2 - J Z]   }
\ee 
For $\varepsilon \ll 1$ we can take the saddle point w.r.t. all fields. Taking a derivative w.r.t. $\hat Z$ and to ${\mathcal W}$ respectively
we obtain the two equations
\be
\label{noise} 
\begin{split}
 \p_t Z &=\p_x^2 Z  + 
 Z^{1/2} \mathcal{W} Z^{1/2}, \quad 
 \mathcal{W} = 2 g Z^{1/2} \hat{Z} Z^{1/2} \, .
\end{split}
\ee
To obtain the third equation we must take a derivative with respect to $Z$, which is more delicate due to the explicit presence of its square root. To this aim we use the identity \eqref{eq:taylor-sqrt-matrix} and the cyclicity of the trace to obtain
\bea 
&& - \partial_t \hat Z = \partial_x^2 \hat  Z +  \int_0^{+\infty} e^{- t Z^{1/2} }  ( \mathcal{W} Z^{1/2}  
\hat{Z} + \hat{Z}  Z^{1/2}   \mathcal{W} ) e^{- t Z^{1/2} } + J 
\eea  
Now if we eliminate the noise using \eqref{noise}, it becomes
\begin{equation}
    \begin{split}
 - \partial_t \hat  Z &= \partial_x^2 \hat  Z + 2 g  \int_0^{+\infty} e^{- t Z^{1/2} } ( Z^{1/2} \hat Z Z \hat Z + 
\hat Z Z \hat Z Z^{1/2} )  e^{- t Z^{1/2} } + J \\
& = \partial_x^2 \hat Z - 2 g  \int_0^{+\infty} \partial_t \left( e^{- t Z^{1/2} } \hat Z Z \hat Z e^{- t Z^{1/2} }  \right) + J 
\end{split}
\end{equation}
Finally we obtain the matrix system
\begin{equation}
  \begin{split}
  \label{eq:sp-app-2} 
 \p_t  Z &=\p_x^2 Z + 2 g 
 Z \hat Z Z \\
 - \partial_t \hat{Z} &= \partial_x^2 \hat{Z} + 2 g \hat{Z} Z \hat Z  + J 
\end{split}  
\end{equation}

For the special case of the average of an observable defined at $t=t_f$ and at $x=0$ one
chooses $J(x,t)=B\delta(x)\delta(t-t_f)$. We will solve explicitly the WNT system in that case in Section~\ref{sec:scattering-kpz-wnt} using its integrability and its representation in terms of the Lax pair \eqref{eq:LaxPairU}.

\section{FD transformation of the matrix SHE} 

We will construct a special time-reversal "FD" (fluctuation-dissipation) transformation on the dynamical action the case $\beta=1$) 
from the pair of real symmetric matrix fields $\{\hat Z,Z \}$,
into another pair $\{\hat Y, Y \}$ of symmetric matrix fields. This transformation is known for the scalar case $d=1$,
see e.g.~\cite{UsWNTFlat2021,smith2018exact,CanetFDTKPZ} 
and we need to find the proper generalisation to the non-commuting matrix case.  We first recall the dynamical action
\begin{equation}
\label{dynS-app-2}
  S[Z,\hat{Z}] =  \int_{t_i}^{t_f} \rmd t \int \rmd x \, 
 \Tr\big[ \hat{Z} ( \p_t Z -\p_x^2 Z ) - g\hat{Z} Z \hat{Z} Z   \big]
\end{equation}

We start by defining, by analogy with the scalar case
\be 
Z(x,t) = Q(x,\tau)^{-1}  \quad , \quad Q(x,\tau) = Z(x,t)^{-1} \quad , \quad 
\tau=t_f+t_i-t 
\ee
and we will determine later the transformation rule for $\hat Z$. 
We use 
\be
\begin{split}
& \p_x Z=-Z (\p_x Q) Z \\
& \p_t Z =- Z (\p_t Q) Z = Z (\p_\tau Q) Z \\
& \p^2_x Z =2 Z (\p_x Q) Z (\p_x Q) Z - Z (\p^2_x Q) Z 
\end{split}
\ee  
Inserting into the dynamical action \eqref{dynS-app-2}, it becomes
\begin{equation}
\begin{split}
    S[Z,\hat{Z}]  &=   \int_{t_i}^{t_f} \rmd \tau \int \rmd x \,  \Tr\bigg[  \hat{Z} ( Z (\p_\tau Q) Z -(2 Z (\p_x Q) Z (\p_x Q) Z- Z (\p^2_x Q) Z) ) - g 
  \hat{Z} Z \hat{Z} Z   \bigg]\\
  &=    \int_{t_i}^{t_f} \rmd \tau  \int \rmd x \,  \Tr\bigg[ Z \hat{Z}Z ( \p_\tau Q-\p_x^2 Q  -2( (\p_x Q) Q^{-1} (\p_x Q) -  \p^2_x Q ) ) - g 
  Z\hat{Z} Z \hat{Z}    \bigg]\\
  \end{split}
\end{equation}
Let us now define the second transformation as
\begin{equation}
    Z(x,t) \hat{Z}(x,t) Z(x,t) = \hat Q(x,\tau) + \mathcal{F}
\end{equation}
where $\mathcal{F}$ is determined subsequently in \eqref{FF}. It can 
equivalently be written as
\begin{equation}
    \hat{Z}(x,t) = Q(x,\tau) \hat Q(x,\tau)  Q(x,\tau) + Q(x,\tau)  \mathcal{F} \, Q(x,\tau)  \label{s48} 
\end{equation}
and will define the new response field $\hat Q(x,\tau)$. Inserting and regrouping 
we see that we can write
\begin{equation}
\begin{split}
    S[\hat Z,Z]&=\hat{S}[\hat Q,Q]+ \int_{t_i}^{t_f} \rmd \tau \int \rmd x \,  \Tr\bigg[2 \hat Q (- g Q\mathcal{F} Q-( (\p_x Q) Q^{-1} \p_x Q -  \p^2_x Q ) ) \\
    &- g \mathcal{F} Q \mathcal{F} Q+\mathcal{F}( \p_\tau Q-\p_x^2 Q  -2( (\p_x Q) Q^{-1} \p_x Q -  \p^2_x Q ) )\bigg]
\end{split}
\end{equation}
The action $\hat{S}$ has exactly the same form as the action $S$ in the time-reversed coordinates $t \to \tau$. Cancelling the linear term in the new response field $\hat Q$ allows to determine ${\cal F}$ as
\begin{equation}
\begin{split}
   -  g\mathcal{F}&= Q^{-1}( (\p_x Q) Q^{-1} \p_x Q -  \p^2_x Q )  Q^{-1}\\
&= -\p_x(Q^{-1}\p_x Q) Q^{-1} \\
&= - Q^{-1}\p_x((\p_x Q) Q^{-1}) \label{FF} 
\end{split}
\end{equation}
Inserting this expression into the action gives 
\begin{equation}
\begin{split}
S[\hat Z,Z]&=\hat{S}[\hat Q,Q]+\frac{1}{g}\int_{t_i}^{t_f} \rmd \tau \int \rmd x \,  \Tr\bigg[
    - g^2 \mathcal{F} Q \mathcal{F} Q+g\mathcal{F}( \p_\tau Q-\p_x^2 Q  -2( (\p_x Q) Q^{-1} \p_x Q -  \p^2_x Q ) )\bigg]\\
    &=\hat{S}[\hat Q,Q]+ \frac{1}{g}\int_{t_i}^{t_f} \rmd \tau \int \rmd x \,  \Tr\bigg[
    g\mathcal{F}( \p_\tau Q  - (\p_x Q) Q^{-1} \p_x Q   )\bigg] \label{Sfin} 
\end{split}
\end{equation}
 The last term is a total spatial derivative, it reads
\begin{equation}
\begin{split}
    \Tr [g\mathcal{F} \p_x Q Q^{-1} \p_x Q] &= - \Tr[Q^{-1}( (\p_x Q) Q^{-1} \p_x Q -  \p^2_x Q )  Q^{-1} \p_x Q Q^{-1} \p_x Q ]\\
    &=\Tr[\p_x( Q^{-1} \p_x Q )  (Q^{-1} \p_x Q)^2 ] \\
    &=\frac{1}{3}\Tr[\p_x( (Q^{-1} \p_x Q )^3 ) ]
\end{split}
\end{equation}
Inserted in the action it is thus a surface term. 
We will assume that one can discard this term for $x$ on the full line, as is the case for the KPZ equation.
If one wants a more controlled setting, one can assume periodic boundary conditions along $x$, in which
case the integral of this term exactly vanishes. Hence we are left with only the first part in \eqref{Sfin}
which can be rearranged as follows
\begin{equation}
    \begin{split}
        \Tr[ g\mathcal{F} \p_\tau Q]         &=\Tr \bigg[\p_x(Q^{-1}\p_x Q) Q^{-1}\p_\tau Q \bigg]  =-\Tr \bigg[Q^{-1}\p_x Q \p_x(Q^{-1}\p_\tau Q) \bigg]  \\
        &=-\Tr \bigg[Q^{-1}\p_x Q \p_\tau(Q^{-1}\p_x Q) \bigg] =-\frac{1}{2}\Tr \bigg[\p_\tau((Q^{-1}\p_x Q)^2) \bigg]  \\
        &=\frac{1}{2}\p_\tau\Tr \bigg[\p_x Q^{-1}\p_x Q \bigg]  \\
    \end{split}
\end{equation}
where 
we have performed an integration by part w.r.t. $x$ in the first line (the $\int \rmd x$ being implicit here, this leads again to a surface term
that we discard) and we have
used the cyclicity of the trace to commute the space and time derivatives going from the first to the second line. The final expression
is a total derivative in time. \\

Let us now summarize what we have achieved. We have defined the following transformation 
of the fields 
\bea  \label{fdt0} 
 Z(x,t) &=& Q(x,\tau)^{-1}  \quad , \quad \tau=t_f+t_i-t \\
 Z(x,t) \hat Z(x,t) &=&  \hat Q(x,\tau) Q(x,\tau)  + \frac{1}{g} \partial_x (Q(x,\tau)^{-1} \partial_x Q(x,\tau)) 
\eea  
where in the last line we have put together \eqref{s48} and \eqref{FF}. The last line can also be written in a symmetrised way
\be 
Z(x,t) \hat Z(x,t) + \frac{1}{2g} \partial_x ( (\partial_x Z(x,t)) Z(x,t)^{-1}) =  \hat Q(x,\tau) Q(x,\tau) +  \frac{1}{2g} \partial_x (Q(x,\tau)^{-1} \partial_x Q(x,\tau)) 
\ee 
which shows that it is an involution. Then, restoring the time boundaries, we have shown that 
\begin{equation} \label{shown} 
\begin{split}
    S[\hat Z,Z]
    &=\hat S[\hat Q,Q] - \frac{1}{2 g} \bigg[ \int  \rmd x \, \Tr \big[(Q^{-1}\p_x Q)^2 \big] \bigg]^{t_f}_{t_i} \\
    &=\hat S[\hat Q,Q] + \frac{1}{2 g} \bigg[ \int  \rmd x \, \Tr \big[(Z^{-1}\p_x Z)^2 \big] \bigg]^{t_f}_{t_i} 
\end{split}
\end{equation}
Note that the boundaries are taken at $t_i$ and $t_f$ because the integration domain must include the initial and terminal conditions.

\section{Invariant measure of the matrix SHE} 
\label{suppmat:sec:invariant-measureMSHE}
To obtain the invariant measure from the FD transformation we need to consider carefully the transformation
of the MSR measure and the ensuing Jacobians. To this aim, and by analogy with the OY and the log Gamma polymers treated below,
we define the fields with arguments on either

\begin{itemize}
    \item the initial boundary $t=t_i$ 
    \item the final boundary $t=t_f$ 
    \item the "bulk" $t \in [t_i^+,t_f^-]$
\end{itemize}

We also need to specify more precisely the transformation rule on the fields (including the time regularisation)
\bea  \label{fdt} 
 Z(x,t)&=& Q^{-1}(x,\tau(t)^-)   \quad , \quad \tau(t)=t_f+t_i^+-t \\
 Z(x,t) \hat Z(x,t) &=&  \hat Q(x,\tau(t)) Q(x,\tau(t))  + \frac{1}{g} \partial_x (Q(x,\tau(t))^{-1} \partial_x Q(x,\tau(t))) 
\eea  
This choice is such that the following properties hold
\begin{equation}
\label{eq:change-var-KPZ-FDT-indices}
\begin{split}
    &\tau([t_i^+,t_f])=[t_i^+,t_f]\, ,  \quad \tau^-([t_i,t_f])=[t_i,t_f] \, ,\\
    & Z(x,t_i) = Q^{-1}(x,t_f) \, , \quad  Z(x,t_f) = Q^{-1}(x,t_i) \, .
\end{split}
\end{equation}
The change of time $\tau$ preserves the time domain for the dynamics $[t_i^+,t_f]$ and the time regularisation $\tau^-$ maps the initial condition onto the final condition.\\

Given the regularisation, we now compute the Jacobian of the transformation $(Z,\hat{Z})\to (Q,\hat{Q})$ in the bulk as well as on the final boundary. Since $ \frac{\delta Z(x,t)}{\delta \hat{Q}(x',\tau)} =0$, the Jacobian is always triangular and the diagonal blocks are local in space-time since the FD transformation is itself local. Thus the Jacobian is the product of the Jacobians of the maps $Z\to Q$ and $\hat{Z} \to \hat{Q}$ (where in the second map the $Z=Q^{-1}$ variables are fixed), respectively.

We will use this Jacobian below for a change of variable in the MSR integration measure, where we will integrate over $Z(x,t)$ in the bulk, while we keep $Z(x,t_i)$ and $Z(x,t_f)$ fixed, and integrate $\hat Z(x,t)$ for $t \in [t_i^+,t_f]$ (i.e., in the bulk together with the final boundary). We now compute
the total Jacobian needed for this change of variables which we call $J_{total}$.\\

Using  \eqref{detg} and \eqref{detinverse} the first contribution reads
\begin{equation}
\begin{split}
   | \frac{\mathcal{D}Z}{\mathcal{D}Q}|_{\rm bulk } &= \prod_{x}\prod_{t=t_i^+}^{t_f^-}|\Det Z(x,t) |^{d+1} 
    \end{split}
\end{equation}
and the second is
\begin{equation}
\begin{split}
    |\frac{\mathcal{D}\hat{Z}}{\mathcal{D}\hat{Q}}|_{\rm bulk+final} &=  \prod_{x} \prod_{t=t_i^+}^{t_f} |\det Z(x,t)|^{-\frac{d+1}{2}}   |\det Q(x,\tau(t))|^{\frac{d+1}{2}} =  \prod_{x} \prod_{t=t_i^+}^{t_f} |\det Z(x,t)|^{-\frac{d+1}{2}}  \prod_{t=t_i}^{t_f^-} |\det Z(x,t)|^{-\frac{d+1}{2}} 
\end{split}
\end{equation}
where the bounds $t_f^-$ and $t_i$ in the last product originate from the identity
\begin{equation}
\begin{split}
    Q^{-1}(x,\tau(t))&=Z(x,t_i+t_f-\tau(t))=Z(x,t_i+t_f-(t_f+t_i^+-t ))=Z(x,t^-)\\
    \end{split}
\end{equation}

The product of the two contributions now reads
\begin{equation}
\label{eq:KPZ-total-jacobian-MSR-invariant}
\begin{split}
    J_{total}&=|\frac{\mathcal{D}\hat{Z}}{\mathcal{D}\hat{Q}}|_{\rm bulk+final}|\frac{\mathcal{D}Z}{\mathcal{D}Q}|_{\rm bulk }  \\
    &= \prod_{x} |\Det Z(x,t_i)|^{-\frac{d+1}{2}}|\Det Z(x,t_f)|^{-\frac{d+1}{2}}
    \end{split}
\end{equation}

We now exponentiate (minus) the FD relation \eqref{shown} and rearrange both sides of the equation. Next, we integrate over all responses fields $\hat{Z}(x,t)$ in the bulk and on the final boundary as well as over all the fields $Z(x,t)$ in the bulk (i.e., keeping $Z(x,t_i)$ and $Z(x,t_f)$ fixed). 
This integration requires the Jacobian \eqref{eq:KPZ-total-jacobian-MSR-invariant} on the r.h.s. One obtains after insertion of any observable $\mathcal{O}(Z_f)$
over the final fields, 
$Z_f = \{ Z(x,t_f) \}$
\be
\label{eq:KPZ-MSR-invariant-measure}
\begin{split}
& \int_{bulk} \mathcal{D}Z\int \mathcal{D}\hat{Z} e^{-S_0[Z,\hat{Z}]} \mathcal{O}(Z_f)   e^{- \frac{1}{2 g} \int  \rmd x \, \Tr \big[(Z(t_i)^{-1}\p_x Z(t_i))^2 \big] }
\\
& = \int_{bulk} \mathcal{D}Q \int  \mathcal{D}\hat{Q} e^{-\hat{S}_0[Q,\hat{Q}]}\mathcal{O}(Z_f)  e^{- \frac{1}{2 g} \int  \rmd x \, \Tr \big[(Z(t_f)^{-1}\p_x Z(t_f))^2 \big] }\times  \prod_{x} |\Det Z(x,t_i)|^{-\frac{d+1}{2}}|\Det Z(x,t_f)|^{-\frac{d+1}{2}}
\end{split}
\ee  
We now multiply both sides of \eqref{eq:KPZ-MSR-invariant-measure} by the following measure over the initial and final fields
\begin{equation}
\label{eq:KPZ-MSR-invariant-measure-integration-partition-func}
    \underbrace{\prod_{x} \mu(\rmd Z(x,t_i))}_{\text{initial values}} \times \underbrace{\prod_{x} \rmd Z(x,t_f) }_{\text{final values}}
\end{equation}
where we recall the definition of the $GL_d$ invariant measure in 
\eqref{eq:measure-psd}. Next we integrate over the variables $Z(x,t_i)$ and $Z(x,t_f)$, obtaining schematically the new equation LHS$=$RHS.
We now consider separately each side of this new equation.

\subsection{RHS}
The integration over the initial fields in the RHS can be done explicitly through the normalisation of the MSR path integral. Indeed, the MSR normalisation reads, where $Q(x,\tau=t_i)$ are fixed

\begin{equation}
    \int\prod_{x}\rmd Q(x,t_f) 
   \int_{\rm bulk} \mathcal{D}Q \int \mathcal{D}\hat{Q} e^{-S_0[Q,\hat{Q}]}=1 
\end{equation}
This implies using the FD transformation \eqref{fdt} and \eqref{detg},\eqref{detinverse} 
\begin{equation}
    \int\prod_{x}\frac{\rmd Z(x,t_i)}{|\Det Z(x,t_i)|^{d+1 }}
    \int_{\rm bulk} \mathcal{D}Q \int \mathcal{D}\hat{Q}  e^{-S_0[Q,\hat{Q}]}=1 
\end{equation}
Using this identity, and inserting the expression for the Jacobian \eqref{eq:KPZ-total-jacobian-MSR-invariant}, the RHS becomes
\begin{equation}
    \text{RHS} = \int \prod_{x} \mu(\rmd Z(x,t_f)) \mathcal{O}(Z_f)e^{- \frac{1}{2 g} \int  \rmd x \, \Tr \big[(Z(t_f)^{-1}\p_x Z(t_f))^2 \big] }
\end{equation}

\subsection{LHS}

We now consider the LHS, and use the expression for the conditional average \eqref{pathP} 
\begin{equation}
\begin{split}
    \text{LHS} &=    \int\prod_{x} \mu(\rmd Z(x,t_i))\iint_{bulk+final}\mathcal{D}Z \mathcal{D}\hat{Z} e^{-S_0[Z,\hat{Z}]} \mathcal{O}(Z_f)   e^{- \frac{1}{2 g} \int  \rmd x \, \Tr \big[(Z(t_i)^{-1}\p_x Z(t_i))^2 \big] }\\
    &=    \int\prod_{x} \mu(\rmd Z(x,t_i)) \E[\mathcal{O}(Z_f)|Z_i]   e^{- \frac{1}{2 g} \int  \rmd x \, \Tr \big[(Z(t_i)^{-1}\p_x Z(t_i))^2 \big] }\\
    \end{split}
\end{equation}

\subsection{Summary and invariant measure}
By equation $LHS=RHS$, we have shown that the matrix geometric Brownian motion measure on the matrix field $Z(x)$
\be \label{invSHE} 
\prod_{x} \mu(\rmd Z(x))
e^{- \frac{1}{2 g} \int  \rmd x \, \Tr \big[(Z(x)^{-1}\p_x Z(x))^2 \big] } 
\ee 
is an invariant measure for the matrix SHE equation. This measure can also be written as
\be
\mu(\rmd Z(x_0)) \prod_{x} \rmd (Z^{-1/2}\p_x Z Z^{-1/2})e^{- \frac{1}{2 g} \int  \rmd x \, \Tr \big[(Z(x)^{-1}\p_x Z(x))^2 \big] }
\ee 
which emphasizes that the measure on the "zero mode" is uniform and that the log-derivative increments of $Z(x)$ along $x$ are independent. See Section.~\ref{subsec:convention-log-derivative} and \eqref{eq:AB-YO-2} for a discussion related to this change of measure.

\subsection{Parametrisations an interpretation of the invariant measure}
\label{suppmat:subsec:invariant-measure-process-MSHE}

Let us first recall that ${\cal P}_d$ can naturally be seen as a Riemannian manifold, with an invariant line element (or arclength)
whose square is given by, see e.g.~\cite{Terras_2016,dolcetti2014some} 
\begin{equation}
\label{eq:line-element}
      (\p_x s_Z)^2 :=\Tr [(Z^{-1} \partial_x Z)^2] 
\end{equation}
{also called Fisher-Rao metric or trace metric.}
This square line element appears in the invariant measure \eqref{invSHE}. 
This measure can be interpreted as the path integral associated to the (geometric) Brownian motion on ${\cal P}_d$, 
for recent studies see \cite{matsumoto2023laplacian,rogers2000diffusions,matrixOYpolymer}.
It can be characterised using the Laplace-Beltrami operator $\Delta_Z=\Tr[(Z\frac{\p}{\p Z})^2]$
and obeys the Feynman-Kac identity \cite[Theorem~6.2]{bar2011wiener}
\be \label{statapp}
\int_{Z(x_1) = Z_1}^{Z(x_2) = Z_2} {\mathcal D}_\mu Z(x)   \exp\left( - \frac{1}{2 g} \int \rmd x \Tr[ (Z(x)^{-1} \partial_x Z(x))^2 ] \right) 
= \langle  Z_2 | e^{ (x_2-x_1) \Delta_Z } | Z_1 \rangle 
\ee 
Note that the Laplace-Beltrami operator is the generator of the process
on positive symmetric matrices defined in Stratanovich as  
$dX = X^{1/2} \partial W X^{1/2}$, see \cite[Section~2.9]{matrixOYpolymer}.
{For other references on diffusions on manifolds see e.g. 
\cite[Chap.~5]{rogers2000diffusions}, \cite{norris1986brownian} and \cite{matsumoto2023laplacian}.
Note that the eigenvalues do not cross and do not reach zero. }

The Gibbs measure related to the energy $\int \rmd x (\p_x s_Z)^2$ 
is also referred to as the Brownian loop measure, see Ref.~\cite{bruned2022geometric}. 
{ In that paper a periodic version ($x$ on a circle) is considered. There
is a subtle question of regularization ambiguity, see \cite[Remark~1.14]{bruned2022geometric}
and \cite{cheng1972quantization,um1974normalization}, which leads to an additional 
factor in the Brownian loop measure of the form exponential of the integral of the scalar curvature $R$. 
Here, however ${\cal P}_d$ has a constant scalar curvature, 
$R = - \frac{d (d-1) (d+2)}{8}$, see Ref.~\cite[Proposition~3.1]{dolcetti2018differential},
(see also \cite{moakher2011riemannian} which however seems to have a sign misprint).
Hence in the present case there should be no ambiguity \cite{HairerPrivate1}}.

The line element \eqref{eq:line-element} and the Laplace-Beltrami operator  
admit different representations depending on the decomposition chosen for the matrix $Z$.

\subsubsection{Using the Iwasawa coordinates}

From Ref.~\cite[Eq.~(1.36)]{Terras_2016}, the invariant measure admits the following expression with the partial Iwasawa coordinates \eqref{eq:iwasawa-coordinates}

\begin{equation}
    \Tr [(Z^{-1} \partial_x Z)^2]  = \Tr [(V^{-1} \partial_x V)^2)] +(w^{-1}\p_x w)^2 + \frac{2}{w} \p_x y^\intercal V \p_x y
\end{equation}
where $\p_x y \in \R^{d-1}$ is the vector of the derivative of the coordinates of $y$. The Laplace-Beltrami operator in the Iwasawa coordinates reads
\begin{equation}
    \Delta_Z=\Delta_V + \frac{1}{2}\Tr (V\p_V) +(w\p_w)^2-\frac{d-1}{2}w\p_w+\frac{1}{2} w \vec{\nabla}_y^\intercal V^{-1}\vec{\nabla}_y
\end{equation}
For the Riemannian geometry of ${\cal P}_d$ using Cholesky decomposition see~\cite{lin2019riemannian}. 
\subsubsection{Using the polar coordinates}

From Ref.~\cite[Eq.~(1.37)]{Terras_2016}, the invariant measure admits the following expression with the polar coordinates \eqref{eq:polar-coordinates}

\be 
\Tr [(Z^{-1} \partial_x Z)^2] =  \sum_{i,j=1}^d \left(  [\partial_x R R^{-1}]_{ij} \right)^2 \frac{(\lambda_i-\lambda_j)^2}{ \lambda_i \lambda_j} 
+ \sum_{i=1}^d (\partial_x \log \lambda_i)^2
\ee 

The Laplace-Beltrami operator in the polar coordinates read 
\begin{equation}
    \Delta_Z=\sum_{i=1}^d \lambda_i^2 \p_{\lambda_i}^2-\frac{d-3}{2}\sum_{i=1}^d \lambda_i \p_{\lambda_i}+\frac{1}{2}\sum_{i<j}\frac{\lambda_i \lambda_j }{(\lambda_i-\lambda_j)^2}\p_{x_{i,j}}^2+\sum_{k=1}^d \left( \sum_{j \neq k}\frac{\lambda_k^2}{\lambda_k-\lambda_j}\right) \p_{\lambda_k}
\end{equation}
where $d x_{ij} = [d R R^{-1}]_{ij}$.

\subsubsection{Expression of the invariant measure for $d=2$}

In the case $d=2$ the invariant measure reads
\be
\propto \prod_x d\lambda_1(x) d\lambda_2(x) d \theta(x) 
\frac{ |\lambda_1(x) - \lambda_2(x)| }{(\lambda_1(x) \lambda_2(x))^{\frac{3}{2}} }
e^{  - \frac{1}{2 g} \int \rmd x   \left( (\partial_x \log \lambda_1(x))^2 + 
(\partial_x \log \lambda_2(x))^2 + 
\frac{2 (\lambda_1(x)-\lambda_2(x))^2}{ \lambda_1(x) \lambda_2(x)} ( \partial_x \theta(x) )^2
\right) } 
\ee
Let us introduce the height fields 
\be 
h_1= \log \lambda_1 \quad , \quad h_2= \log \lambda_2 \quad , \quad H = \frac{h_1+h_2}{2} 
\quad , \quad h = \frac{h_1-h_2}{2} 
\ee 
The invariant measure then factorizes into a Brownian measure 
$\propto \prod_x dH(x) e^{- \frac{1}{g} \int \rmd x (\partial_x H(x))^2 }$, 
for $H(x)$ which decouples, and a joint measure for the two fields $h(x)$ 
and $\theta(x)$ which remain coupled,
\be 
\propto \prod_x dh(x) d\theta(x) |\sinh h(x) | e^{- \frac{1}{2 g} \int \rmd x \left( 2 (\partial_x h(x))^2  + 8 \sinh^2(h(x)) ( \partial_x \theta(x) )^2 
\right) }
\ee 
Qualitatively one sees that for a given $h$ the "diffusion coefficient" of the angle variable $\theta$
is $D \sim 1/\sinh^2(h)$. Hence when $|h| \gg 1$ the variable $\theta$ does not vary much with $x$, 
while when $|h| \ll 1$, $\theta$ varies a lot. 
Also the entropy of the fluctuations of $\theta$ leads to an effective potential for $h$, very qualitatively 
$\sim - \frac{1}{2} \log D \sim |h|$, i.e., a linear attraction. To see that more
precisely one may
perform the Gaussian integral over $\theta(x)$ leading to an effective action for $h(x)$
$\propto \prod_x dh(x) |\sinh h(x) | e^{- S_{\rm eff}[h]}$
with 
\be 
S_{\rm eff}[h] =  \int \rmd x \left( \frac{1}{g} (\partial_x h(x))^2   - \log|\sinh h(x)| \right) 
+ \frac{1}{2} {\rm Tr } \log ( - \partial_x \sinh^2(h(x)) \partial_x ) 
\ee 
The first potential term originates from the repulsion between the eigenvalues (i.e.,
between the two Brownians $h_1,h_2$). The second term is a functional
determinant and qualitatively leads to an attraction between the eigenvalues (very naively
a linear potential $\sim |h(x)|$ at large $|h(x)|$ if one ignores the gradients). Note that we have not 
taken into account the periodicity in $\theta$. A more precise study of this action
is left for the future. Let us note that stochastic analysis of 
the Brownian motion on ${\cal P}_d$ leads to a linear drift $h \sim t$,
see~\cite[Eq.~(1.1)]{matsumoto2023laplacian}.

\section{$GL_d(\R)$ gauge invariance of the Lax pair and the scattering matrix}

The WNT system \eqref{eq:sp-app-2} is invariant by the transformation
\be 
\label{eq:app-gauge-transform-field}
Z(x,t) \to X^\intercal Z(x,t) X \quad , \quad \hat Z \to X^{-1} \hat Z(x,t) (X^\intercal)^{-1}
\ee 
for any $X \in GL_d(\R)$. This invariance holds for the equations, and the boundary conditions have to be modified accordingly. In terms of the Lax pair \eqref{eq:LaxPairU}, this translates into 

\begin{equation}
    U_1=
\begin{pmatrix}
-\I \frac{k}{2} I_d  & - g \hat Z\\  Z & \I  \frac{k}{2} I_d 
\end{pmatrix} \to 
\begin{pmatrix}
    X^{-1} & 0 \\
    0 & X^\intercal \\
\end{pmatrix}
\begin{pmatrix}
-\I \frac{k}{2} I_d  & - g \hat Z\\  Z & \I  \frac{k}{2} I_d 
\end{pmatrix}
\begin{pmatrix}
    X & 0 \\
    0 & (X^\intercal)^{-1} \\
\end{pmatrix}
\end{equation}
and 
\begin{equation}
    U_2= 
    \begin{pmatrix}
     \frac{1}{2} k^2 I_d - g \hat Z Z & g (\partial_x - \I k) \hat Z\\   
      (\partial_x +  \I k) Z & \frac{1}{2} k^2 I_d - g Z \hat Z
    \end{pmatrix}
    \to 
    \begin{pmatrix}
    X^{-1} & 0 \\
    0 & X^\intercal \\
\end{pmatrix}
 \begin{pmatrix}
     \frac{1}{2} k^2 I_d - g \hat Z Z & g (\partial_x - \I k) \hat Z\\   
      (\partial_x +  \I k) Z & \frac{1}{2} k^2 I_d - g Z \hat Z
    \end{pmatrix}
\begin{pmatrix}
    X & 0 \\
    0 & (X^\intercal)^{-1} \\
\end{pmatrix}
\end{equation}

A gauge transformation on the Lax matrices is defined as the map $(U_1,U_2)\mapsto (\tilde{U}_1,\tilde{U}_2) $ involving an invertible gauge $\mathcal{G}$ so that

\begin{equation}
\label{eq:definition-gauge-transform}
    \tilde{U}_1=\mathcal{G}^{-1}U_1 \mathcal{G}-\mathcal{G}^{-1}\p_x \mathcal{G}, \quad \tilde{U}_2=\mathcal{G}^{-1}U_2 \mathcal{G}-\mathcal{G}^{-1}\p_t \mathcal{G}
\end{equation}
This indicates that the $GL_d(\R)$ transformation on the fields $\{Z,\hat{Z} \}$ act as a gauge transformation on the Lax pair and thus that the Lax matrices are defined up to a block diagonal $GL_d(\R)$ gauge matrix.

\section{Scattering} 
\label{sec:scattering-kpz-wnt}

We solve in this section the scattering problem for the WNT of the matrix KPZ equation. We will call ${\phi}_{1,2}(x,t)$ and $\bar{\phi}_{1,2}(x,t)$ the two components of $\phi, \bar{\phi}$. Recalling that the asymptotics of the Lax problem are given as 
\begin{equation}
    \bar \phi \underset{x \to -\infty}{\simeq}
    \begin{pmatrix}
        0 \\
        -e^{\I k x/2}I_d
    \end{pmatrix}, \;
    \bar \phi \underset{x \to +\infty}{\simeq}
    \begin{pmatrix}
\tilde{b}(k,t)e^{-\frac{\I  kx}{2}}\\ -\tilde{a}(k,t)e^{\frac{\I  kx}{2}}
\end{pmatrix}, \quad 
     \phi \underset{x \to -\infty}{\simeq}
    \begin{pmatrix}
        e^{-\I k x/2}I_d \\
        0
    \end{pmatrix}, \;
     \phi \underset{x \to +\infty}{\simeq}
    \begin{pmatrix}
a(k,t)e^{-\frac{\I  kx}{2}}\\b(k,t)e^{\frac{\I  kx}{2}}
\end{pmatrix} 
\end{equation}
We will subsequently solve the first equation of the Lax pair given in the text,  $\partial_x \vec v= U_1 \vec v$  with $U_1$ given in \eqref{eq:LaxPairU} and with the successive choices $\Vec{v}=e^{- k^2 t/2} {\bar{\phi}}$ and $\Vec{v}=e^{ k^2 t/2} {{\phi}}$.

\subsection{Scattering problem: general $t$}
\label{sec:scattgeneral} 

Here, we obtain some relations from the scattering problem at any $t$ always valid for the $\{ P,Q \} $ system 
(for arbitrary boundary conditions, i.e., beyond the WNT). In the case of the WNT with the boundary 
conditions \eqref{eq:init-cond-wnt-kpz}, when specified to $t=1$ they lead to the same results as in the previous
section. When specified to $t=0$ they give some  formula for $b(k)$ 
for general initial condition. \\

{\bf Scattering for $\bar{\phi}$}.
Let us return to the $\partial_x$ equation of the Lax pair for $\phi(x,t)$, which we write at a fixed $t$, in the form
(here we also indicate the dependence in $k$) 
\begin{equation}
\begin{split}
\partial_x [e^{\I \frac{k}{2} x}\phi_1(x,t,k)]&=- g \hat{Z}(x,t) \phi_2(x,t,k)e^{\I \frac{k}{2} x}, \qquad \partial_x [e^{-\I \frac{k}{2} x}\phi_2(x,t,k)] = Z(x,t)\phi_1(x,t,k)e^{-\I \frac{k}{2} x}
\end{split}
\label{eq:SuppMatLaxPairEqGeneral}
\end{equation}
Integrating the first equation of \eqref{eq:SuppMatLaxPairEqGeneral} from $x=-\infty$ to $x=+\infty$, and using that $a(k,t)=a(k)$ we have
\begin{equation}
a(k)-I_d=-g \int_{\R} \rmd x \, \hat{Z}(x,t)\phi_2(x,t,k)e^{\I \frac{k}{2} x}
\end{equation}
and from $-\infty$ and a value $x$
\begin{equation}
e^{\I \frac{k}{2}  x}\phi_1(x,t,k)=I_d- g \int_{-\infty}^{x} \rmd x' \, \hat{Z}(x',t)\phi_2(x',t,k)e^{\I \frac{k}{2} x'}
\end{equation}

Integrating the second equation of \eqref{eq:SuppMatLaxPairEqGeneral} between $-\infty$ and $+\infty$, we have 
\begin{equation}
b(k) e^{-k^2 t} =  \int_{\R} \rmd x Z(x,t)\phi_1(x,t,k)e^{-\I \frac{k}{2} x}
\end{equation}
and from $-\infty$ and a value $x$
\begin{equation}
e^{-\I \frac{k}{2} x}\phi_2(x,t,k)= \int_{-\infty}^{x} \rmd x' \, Z(x',t)\phi_1(x',t,k)e^{-\I \frac{k}{2} x'}
\end{equation}
Inserting, this gives two integral equations for $\phi_{1,2}$
\begin{equation}
\begin{split}
e^{\I \frac{k}{2} x}\phi_1(x,t,k)&=I_d-g  \int_{-\infty}^{x} \rmd x' \int_{-\infty}^{x'} \rmd x'' \, \hat{Z}(x',t) Z(x'',t)\phi_1(x'',t,k)e^{\I k x'-\I \frac{k}{2} x''}\\
e^{-\I \frac{k}{2} x}\phi_2(x,t,k)&=\int_{-\infty}^{x} \rmd x' \, e^{-\I k x'} Z(x',t)\left(I_d- g \int_{-\infty}^{x'} \rmd x'' \, \hat{Z}(x'',t)\phi_2(x'',t,k)e^{\I \frac{k}{2} x''}\right)
\end{split}
\end{equation}
Iteration of these equations gives a series representation for $b(k)$ as a sum of alternating products of terms $Z (\hat{Z} Z)^{n}$, $n \geq 0$, integrated over ordered sectors as
\be \label{agaa} 
 b(k) e^{-k^2 t} =  \sum_{n=0}^\infty (-g)^n \int_{x_{2n+1}<\dots < x_1} \prod_{j=1}^{2n+1}  \rmd x_j e^{ \I k (\sum_{j=1}^n x_{2j}- \sum_{j=0}^n x_{2j+1})}Z(x_1,t)\hat{Z}(x_2,t)\dots \hat{Z}(x_{2n},t)Z(x_{2n+1},t) 
\ee
as well as a a series representation for $a(k)$ as a sum of alternating products of terms $(\hat{Z} Z)^{n}$, $n \geq 0$, integrated over ordered sectors as
\begin{equation} \label{agaa2} 
a(k)= \sum_{n=0}^\infty (-g)^n \int_{x_{2n}<\dots < x_1} \prod_{j=1}^{2n} \rmd x_j e^{\I k  (\sum_{j=0}^{n-1} x_{2j+1}-\sum_{j=1}^n x_{2j})}\hat{Z}(x_1,t) Z(x_2,t) \dots \hat{Z}( x_{2n-1},t)Z(x_{2n},t)
\end{equation}
These relations are valid for any $t$, and any boundary condition for the $\{ Z,\hat{Z} \} $ system (beyond its application to WNT).\\

{\bf Scattering for ${\phi}$}.
Next one has also
\begin{equation}
\begin{split}
\partial_x [e^{\I \frac{k}{2} x}\bar \phi_1(x,t,k)]&=- g \hat{Z}(x,t) \bar \phi_2(x,t,k)e^{\I \frac{k}{2} x}, \qquad \partial_x [e^{-\I \frac{k}{2} x} \bar \phi_2(x,t,k)] = Z(x,t) \bar \phi_1(x,t,k)e^{-\I \frac{k}{2} x}
\end{split}
\label{eq:SuppMatLaxPairEqGeneral2}
\end{equation}
Integrating the second equation of \eqref{eq:SuppMatLaxPairEqGeneral2} from $x=-\infty$ to $x=+\infty$, and using that $\tilde a(k,t)=\tilde a(k)$ we have
\begin{equation}
- \tilde a(k)+ I_d= \int_{\R} \rmd x \, Z(x,t) \bar \phi_1(x,t,k)e^{-\I \frac{k}{2} x}
\end{equation}
and from $-\infty$ and a value $x$
\begin{equation}
e^{-\I \frac{k}{2} x} \bar \phi_2(x,t,k) = - I_d + \int_{-\infty}^{x} \rmd x' \, Z(x',t) \bar \phi_1(x',t,k)e^{-\I \frac{k}{2} x'} 
\end{equation}
Integrating the first equation of \eqref{eq:SuppMatLaxPairEqGeneral2} between $-\infty$ and $+\infty$, we have 
\begin{equation}
\tilde b(k) e^{k^2 t} = - g  \int_{\R} \rmd x \hat{Z}(x,t) \bar \phi_2(x,t,k)e^{\I \frac{k}{2} x}
\end{equation}
and from $-\infty$ and a value $x$
\begin{equation}
e^{\I \frac{k}{2} x}\bar \phi_1(x,t,k) = - g  \int_{-\infty}^{x} \rmd x' \, \hat{Z}(x',t) \bar \phi_2(x',t,k)e^{\I \frac{k}{2} x'}
\end{equation}

Inserting, this gives two integral equations for $\bar \phi_{1,2}$
\begin{equation}
\begin{split}
e^{\I \frac{k}{2} x} \bar \phi_1(x,t,k)&= - g  \int_{-\infty}^{x} \rmd x' \, \hat{Z}(x',t) 
e^{\I k x'} 
\left( - I_d + \int_{-\infty}^{x'} \rmd x'' \, Z(x'',t) \bar \phi_1(x'',t,k)e^{-\I \frac{k}{2} x''} \right)  \\
e^{-\I \frac{k}{2} x} \bar \phi_2(x,t,k)&=
- I_d - g \int_{-\infty}^{x} \rmd x' \, Z(x',t) e^{-\I k x'} 
\left(   \int_{-\infty}^{x'} \rmd x'' \, \hat{Z}(x'',t) \bar \phi_2(x'',t,k)e^{\I \frac{k}{2} x''} \right) 
\end{split}
\end{equation}

Iteration of these equations gives a series representation for $\tilde b(k)$ as a sum of alternating products of terms $\hat{Z} (Z \hat{Z})^{n}$, $n \geq 0$, integrated over ordered sectors as
\be \label{agaa4} 
\tilde b(k) e^{k^2 t} =  g \sum_{n=0}^\infty (-g)^n \int_{x_{2n+1}<\dots < x_1} \prod_{j=1}^{2n+1}  \rmd x_j e^{ \I k (\sum_{j=0}^n x_{2j+1} - \sum_{j=1}^n x_{2j} )}\hat{Z}(x_1,t)Z(x_2,t)\dots Z(x_{2n},t)\hat{Z}(x_{2n+1},t)
\ee
as well as a a series representation for $\tilde a(k)$ as a sum of alternating products of terms $(Z \hat{Z})^{n}$, $n \geq 0$, integrated over ordered sectors as
\begin{equation} \label{agaa5} 
\tilde a(k)= \sum_{n=0}^\infty (-g)^n \int_{x_{2n}<\dots < x_1} \prod_{j=1}^{2n} \rmd x_j e^{\I k (\sum_{j=1}^{n} x_{2j}-\sum_{j=0}^{n-1} x_{2j+1})}Z(x_1,t) \hat{Z}(x_2,t) \dots Z(x_{2n-1},t)\hat{Z}(x_{2n},t)
\end{equation}
These relations are valid for any $t$, and any boundary condition for the $\{ Z,\hat{Z} \} $ system (beyond its application to WNT). Let us apply them to the WNT with the boundary conditions \eqref{eq:init-cond-wnt-kpz}. \\

{\bf Scattering at $t=1$}. Setting $t=1$  we see that $\hat{Z}(x,1)= B \delta(x)$
implies that only the first two terms, $n=0$ and $n=1$, survive in the different series. Indeed for $n \geq 2$ the $\delta$ function
implies that in the integral all odd $x_{2n+1}=0$ and the integration over the even $x_{2n}$
will be restricted to a vanishing small interval, leading to a vanishing result since $Z(x,1)$ is a smooth function. Hence, from $\hat{Z}(x,1)= B \delta(x)$ we obtain 
\begin{itemize}
    \item from \eqref{agaa2} for $k \in \mathbb{H}^+$ 
    \be 
    \begin{split}
a(k) &= I_d - g \int_{x_2<x_1}  \rmd x_1  \rmd x_2 \hat{Z}(x_1,1) Z(x_2,1) e^{\I k (x_1-x_2) } \\
&= I_d - g B  \int_{x_2<0}  \rmd x_2 Z(x_2,1) e^{- \I k x_2 }\\
&= I_d - g B  Z_-(k)
\end{split}
\ee 
\item from \eqref{agaa5} for $k \in \mathbb{H}^-$
\be 
\begin{split}
\tilde a(k) &= I_d - g \int_{x_2<x_1}  \rmd x_1  \rmd x_2 Z(x_1,1) \hat{Z}(x_2,1) e^{\I k (x_2-x_1) }\\
&= I_d - g   \int_{x_1>0}  \rmd x_1 Z(x_1,1) e^{- \I k x_1 } B \\
&= I_d - g  Z_+(k) B 
\end{split}
\ee 
    \item from \eqref{agaa}
\be\label{agaanew} 
\begin{split}
& b(k) e^{-k^2} = \int_\R \rmd x_1 e^{- \I k x_1} Z(x_1,1) -g 
 \int_{x_{3}<x_{2}< x_1} \prod_{j=1}^{3}  \rmd x_j e^{ \I k (x_2- x_1-x_3)}
 Z(x_1,1) \hat{Z}(x_2,1) Z(x_{3},1)  \\
 & = \int_\R \rmd x_1 e^{- \I k x_1} Z(x_1,1) -g \left( \int_{x_1>0} \rmd x_1   Z(x_1,1) e^{ - \I k x_1} \right) \, B \, \left(  \int_{x_3<0} \rmd x_3 Z(x_{3},1) e^{ - \I k x_3} \right) \\
 & = Z_+(k)+Z_-(k) -g Z_+(k) B Z_-(k) \\
 \end{split}
\ee
\item from \eqref{agaa4} only the first term $n=0$ remains
\be 
\tilde b(k) e^{k^2} =  g \int_{\R} \rmd x_1 \hat{Z}(x_1,1) = g B 
\ee 
\end{itemize}

We summarise the results obtained in Table~\ref{table:scattering-t1}.\\

\begin{table}[  t!]
    \centering
    \begin{tabular}{p{4cm} p{6cm}  }
    \hline 
        \hline 
        &\\[-1.5ex]
        scattering coefficient & expression  \\[1.5ex]
        \hline &\\[-1.5ex]
      $\tilde a(k)$   & $ I_d- g Z_+(k) B $  \\[1ex]
      $ a(k)$   & $ I_d- g B Z_-(k) $  \\[1ex]
      $ \tilde{b}(k)$   & $ g e^{-k^2} B $  \\[1ex]
      $ {b}(k)e^{-k^2}$   & $Z_+(k) + Z_-(k) - g Z_+(k) B Z_-(k)   $  \\[1ex]
        \hline
    \end{tabular}
    \caption{Summary of the scattering analysis at $t=1$. Here $Z_\pm(k)$ refers to $Z(t=1)$}
    \label{table:scattering-t1}
\end{table}

{\bf Scattering at $t=0$}. For the droplet initial condition $Z(x,0)= A \delta(x)$, by setting $t=0$ 
one obtains similarly
\begin{itemize}
    \item for \eqref{agaa2} for $k \in \mathbb{H}^+$
    \be 
    \begin{split}
a(k) &= I_d - g \int_{x_2<x_1}  \rmd x_1  \rmd x_2 \hat{Z}(x_1,0) Z(x_2,0) e^{\I k (x_1-x_2) } \\
&= I_d - g   \int_{x_1>0}  \rmd x_1 \hat{Z}(x_1,0) e^{ \I k x_1 } A \\
&=I_d -g \hat{Z}_+(k)A
\end{split}
\ee 
\item for \eqref{agaa5} for $k \in \mathbb{H}^-$
\be 
\begin{split}
\tilde a(k) &= I_d - g \int_{x_2<x_1}  \rmd x_1  \rmd x_2 Z(x_1,0) \hat{Z}(x_2,0) e^{\I k (x_2-x_1) }\\
&= I_d - g  A  \int_{x_2<0}  \rmd x_2 \hat{Z}(x_2,0) e^{\I k x_2 } \\
&= I_d -g A \hat{Z}_-(k)
\end{split}
\ee  
\item for \eqref{agaa4}
\be \label{agaa4new2} 
\begin{split}
& \tilde b(k) = g \int_{\R} \rmd x_1 \hat{Z}(x_1,0) 
- g^2 \int_{x_{3}<x_{2}< x_1} \prod_{j=1}^{3}  \rmd x_j e^{ \I k (x_1 + x_3  - x_2 )}\hat{Z}(x_1,0)Z(x_2,0) \hat{Z}(x_{3},0) \\
& = g \int_{\R} \rmd x_1 \hat{Z}(x_1,0)  
- g^2 
\left( \int_{x_1>0}  \rmd x_1 \, \hat{Z}(x_1,0) e^{  \I k x_1} \right) \, A \, \left(  \int_{x_3<0} \rmd x_3 \,  \hat{Z}(x_{3},0) e^{  \I k x_3} \right) \\
&= g \hat{Z}_+(k)+g \hat{Z}_-(k)-g^2 \hat{Z}_+(k)A\hat{Z}_-(k)
\end{split}
\ee

\item for \eqref{agaa} only the first term $n=0$ remains
\be 
b(k) = \int_\R \rmd x_1 e^{- \I k x_1} Z(x_1,0) = A 
\ee 
\end{itemize}

We summarise the results obtained in Table~\ref{table:scattering-t0}.\\

Additionally, we note the relation if the solutions are even in $x$, i.e., $\hat{Z}(x,t)=\hat{Z}(-x,t)$
\begin{equation}
    \hat{Z}_+(k)=(\hat{Z}_-(k^*))^*
\end{equation}
This implies that
\begin{equation}
\label{eq:scattering-coeff-complex-conj}
    a(k)=(\tilde{a}(k^*))^\dagger
\end{equation}
We expect $ {a}(k)$ to be analytic in the upper-half plane $\mathbb{H}^+$ and $ \tilde{a}(k)$ in the lower-half plane $\mathbb{H}^-$.

\begin{table}[  t!]
    \centering
    \begin{tabular}{p{4cm} p{6cm}  }
    \hline 
        \hline 
        &\\[-1.5ex]
        scattering coefficient & expression  \\[1.5ex]
        \hline &\\[-1.5ex]
      $\tilde a(k)$   & $ I_d - g A \hat{Z}_-(k) $  \\[1ex]
      $ a(k)$   & $ I_d - g \hat{Z}_+(k) A  $  \\[1ex]
      $ \tilde{b}(k)$   & $ g \hat{Z}_+(k) +g \hat{Z}_-(k) -g^2 \hat{Z}_+(k) A \hat{Z}_-(k) $  \\[1ex]
      $ {b}(k)$   & $A  $  \\[1ex]
        \hline
    \end{tabular}
    \caption{Summary of the scattering analysis at $t=0$. Here $\hat Z_\pm(k)$ refers to $\hat Z(t=0)$.}
    \label{table:scattering-t0}
\end{table}

\subsection{Transformation of the scattering coefficients from the $GL_d(\R)$ invariance}
From the explicit expressions of the scattering coefficients \eqref{agaa} \eqref{agaa2} \eqref{agaa4} and \eqref{agaa5}, the gauge transformation on the fields $\{Z,\hat{Z} \}$ \eqref{eq:app-gauge-transform-field} translate for the scattering coefficients into the invariance upon

\begin{equation}
    b(k)\to X^\intercal b(k) X, \; a(k)\to X^{-1}a(k) X, \; \tilde{b}(k)\to X^{-1}\tilde{b}(k)(X^\intercal)^{-1}, \; \tilde{a}(k) \to X^\intercal \tilde{a}(k) (X^\intercal)^{-1}
\end{equation}
for any $X\in  GL_d(\R)$.

\subsection{Inverse scattering for the large deviation problem}
From Table~\ref{table:scattering-t0}, using the fact that $A$ is invertible, we close the equation involving the scattering coefficients and $A$ as follows
\begin{equation}
\label{eq:normalisation_relation}
    \tilde{b}(k)=A^{-1}-a(k)A^{-1}\tilde{a}(k) \quad \Longleftrightarrow \quad A^{1/2}BA^{1/2}e^{-k^2}=I_d-A^{1/2}a(k)A^{-1}\tilde{a}(k) A^{1/2}
\end{equation}
Before considering the complete matrix Riemann-Hilbert problem we first recall the solution 
in the scalar case which already provides a solution for the determinant of the scattering matrix coefficients.

\subsubsection{Scalar Riemann-Hilbert analysis}
From \eqref{eq:normalisation_relation}, we deduce that 
\begin{equation}
\label{eq:normalisation-scattering}
\begin{split}
    \det(a(k))\, \det(\tilde{a}(k))
    &=\det (I_d-gA^{1/2}BA^{1/2}e^{-k^2})\\
    \end{split}
\end{equation}
As we expect $\det {a}(k)$ to be analytic in the upper-half plane $\mathbb{H}^+$ and $\det \tilde{a}(k)$ in the lower-half plane $\mathbb{H}^-$, a scalar Riemann Hilbert analysis \cite{tsai2023integrability,UsWNTCrossover,ablowitz2004discrete} gives us that
\begin{equation}
\label{eq:solution-scalar-RH-det}
    \det(a(k))=\begin{cases}
        e^{\varphi(k)}, \quad  & k \in \mathbb{H}^+\\
        G(k)^{1/2}e^{\tilde{\varphi}(k)}, \quad  & k \in \R\\
        G(k)e^{\varphi(k)}, \quad  & k \in \mathbb{H}^-\\
    \end{cases}, \qquad 
    \det(\tilde{a}(k))=\begin{cases}
        G(k) e^{-\varphi(k)}, \quad  & k \in \mathbb{H}^+\\
        G(k)^{1/2}e^{-\tilde{\varphi}(k)}, \quad  & k \in \R\\
        e^{-\varphi(k)}, \quad  & k \in \mathbb{H}^-\\
    \end{cases}
\end{equation}
where $G(k)=\det (I_d-g A^{1/2}BA^{1/2} e^{-k^2})$ and
\begin{equation}
\label{eq:riemann-hilbert1}
\begin{split}
    \varphi(k)&= \int_\R \frac{\rmd q}{2\I \pi} \frac{k}{q^2-k^2}\log \det (I_d-g A^{1/2}BA^{1/2} e^{-q^2})= \int_\R \frac{\rmd q}{2\I \pi} \frac{k}{q^2-k^2}\Tr \log (I_d-g A^{1/2}BA^{1/2} e^{-q^2})\\
\end{split}
\end{equation}
and when $k$ is real, we consider the integral with a principal value
\begin{equation}
\begin{split}
   \tilde{\varphi} (k)&= \dashint_\R \frac{\rmd q}{2\I \pi} \frac{k}{q^2-k^2}\log \det (I_d-g A^{1/2}BA^{1/2} e^{-q^2})\\
\end{split}
\end{equation}

Note that the above solutions depend only on the matrix $B'=A^{1/2}BA^{1/2}$ as discussed in the main text.
As in the case of the WNT for the (scalar) KPZ equation, solitonic solution can appear depending on the situation.
For each eigenspace of the matrix $A^{1/2}BA^{1/2}$, denoting the corresponding eigenvalue by $b_i' $, the zeroes of \eqref{eq:normalisation-scattering} are given by
\begin{equation}
    e^{\kappa^2}=gb_i'
\end{equation}
For $g b'_i\in (0,1]$ there exist two purely imaginary solutions $\{ \I \kappa^{(i)}_0 \in \mathbb{H}^+$, $-\I \kappa^{(i)}_0 \in \mathbb{H}^- \}$. The solitonic solutions of the inverse scattering problem yielding physical solutions to the large deviation problem are then constructed as follows.  We identify the set of eigenvalues which verify $gb_i'\in (0,1]$, i.e., $\mathcal{S}=\{i \in [1,d] \mid  gb_i'\in (0,1] \}$ and $|\mathcal{S}|=n$. We then select a subset $\mathcal{S}'\subseteq \mathcal{S}$ so that

\begin{equation}
\label{eq:solution-scalar-RH-det-soliton}
    \det(a(k))= \prod_{i \in \mathcal{S}'} \frac{k-\I \kappa_0^{(i)}}{k+\I \kappa_0^{(i)}} \times \begin{cases}
        e^{\varphi(k)}, \quad  & k \in \mathbb{H}^-\\
        G(k)^{1/2}e^{\tilde{\varphi}(k)}, \quad  & k \in \R\\
        G(k)e^{\varphi(k)}, \quad  & k \in \mathbb{H}^+\\
    \end{cases}
\end{equation}
and
\begin{equation}
    \det(\tilde{a}(k))=\prod_{i \in \mathcal{S}'} \frac{k+\I \kappa_0^{(i)}}{k-\I \kappa_0^{(i)}} \times \begin{cases}
        G(k) e^{-\varphi(k)}, \quad  & k \in \mathbb{H}^-\\
        G(k)^{1/2}e^{-\tilde{\varphi}(k)}, \quad  & k \in \R\\
        e^{-\varphi(k)}, \quad  & k \in \mathbb{H}^+\\
    \end{cases}
\end{equation}
Note that
\begin{equation}
\label{eq:riemann-hilbert2}
    \prod_{i \in \mathcal{S}'} \frac{k-\I \kappa_0^{(i)}}{k+\I \kappa_0^{(i)}}= \exp \left(-2\I \sum_{i \in \mathcal{S}'} \arctan(\frac{\kappa_0^{(i)}}{k})\right)
\end{equation}
The sum over $\mathcal{S}'$ can be seen as a partial trace over the solitons induced by the eigenspaces of the matrix $A^{1/2}BA^{1/2}$ allowing the spontaneous generation of solitons.

\subsubsection{Large $k$ expansion}
From Table~\ref{table:scattering-t1} we find that at large $k$, using $Z_{\pm}(k)\underset{k\to \infty}{\simeq} \pm\frac{1}{\I k}Z(0,1)$, we have that
\begin{equation}
\label{eq:large-k-exp-scattering}
    a(k)\underset{k\to \infty}{\simeq} I_d +\frac{g}{\I k}BZ(0,1), \quad \tilde{a}(k)\underset{k\to \infty}{\simeq} I_d -\frac{g}{\I k}Z(0,1)B
\end{equation}
in their domain of analyticity. Hence the log-determinant behaves at large $k$ as
\begin{equation} \label{det} 
    \log \Det (a(k)) \simeq \frac{g}{\I k}\Tr[BZ(0,1)], \quad \log \Det (\tilde{a}(k)) \simeq -\frac{g}{\I k}\Tr[Z(0,1)B]
\end{equation}
which leads to after the large $k$ expansion of \eqref{eq:riemann-hilbert1} and \eqref{eq:riemann-hilbert2}
\begin{itemize}
    \item In the absence of soliton, identifying the leading large $k$ behavior in \eqref{det}, \eqref{eq:solution-scalar-RH-det} and \eqref{eq:riemann-hilbert1},
    one obtains
    \begin{equation}
\begin{split}
    g\Tr[BZ(0,1)]&=-\int_\R \frac{\rmd q}{2 \pi} \Tr \log (I_d-g A^{1/2}BA^{1/2} e^{-q^2})\\
    &=\int_\R \frac{\rmd q}{2 \pi} \Tr \, \mathrm{Li}_1 (g A^{1/2}BA^{1/2} e^{-q^2}) = \frac{1}{\sqrt{4\pi}} \Tr \, \mathrm{Li}_{3/2} (g A^{1/2}BA^{1/2} )
\end{split}
\end{equation}
\item In the presence of solitons
\begin{equation}
\begin{split}
    g\Tr[BZ(0,1)]    &= \frac{1}{\sqrt{4\pi}} \Tr \, \mathrm{Li}_{3/2} (g A^{1/2}BA^{1/2} )+2\sum_{i \in \mathcal{S}'}\kappa_0^{(i)}
\end{split}
\end{equation}
\end{itemize}

This gives only a partial information on $Z(0,1)$. To perform the Legendre transform in order 
to obtain the large deviation function (see below) , we need more information. To
this aim we now turn to the full matrix problem. 

\subsubsection{Matrix Riemann-Hilbert analysis }

Recalling that $A$ is invertible and positive definite symmetric, the matrix scattering problem is 

\begin{equation}
    A^{1/2}a(k)A^{-1}\tilde{a}(k) A^{1/2}=I_d-A^{1/2}BA^{1/2}e^{-k^2}
\end{equation}

Recalling from Eq.~\eqref{eq:scattering-coeff-complex-conj} that if the solutions of the WNT equations are even in space $Z(x,t)=Z(-x,t)$ and similarly for $\hat{Z}$ then
\begin{equation}
    a(k)=(\tilde{a}(k^*))^{\dagger}
\end{equation}
Defining $X(k)=\frac{1}{A^{1/2}}\tilde{a}(k)A^{1/2}$ and $(X(k^*))^\dagger=A^{1/2}a(k)\frac{1}{A^{1/2}}$, we rewrite the Riemann-Hilbert problem as
\begin{equation}
    (X(k^*))^\dagger X(k) =I_d-A^{1/2}BA^{1/2}e^{-k^2}
\end{equation}
with $X(k)$ being analytic in $\mathbb{H}^-$. We conjecture that the solution to the matrix Riemann-Hilbert problem reads

\begin{equation}
    X(k)=\frac{1}{A^{1/2}}\tilde{a}(k)A^{1/2}=\begin{cases}
         e^{- \varphi(k)}(I_d-A^{1/2}BA^{1/2}e^{-k^2}), \quad  & k \in \mathbb{H}^+\\
        e^{- \tilde{\varphi}(k)}\sqrt{I_d-A^{1/2}BA^{1/2}e^{-k^2}}, \quad  & k \in \R\\
        e^{- \varphi(k)}, \quad  & k \in \mathbb{H}^-\\
    \end{cases}
\end{equation}
and
\begin{equation}
\label{eq:solution-matrix-RH-det}
   (X(k^*))^\dagger= A^{1/2}a(k)\frac{1}{A^{1/2}}=\begin{cases}
         e^{ \varphi(k)}, \quad  & k \in \mathbb{H}^+\\
        \sqrt{I_d-A^{1/2}BA^{1/2}e^{-k^2}}e^{ \tilde{\varphi}(k)}, \quad  & k \in \R\\
        (I_d-A^{1/2}BA^{1/2}e^{-k^2}) e^{ \varphi(k)}, \quad  & k \in \mathbb{H}^-\\
    \end{cases}
\end{equation}
with a matrix-valued phase
\begin{equation}
\label{eq:riemann-hilbert1-matrix}
    \begin{split}
        \varphi(k)&= \int_\R \frac{\rmd q}{2\I \pi} \frac{k}{q^2-k^2}\log (I_d-g A^{1/2}BA^{1/2} e^{-q^2})
    \end{split}
\end{equation}
The conjugacy relation \eqref{eq:scattering-coeff-complex-conj} implies that $\varphi(k)=-(\varphi(k^*))^\dagger$ which we explicitly verify. Furthermore for the spectral parameter on the real axis $k\in \R$, the phase is defined from its principal value.
\begin{equation}
    \begin{split}
        \tilde{\varphi}(k)&= \dashint_\R \frac{\rmd q}{2\I \pi} \frac{k}{q^2-k^2}\log (I_d-g A^{1/2}BA^{1/2} e^{-q^2})
    \end{split}
\end{equation}

The rationale behind the conjecture is as follows. 

\begin{itemize}
    \item For $k \to \infty$, $\varphi(k)\to 0$, ensuring that the scattering coefficients are asymptotically the identity.
    \item The Riemann-Hilbert equation \eqref{eq:normalisation-scattering} is verified for all $k\in \C$.
    \item The solution $X(k)$ is continuous when $k$ approaches the real axis as is easily seen in the polar coordinates. Indeed let us introduce
    the polar decomposition 
    \be 
    B'= A^{1/2}BA^{1/2}= R'^{-1} \Lambda' R'   \quad , \quad \Lambda' = {\rm diag}(b'_i) 
    \ee
    where $R'$ is a rotation matrix in $\mathrm{SO}(d)$. Then 
    one has
\be 
\varphi(k) = \int_\R \frac{\rmd q}{2\I \pi} \frac{k}{q^2-k^2} R'^{-1} \log (I_d-g \Lambda' e^{-q^2}) R' = R'^{-1}\left[\int_\R \frac{\rmd q}{2\I \pi} \frac{k}{q^2-k^2}  \log (I_d-g \Lambda' e^{-q^2}) \right]R'
\ee 
and the same holds for $\tilde{\varphi}(k)$. Using the Sokhotski–Plemelj formula on the diagonal matrix for each subspace related to the eigenvalue $b'_i$, this implies for $k \in \R$
\be 
\varphi(k + \I 0^+) = \tilde{\varphi}(k) + \frac{1}{2} \log (I_d-g A^{1/2}BA^{1/2} e^{-k^2}) 
\ee 
and that the matrices on the right hand side commute as they are diagonalisable in the same basis using $R'$. Therefore their exponentials commute. 
\end{itemize}

The matrix RH problem also admits solitonic solution which we also conjecture as follows. For each eigenspace of the matrix $A^{1/2}BA^{1/2}$, denoting the corresponding eigenvalue by $b_i' $, the zeroes of \eqref{eq:normalisation-scattering} are given by
\begin{equation}
    e^{\kappa^2}=gb_i'
\end{equation}
For $g b'_i\in (0,1]$ there exist two purely imaginary solutions $\{ \I \kappa^{(i)}_0 \in \mathbb{H}^+$, $-\I \kappa^{(i)}_0 \in \mathbb{H}^- \}$. The solitonic solutions of the inverse scattering problem yielding physical solutions to the large deviation problem are then constructed as follows.  We identify the set of eigenvalues which verify $gb_i'\in (0,1]$, i.e., $\mathcal{S}=\{i \in [1,d] \mid  gb_i'\in (0,1] \}$ and $|\mathcal{S}|=n$. We then select a subset $\mathcal{S}'\subseteq \mathcal{S}$ and construct a soliton matrix 
\begin{equation}
    \kappa = R'^{-1} \mathrm{Diag}\left(\frac{k+\I \kappa_0^{(i)}}{k-\I \kappa_0^{(i)}}\right) R'
\end{equation}
where $\kappa_0^{(i)}=0$ if $i \notin \mathcal{S}'$ and

\begin{equation}
    \frac{1}{A^{1/2}}\tilde{a}(k)A^{1/2}=\kappa \times \begin{cases}
         e^{- \varphi(k)}(I_d-A^{1/2}BA^{1/2}e^{-k^2}), \quad  & k \in \mathbb{H}^+\\
        e^{- \tilde{\varphi}(k)}\sqrt{I_d-A^{1/2}BA^{1/2}e^{-k^2}}, \quad  & k \in \R\\
        e^{- \varphi(k)}, \quad  & k \in \mathbb{H}^-\\
    \end{cases}
\end{equation}
and
\begin{equation}
   A^{1/2}a(k)\frac{1}{A^{1/2}}=\kappa^{-1} \times \begin{cases}
         e^{ \varphi(k)}, \quad  & k \in \mathbb{H}^+\\
        \sqrt{I_d-A^{1/2}BA^{1/2}e^{-k^2}}e^{ \tilde{\varphi}(k)}, \quad  & k \in \R\\
        (I_d-A^{1/2}BA^{1/2}e^{-k^2}) e^{ \varphi(k)}, \quad  & k \in \mathbb{H}^-\\
    \end{cases}
\end{equation}

\subsubsection{Large $k$ expansion of the matrix RH problem and large deviation function}

We recall the large $k$ expansion  \eqref{eq:large-k-exp-scattering} of the scattering coefficients

\begin{equation}
\label{eq:asymptotics-scattering-coeff-large-k}
    A^{1/2}a(k)\frac{1}{A^{1/2}}\underset{k\to \infty}{\simeq} I_d +\frac{g}{\I k}A^{1/2}BZ(0,1)\frac{1}{A^{1/2}}, \quad \frac{1}{A^{1/2}}\tilde{a}(k)A^{1/2}\underset{k\to \infty}{\simeq} I_d -\frac{g}{\I k}\frac{1}{A^{1/2}}Z(0,1)BA^{1/2}
\end{equation}
in their respective domain of analyticity.\\

\begin{itemize}
    \item In the absence of soliton, identifying the leading large $k$ behavior in \eqref{eq:asymptotics-scattering-coeff-large-k}, \eqref{eq:solution-matrix-RH-det} and \eqref{eq:riemann-hilbert1-matrix}, one obtains from $a(k)$
    \begin{equation}
    \label{eq:saddle-point-derivative-1}
    g A^{1/2}BZ(0,1)A^{-1/2}=-  \int_\R \frac{\rmd q}{2 \pi}\log (I_d-g A^{1/2}BA^{1/2} e^{-q^2})
\end{equation}
and from $\tilde{a}(k)$
\begin{equation}
\label{eq:saddle-point-derivative-2}
    g A^{-1/2}Z(0,1)B A^{1/2}=-  \int_\R \frac{\rmd q}{2 \pi}\log (I_d-g A^{1/2}BA^{1/2} e^{-q^2})
\end{equation}
so that with $B'=A^{1/2}BA^{1/2}$ we finally obtain 
\begin{equation} \label{Z01solu}
    Z(0,1)= \frac{1}{\sqrt{4\pi}g} A^{1/2} B'^{-1} \mathrm{Li}_{3/2}(g B') A^{1/2} = \frac{1}{\sqrt{4\pi}g} A^{1/2} R'^{-1} \Lambda'^{-1} \mathrm{Li}_{3/2}(g \Lambda') R' A^{1/2} 
\end{equation}
where $ \mathrm{Li}_{3/2}$ denotes the polylogarithm of index $3/2$ and we also note that $ B'^{-1} $ commutes with $\mathrm{Li}_{3/2}(B')$ so that \eqref{eq:saddle-point-derivative-1} and \eqref{eq:saddle-point-derivative-2} are consistent.

\item In the presence of solitons
    \begin{equation}
    \label{eq:saddle-point-derivative-soliton-1}
    \begin{split}
    g A^{1/2}BZ(0,1)A^{-1/2}&=-  \int_\R \frac{\rmd q}{2 \pi}\log (I_d-g A^{1/2}BA^{1/2} e^{-q^2})+2  R'^{-1} \mathrm{Diag}\left(\kappa_0^{(i)}\right) R'\\
    \end{split}
\end{equation}
and thus 

\begin{equation}
\label{eq:saddle-point-derivative-soliton-2}
    Z(0,1) =  \frac{1}{g} A^{1/2} R'^{-1} \Lambda'^{-1} \left[\frac{\mathrm{Li}_{3/2}(g \Lambda')}{\sqrt{4\pi}}+2 \,  \mathrm{Diag}\left(\kappa_0^{(i)}\right) \right] R' A^{1/2} 
\end{equation}

\end{itemize}

\subsubsection{Legendre transform considerations}

Let us recall that the PDF ${\cal P}_A(Z)$ of $Z={\cal Z}(0,1)$ with initial condition ${\cal Z}(x,0)=A \delta(x)$ satisfies the large deviation principle
\be 
{\cal P}_A(Z) \sim e^{- \frac{1}{\varepsilon} \hat \Phi_A(Z)}
\ee 
where $\hat \Phi_A(Z)$ is the rate function. This implies the following large deviation form for the observable
\bea \label{defPsiA}
\langle e^{ \frac{1}{\varepsilon} \Tr [B {\cal Z}(0,1)] }  \rangle \sim e^{- \frac{1}{\varepsilon} \Psi_A(- g B)} 
\eea 
where the minus sign in the argument of $\Psi_A$ ensures consistency with previous conventions of Ref.~\cite{UsWNTDroplet2021}
in the scalar case $d=1$,
setting $g B=-z$. The function $\Psi_A$ can be related to $\hat \Phi_A$ by a saddle point evaluation of the
expectation value in \eqref{defPsiA} 
\be 
 \Psi_A(-g B) =  
\min_{Z \in {\cal P}_d} [ 
 \hat \Phi_A(Z) - \Tr[B Z] ] 
\ee 
This is a matrix Legendre transform, where the minimizer $Z=Z_B$ is the solution of the matrix equation
\begin{equation}
    \frac{\rmd \hat{\Phi}_A(Z)}{\rmd Z}|_{Z=Z_B} =B, \quad \frac{\rmd \hat{\Phi}_A(Z)}{\rmd Z_{ij}}|_{Z=Z_B} =B_{ij}
\end{equation}
We can invert the Legendre transform and obtain the dual variational problem

\be \label{derPsi}
\frac{ \rmd \Psi_A(-gB) }{\rmd B_{ij}} = g (Z_B)_{ij} 
\ee
To determine $\Psi_A$ we can now identify the optimal value $Z_B$ with $Z(0,1)$ from the solution of the WNT equations
obtained above in \eqref{Z01solu} (through the solution to the scattering problem). One can then
integrate \eqref{derPsi} and obtain
\begin{equation}
\label{eq:large-dev-generating-func}
        \Psi_A(-gB) = -\Tr\left[\frac{\mathrm{Li}_{5/2}(gA^{1/2}BA^{1/2})}{\sqrt{4\pi}}\right] = -\Tr\left[\frac{\mathrm{Li}_{5/2}(g \Lambda')}{\sqrt{4\pi}} \right] 
        = -  \frac{1}{\sqrt{4\pi}} \sum_{i=1}^d \mathrm{Li}_{5/2}(g b'_i) = \sum_{i=1}^d \Psi_{\rm KPZ}(-g b'_i) 
\end{equation}
where we recall that the $b'_i$ are the eigenvalues of $B'=A^{1/2} B A^{1/2}$. 
This can be checked using the formula, valid for $B$ symmetric and $A \in {\cal P}_d$ and differentiable matrix function $f$
\begin{equation}
    \frac{\delta}{\delta B_{ij}} \Tr [f(A^{1/2}BA^{1/2})] = \left(A^{1/2}f'(A^{1/2}BA^{1/2})A^{1/2}\right)_{ij}
\end{equation}
and using that $\Psi_A(0)=0$ and $\frac{d}{\rmd x} \mathrm{Li}_{s}(x)= \mathrm{Li}_{s-1}(x)/x$. Note that 
$\Psi_{\rm KPZ}(z)= -  \frac{1}{\sqrt{4\pi}} \mathrm{Li}_{5/2}(-z) $ here is
the non-solitonic branch (main branch) of the rate function for the KPZ equation. 

\begin{remark}
    
From the invariance of the equations of the weak-noise theory under the action of $GL_d(\R)$ one has
\begin{equation}
    \Psi_A(-gB) = \Psi_{I_d}(-gA^{1/2}BA^{1/2}) = \Psi_{I_d}(-g B')
\end{equation}
as mentionned in the text.
\end{remark}
\begin{remark}
    The formula \eqref{eq:large-dev-generating-func} was obtained in the case where there is no soliton. In the presence of soliton, a continuation is required as in \eqref{eq:saddle-point-derivative-soliton-2} as
    \begin{equation}
    \label{eq:large-dev-generating-func-with-soliton}
        \Psi^{(continued)}_A(-gB) =   \sum_{i=1}^d \Psi^{(continued)}_{\rm KPZ}(-g b'_i), \quad \Psi^{continued}_{\rm KPZ}(z)=\Psi_{\rm KPZ}(z)+ \frac{4}{3}(-\log(-z))^{3/2}
    \end{equation}
    where $g b'_i$ varies in $(0,1]$.
\end{remark}

Upon inversion of the Legendre transform one obtains the 
rate function $\hat \Phi_A(Z)$ in the form of the following parametric system, where the matrix $B$ acts as the varying parameter
\begin{equation}
\begin{cases}
Z= \frac{1}{\sqrt{4\pi}g} A^{1/2} (A^{1/2}BA^{1/2})^{-1} \mathrm{Li}_{3/2}(g A^{1/2}BA^{1/2}) A^{1/2}\\
   \hat \Phi_A(Z) =  -\Tr\left[\frac{\mathrm{Li}_{5/2}(gA^{1/2}BA^{1/2})}{\sqrt{4\pi}}\right] + \Tr \left[ \frac{\mathrm{Li}_{3/2}(gA^{1/2}BA^{1/2})}{\sqrt{4\pi }g}\right]
   \end{cases}
\end{equation}
One can check that it has the following invariance under the action of $X\in GL_d(\R)$
\begin{equation}
\hat \Phi_A(Z)= \hat \Phi_{X^{\intercal} A X}(X^{\intercal} Z X ) = 
\hat \Phi_{I_d}(A^{-1/2} Z A^{-1/2})
\end{equation}    
where we chose $X=A^{-1/2}$ in the last equality. Hence we can focus on the function $\hat \Phi_{I_d}(Z)$ which reads in parametric form
\begin{equation}
\begin{cases}
Z= \frac{1}{\sqrt{4\pi}g}  B^{-1} \mathrm{Li}_{3/2}(g B) \\
   \hat \Phi_{I_d}(Z) =  -\Tr\left[\frac{\mathrm{Li}_{5/2}(gB)}{\sqrt{4\pi}}\right] + \Tr \left[ \frac{\mathrm{Li}_{3/2}(gB)}{\sqrt{4\pi }g}\right]
   \end{cases}
\end{equation}
This means that to compute $\hat \Phi_{I_d}(Z)$ for a given matrix $Z$, one first find
the eigenvalues $\lambda_i$ of $Z$, and its
polar decomposition $Z= R^{-1} \Lambda R$ with $\Lambda = {\rm diag}(\lambda_i)$ and $R$
a ${\rm SO}(d)$ rotation matrix. The matrix $B$ is then equal to $B= R^{-1} {\rm diag}(b_i) R$ 
where the $b_i$ and $\hat \Phi_{I_d}(Z)$ are given by
\be 
\label{eq:parametric-phi-hat-without-soliton}
\begin{cases}
\lambda_i =  \frac{1}{\sqrt{4\pi}}   \frac{\mathrm{Li}_{3/2}(g b_i)}{g b_i} =  \Psi_{\rm KPZ}'(z)|_{z=- g b_i} , \quad i=1, \dots, d\\
    \hat \Phi_{I_d}(Z) = \sum_{i=1}^d \left[ - \frac{\mathrm{Li}_{5/2}(gb_i)}{\sqrt{4\pi}} +\frac{\mathrm{Li}_{3/2}(gb_i)}{\sqrt{4\pi }g}\right]
    = \sum_{i=1}^d [ \Psi_{\rm KPZ}(z) - z \Psi_{\rm KPZ}'(z)]_{z=- g b_i}
\end{cases}
\ee 
In conclusion the PDF ${\cal P}_A(Z)$ in the large deviation regime is a product
measure over the eigenvalues of $A^{-1/2} Z A^{-1/2}$, which we denote $\lambda_i[A^{-1/2} Z A^{-1/2}]$, i.e., 
one can write at leading order
\bea 
\label{eq:phi-kpz-matrix}
{\cal P}(Z) \sim  \prod_{i=1}^d \, e^{ - \frac{1}{\varepsilon} \hat \Phi_{\rm KPZ}(\lambda_i[A^{-1/2} Z A^{-1/2}])}
\eea 
Note that the PDF of ${\cal P}_A(Z)$ may also contain level repulsion terms which are subdominant
in the large deviation regime.\\

The system \eqref{eq:parametric-phi-hat-without-soliton} is valid for  $\lambda_i \leq \lambda_c = \frac{\zeta(3/2)}{\sqrt{4\pi}}$ and does not admit a solution in the case $\lambda_i > \lambda_c$. The rate function $\hat \Phi_{I_d}(Z)$ when some of the $\lambda_i$ are larger than $\lambda_c$ is obtained 
by including the solitonic contribution, and replace in the corresponding sector $\Psi_{\rm KPZ}(z) \to \Psi^{(continued)}_{\rm KPZ}(z) $ defined in formula \eqref{eq:large-dev-generating-func-with-soliton}. This phenomenon was interpreted in  \cite{UsWNTDroplet2021} as a spontaneous generation of soliton in the WNT system.

\subsection{Mapping of the large deviation problem to the matrix classical Heisenberg chain}

Following Ref.~\cite{ishimori1982relationship}, we map the matrix NLS problem to a matrix classical Heisenberg chain as follows. We choose a solution of the Lax system for the choice of the spectral parameter $k=0$, i.e., solve for the $2d \times 2d$ matrix $\mathsf{g}$ the system
\begin{equation}
    \p_x \mathsf{g} = \begin{pmatrix}
0  & - g \hat Z\\  Z &0
\end{pmatrix}  \mathsf{g}, \quad \p_t \mathsf{g} =
    \begin{pmatrix}
     - g \hat Z Z & g \partial_x  \hat Z\\   
      \partial_x  Z &  g Z \hat Z
    \end{pmatrix} \mathsf{g}
\end{equation}
and use $\mathsf{g}$ as a gauge to obtain new Lax matrices (see Eq.~\eqref{eq:definition-gauge-transform})
\begin{equation}
    \tilde{U}_1=\mathsf{g}^{-1}U_1 \mathsf{g}-\mathsf{g}^{-1}\p_x \mathsf{g}, \quad \tilde{U}_2=\mathsf{g}^{-1}U_2 \mathsf{g}-\mathsf{g}^{-1}\p_t \mathsf{g}
\end{equation}
We further define a spin matrix $\mathsf{S}$ as
\begin{equation}
\mathsf{S}=\mathsf{g}^{-1}\begin{pmatrix}
        I_d & 0\\
        0& -I_d
    \end{pmatrix} \mathsf{g}
\end{equation}
This ensures the following properties  $\mathsf{S}^2=I_{2d}$, $\Tr \,  \mathsf{S}=0$, $\Det \mathsf{S}=(-1)^d$ as well as
\begin{equation}
    \mathsf{S} \p_x \mathsf{S} =- \p_x \mathsf{S} \mathsf{S} =2\mathsf{g}^{-1}\begin{pmatrix}
0  & - g \hat Z\\  Z &0
\end{pmatrix}\mathsf{g}, \quad 
    \tilde{U}_1 = -\frac{\I k}{2}\mathsf{S}, \quad \tilde{U}_2= \frac{k^2}{2}\mathsf{S}+\frac{\I k}{2} \mathsf{S} \p_x \mathsf{S}
\end{equation}
The compatibility then gives the matrix-valued classical Heisenberg chain.
\begin{equation}
    \p_t \mathsf{S} = \frac{1}{2}[\p_x^2 \mathsf{S},\mathsf{S}]=\frac{1}{2}\p_x [\p_x \mathsf{S},\mathsf{S}]
\end{equation}

\section{Field theory of the matrix \LGPol}

The matrix generalisation of the \LGPol was introduced recently in Ref.~\cite{MatrixWhittaker2023} and studied in the context of Matrix Whittaker Processes. It is described by the following recursion on a partition function $Z_{n,m} \in {\mathcal P}_d$
\be 
\label{eq:app-log-gamma-def}
Z_{n,m} = ( Z_{n-1,m}  + Z_{n,m-1} )^{1/2} V_{n,m}  ( Z_{n-1,m}  + Z_{n,m-1} )^{1/2} 
\ee 
This recursion is applied on the rectangle $1 \leq n \leq N-1$, $1 \leq m \leq M-1$. We have taken the convention that the top-right coordinate for the polymer lies at $(N-1,M-1)$ as depicted in Fig.~\ref{fig:loggamma-polymer-lightcone-wnt}. The random matrices $V_{n,m} \in {\mathcal P}_d$ are distributed with an  inverse Wishart (iW) law

\bea 
\label{eq:def-inverse-wishart-pdf}
\rmd P^{iW}_{\alpha,g}[V]= \frac{(2g)^{-\alpha d}}{\Gamma_d(\alpha) } |\det V|^{-\alpha  } e^{- \frac{1}{2g} \Tr \, V^{-1} } \mu(\rmd V)
\eea 
with $\alpha > \frac{d-1}{2}$ and 
\be
\label{eq:normalisation-inverse-wishart}
\Gamma_d(\alpha) = \pi^{\frac{d(d-1)}{4}} \prod_{k=1}^d \Gamma(\alpha - \frac{k-1}{2})  \,.
\ee  
Whenever we choose $g=1/2$, as we do in this Section, we will choose the short-hand notation $\rmd P^{iW}_{\alpha,g=1/2}[V] =\rmd P^{iW}_\alpha[V]$.\\

The evolution of the partition function starts from a list of "initial values"
\begin{equation}
\label{eq:log-gamma-initial-values}
    \begin{split}
 Z_i &= ( Z_{0,0}, \; ( Z_{n,0} )_{ n\in [1,N-1]}, \; ( Z_{0,m})_{ m \in [1,M-1]} )\in \mathcal{P}_d^{N+M-1} 
    \end{split}
\end{equation}
and ends with a list of a "terminal values"
\begin{equation}
\label{eq:log-gamma-terminal-values}
    \begin{split}
         Z_f &= (  Z_{N-1,M-1}, (Z_{n,M-1})_{n\in [1,N-2]}, (Z_{N-1,m})_{m\in [1,M-2]} )\in \mathcal{P}_d^{N+M-3} 
    \end{split}
\end{equation}
Note that the two corners $Z_{N-1,0}$ and $Z_{0,M-1}$ are not included in $Z_f$, since the recursion is not applied at these points. 
Below, we will also need the list of "final values" which are the terminal values complemented with the corners, 
\begin{equation}
\label{eq:log-gamma-final-values}
    \begin{split}
         Z'_f &= (  Z_{N-1,M-1}, (Z_{n,M-1})_{n\in [1,N-1]}, (Z_{N-1,m})_{m\in [1,M-1]} )\in \mathcal{P}_d^{N+M-1} 
    \end{split}
\end{equation}

An example of initial values is the polymer defined from a single source as

\begin{equation} \label{PPinitial} 
   Z_i= \begin{cases}
       Z_{0,m}=0, \quad & 0\leq m \leq M-1\\ 
       Z_{1,0}= I_d \\
       Z_{n,0}=0 \quad & 2 \leq n \leq N-1
    \end{cases}
\end{equation}
so that $Z_{n,m}$ is referred to as the point-to-point partition function. Subsequently we will also study (non-deterministic) invariant boundary conditions.  \\

\begin{remark} 
    For $d \geq 1$, the matrix partition sum $Z_{n,m}$ does not have an obvious interpretation as a simple sum over paths of some products of random matrices. However taking the trace of
    \eqref{eq:app-log-gamma-def} one obtains
\be 
\Tr [Z_{n,m}] = \Tr [V_{n,m}  Z_{n-1,m}]  +  \Tr [V_{n,m} Z_{n,m-1}] 
\ee
Hence $\Tr [Z_{n,m}]$ is still equal to the sum of traces of ordered products of the $V_{n,m}$ over paths terminating at $(n,m)$.
For $d=1$ one recovers the usual interpretation of a partition sum for the scalar \LGPol~. The identity $\det (A^{1/2} B A^{1/2} - z I_d)= \det (B A - z I_d)$ shows that the characteristic polynomial of $Z_{n,m}$ can be written as $\det( Z_{n,m}  -  z I_d) = \det( V_{n,m}  (Z_{n-1,m}  + Z_{n,m-1}) -  z I_d)$
but it does not seem  to lead to a path decomposition interpretation beyond the trace.
\end{remark}

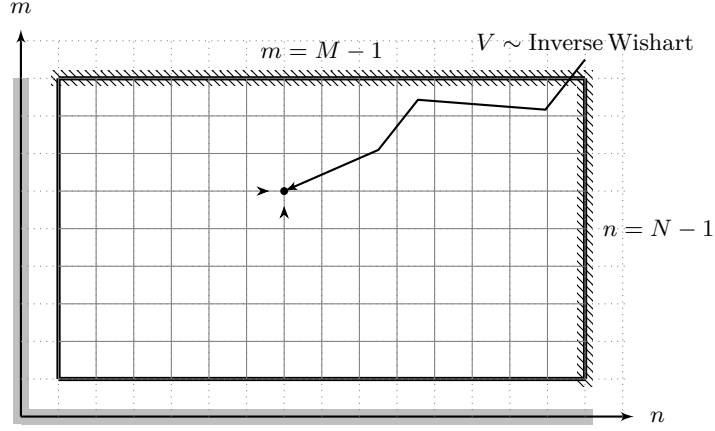
\begin{figure}[t!]
\begin{tikzpicture}[scale=0.5]

\fill[gray!50] (0,-0.2) rectangle (15.2,0.2);
\fill[gray!50] (-0.2,-0.2) rectangle (0.2,9);

\fill[pattern=north west lines, pattern color=black] (0.8,8.8) rectangle (15,9.2);
\fill[pattern=north west lines, pattern color=black] (14.8,0.8) rectangle (15.2,9.2);
% 
%% axis
\draw[->, thick,>=latex'] (0, 0) -- (0, 10.3);
\draw[->, thick,>=latex'] (0,0)--( 16.3, 0);

% %% light cone for z
% \draw[ultra thick] (0,1) -- (8,9);
% \draw[ultra thick] (0,1) -- (7,1);

% %% light cone for ztilde
% \draw[ultra thick] (8,9) -- (15,9);
% \draw[ultra thick] (7,1) -- (15,9);

%% Light cone for z
\draw[ultra thick] (1,1) -- (1,9) ;
\draw[ultra thick] (1,1) -- (15,1) ;

%% Light cone for ztilde
\draw[ultra thick] (1,9) -- (15,9) node[midway, above, yshift=0.1cm] {$m=M-1$};
\draw[ultra thick] (15,1) -- (15,9) node[midway, right, xshift=0.1cm] {$n=N-1$};

\foreach \k in {1,2, ..., 16}
	{\draw[gray, dotted] (\k, 0) -- (\k, 10.1);}
\foreach \k in {1,2, ..., 10}
	{\draw[gray, dotted] (0,\k) -- (16.1, \k);}

\clip (-2, -1) rectangle (19.5, 11.3);

\foreach \k in { 1,2, ..., 9}
	{\draw[gray] (1,\k) -- (15, \k);}
\foreach \k in{ 1, 2, ..., 15}
	{\draw[gray] (\k, 1) -- (\k,9);}

 \draw[black, thick, decorate, decoration={zigzag, segment length=30mm, amplitude=4mm, post length=6mm}, ->,>=latex'] (15,9.5) node[above] {$V\sim \mathrm{Inverse \, Wishart}$} -- (7,6);

\draw[-Stealth] (7,5.4) -- (7,5.6);
\draw[-Stealth] (6.4,6) -- (6.6,6);
\fill (7,6) circle(3pt);

 % \draw(15,9) node[anchor=west]{$(N-1, M-1)$};

\node[above] at (0,10.5) {$m$};
\node[right] at (16.5,0) {$n$};

\end{tikzpicture}
\caption{The black thick lines define the rectangle where the recursion of the \LGPol is applied. The gray band corresponds to the initial values of the partition function $Z_i$ (see Eq.~\eqref{eq:log-gamma-initial-values}) and the band filled with diagonal lines corresponds to the terminal values of the partition function $Z_f$ (see Eq.~\eqref{eq:log-gamma-terminal-values}). The two arrowheads describe how the recursion propagates on the lattice. The response field $\hat{Z}_{n,m}$ is non-zero only in the black thick rectangle where the recursion is enforced. We denote by $\mathcal{C}=\{(n,m) | n\in [1,N-1], m\in [1,M-1] \}$ all lattice sites within the black thick rectangle (including the boundary) - and $\mathcal{C}_{bulk}=\{(n,m) | n\in [1,N-2], m\in [1,M-2] \}$ all lattice sites minus the final points described by the band filled with diagonal lines.}
\label{fig:loggamma-polymer-lightcone-wnt}
\end{figure}

Let us denote $Z^R$ the space time matrix field $Z^R= \{ Z^R_{n,m} \}_{(n,m)\in \mathcal{C}}$ 
which is solution of the recursion \eqref{eq:app-log-gamma-def} with a given fixed initial condition $Z_i$, for 
a given noise $V$. 
Using the MSR method, 
the expectation value over the noise $V$ of any function ${\cal O}[Z^R]$ can be written as a multiple integral involving a (symmetric) response matrix field $\hat Z$
\be 
\label{eq:average-observable-msr-log-gamma}
\E_V[\mathcal{O}(Z^R)] = \iiint {\cal D}_\mu V  {\cal D}\hat Z  {\cal D} Z \, {\cal O}[Z] \, 
e^{-  \hat{S}_0[Z,\hat{Z},V] }
\ee 
Equivalently, we will denote $\overline{\mathcal{O}(Z)}=\E_V[\mathcal{O}(Z^R)]$. The integration is performed over all fields with indices included in $\mathcal{C}$ (as defined in \ref{fig:loggamma-polymer-lightcone-wnt}) with a fixed $Z_i$ and where the action reads
\be 
\label{eq:app-log-gamma-action-three-fields}
\begin{split}
&\hat{S}_0[Z,\hat{Z},V] =\\
&\sum_{(n,m)\in \mathcal{C}}  \Tr \left[ \hat{Z}_{n,m} \big(Z_{n,m}-
 ( Z_{n-1,m}  + Z_{n,m-1} )^{1/2} V_{n,m}  ( Z_{n-1,m}  + Z_{n,m-1} )^{1/2} \big)\right] + \alpha  \log \det V_{n,m} + 
  \Tr \, V_{n,m}^{-1} 
  \end{split}
\ee

The MSR integral which appears in the r.h.s of \eqref{eq:average-observable-msr-log-gamma} is normalised
to unity, as a consequence of the identity $
\iint \rmd Z \rmd \hat Z e^{ {\rm Tr} \hat Z (Z - Z_0)}  = 1 $, consistent with $\E_V[1] =1$
(upon choosing $\mathcal{O}[\cdot]=1$). We recall that the different measures are defined in Section~\ref{subsec:definition-measures}. Equivalently, the right hand side of \eqref{eq:average-observable-msr-log-gamma} will be denoted 
\begin{equation}
\label{eq:MSR-log-gamma-expectation-value}
    \E_V[\mathcal{O}(Z^R)]=\E[\mathcal{O}(Z)|Z_i]
\end{equation}
to express the fact that the path integral provides an expectation conditioned to the initial list. Thus, one can complement the path integral by a convolution with a measure on the initial partition function $Z_i$.\\

The integration over the response matrix field $\hat Z_{n,m}$ enforces the recursion \eqref{eq:app-log-gamma-def} for a given $(n,m)\in \mathcal{C}$. The noise can be integrated out but contrary to the Wishart case (for the strict-weak polymer, see Section~\ref{sec:strict-weak-app}), the characteristic function of an inverse Wishart gives a matrix Bessel function which is not as explicit as in the Wishart case, hence we do not explore this path further.

\begin{remark}
 Our recursion relations \eqref{eq:app-log-gamma-def} are identical to the ones introduced in 
Ref.~\cite{MatrixWhittaker2023} for the matrix log Gamma polymer in Eqs.~(1.8)--(1.9),
where the partition sums $Z_{n,m}$ are called $\{ Z^i(n) \}_{i\geq 1,n \geq 0}$. They
were studied there however only for the analog of the point-to-point polymer geometry, i.e., with
initial conditions $Z^i(0)=\delta_{i1} I_d$. In that paper, the matrix Whittaker process was introduced.
It is a certain Markov processes on triangular arrays 
$\{ X_j^i(n) \}_{1 \leq i \leq j \leq N}$ of matrices in ${\cal P}_d$ (where $n$ is a discrete "time"). It is such that its restriction to the right edge of the triangle
identifies with the point-to-point matrix \LGPol, i.e., $X_1^i(n)=Z^i(n)$. Similarly, according to Remark~3.4 there, the left edge $X_i^i(n)$ identifies with the matrix generalisation of the strict-weak polymer in a point-to-point geometry, see the Remark~3.4 there and the recursion~(3.6) which is identical to our recursion \eqref{eq:recursion-matrix-SW}, whose
scalar version was introduced in  \cite{o2015tracy,corwin2015strict}. Finally the "bottom edge" process $\{ X_i^N(n) \}_{i=1,\dots,N}$ is also shown to be an autonomous process.  In Corrolary~4.10 it is shown that the fixed time law of the $\{ Z^i(n) \}$ is given by a marginal of 
the matrix Whittaker measure, defined in that paper (and which gives the fixed time law of the triangular array) and which can be expressed in terms of matrix Whittaker functions, also defined there. This is a matrix generalisation of \cite{Corwin_2014}. 
\end{remark}

\subsection{Weak-noise theory and saddle point equations for the matrix \LGPol}

To investigate the weak-noise theory of the matrix \LGPol, one needs to first proceed to the change of variable
\begin{equation}
    V_{n,m} \to \alpha^{-1} V_{n,m} , \quad 
Z_{n,m} \to \alpha^{1-n-m} Z_{n,m}  \quad , \quad 
\hat{Z}_{n,m} \to  \alpha^{n+m} \hat{Z}_{n,m} 
\end{equation}
which leaves the recursion relation \eqref{eq:app-log-gamma-def} unchanged.
This ensures that the whole action \eqref{eq:app-log-gamma-action-three-fields} is now proportional to $\alpha$, 
i.e., $\hat{S}_0[Z,\hat{Z},V] \to \alpha  \hat{S}_0[Z,\hat{Z},V]$. The weak-noise limit
is defined by taking $\alpha$ large, i.e., $\alpha \gg 1$. In that limit
the action becomes "classical", i.e., the dynamics is governed by the saddle point equations associated to $S_0$. 
This corresponds to a weak-noise large deviation regime. One can for instance
consider an observable of the type (expressed in terms of these rescaled matrix fields) 
\be 
G^\alpha[J] = \overline{\exp\left( \alpha \sum_{n,m} \Tr \left[ J_{n,m} Z_{n,m} \right] \right)}
\ee 
where we recall that the overline denotes the expectation value over the noise and 
where $J$ is an arbitrary matrix source field. This observable can be expressed as the lattice path integral
\begin{equation}
    G^\alpha[J] = \iiint \mathcal{D}_\mu {V}\mathcal{D} {Z}\mathcal{D} \hat{Z} e^{-\alpha S_J[V,Z,\hat{Z}]}
\end{equation}
where $S_J = S_0 - \sum_{n,m} \Tr [ J_{n,m} Z_{n,m} ]$. In the large $\alpha$ limit
it takes the form
\be 
G^\alpha[J] \underset{\alpha \gg 1}{\sim} e^{- \alpha S_J^*}
\ee 
where $S_J^*$ is the value of the action $S_J[V,Z,\hat{Z}]$ at the saddle point for the fields. Let us now derive the saddle point equations associated to $S_0$ (and $S_J$). We start with the saddle point with respect to the response field, which yields
\be 
\label{supp:WNT-matrix-LG-1}
Z_{n,m} = ( Z_{n-1,m}  + Z_{n,m-1} )^{1/2} V_{n,m}  ( Z_{n-1,m}  + Z_{n,m-1} )^{1/2} 
\ee 
The identity \eqref{eq:inversion-identity} allows to invert Eq.~\eqref{supp:WNT-matrix-LG-1} as
\begin{equation}
\label{eq:matrix-log-gamma-identity}
    ( Z_{n-1,m}  + Z_{n,m-1} )^{1/2}=V_{n,m}^{-1/2}(V_{n,m}^{1/2}Z_{n,m}V_{n,m}^{1/2})^{1/2}V_{n,m}^{-1/2}
\end{equation}
or if we square this identity
\begin{equation}
\label{eq:matrix-log-gamma-identity-square}
    Z_{n-1,m}  + Z_{n,m-1} =V_{n,m}^{-1/2}(V_{n,m}^{1/2}Z_{n,m}V_{n,m}^{1/2})^{1/2}V_{n,m}^{-1}(V_{n,m}^{1/2}Z_{n,m}V_{n,m}^{1/2})^{1/2}V_{n,m}^{-1/2}
\end{equation}
The saddle point with respect to the noise reads
\begin{equation}
\label{eq:matrix-log-gamma-diagonalisation-noise}
    V_{n,m}^{-1}-V_{n,m}^{-2}=( Z_{n-1,m}  + Z_{n,m-1} )^{1/2}\hat{Z}_{n,m}( Z_{n-1,m}  + Z_{n,m-1} )^{1/2}
\end{equation}
Injecting \eqref{eq:matrix-log-gamma-identity} in \eqref{eq:matrix-log-gamma-diagonalisation-noise}, we obtain
\begin{equation}
    I_d-V_{n,m}^{-1}=(V_{n,m}^{1/2}Z_{n,m}V_{n,m}^{1/2})^{1/2}V_{n,m}^{-1/2}\hat{Z}_{n,m}    V_{n,m}^{-1/2}(V_{n,m}^{1/2}Z_{n,m}V_{n,m}^{1/2})^{1/2}
\end{equation}
Injecting the expression of $V_{n,m}^{-1}$ in the right hand side of Eq.~\eqref{eq:matrix-log-gamma-identity-square}, we find the first equation of the saddle point system

\begin{equation}
\label{eq:matrix-log-gamma-SP1}
\begin{split}
    Z_{n-1,m}  + Z_{n,m-1} &=Z_{n,m}-Z_{n,m} \hat{Z}_{n,m}Z_{n,m}\\
\end{split}
\end{equation}

The last saddle point equation is obtained by differentiating with respect to the field $Z$ using the expansion \eqref{eq:saddle-point-sqrt}. Calling $( Z_{n,m}  + Z_{n+1,m-1} )^{1/2}=Z_1^{1/2}$ and $( Z_{n-1,m+1}  + Z_{n,m} )^{1/2}=Z_2^{1/2}$, the saddle point reads
\begin{equation}
\label{eq:matrix-log-gamma-SP2-intermediate}
\begin{split}
   \hat{Z}_{n,m}=& \int_0^\infty \rmd t\, e^{-tZ_1^{1/2}}[\hat{Z}_{n+1,m}Z_1^{1/2}V_{n+1,m}+V_{n+1,m}Z_1^{1/2}\hat{Z}_{n+1,m}] e^{-tZ_1^{1/2}}\\
   &+\int_0^\infty \rmd t\, e^{-tZ_2^{1/2}}[\hat{Z}_{n,m+1}Z_2^{1/2}V_{n,m+1}+V_{n,m+1}Z_2^{1/2}\hat{Z}_{n,m+1}] e^{-tZ_2^{1/2}}
   \end{split}
\end{equation}

We use Eqs.~\eqref{supp:WNT-matrix-LG-1} and \eqref{eq:matrix-log-gamma-SP1} to rewrite the integrand of \eqref{eq:matrix-log-gamma-SP2-intermediate}. We have that 
\begin{equation}
\begin{split}
    Z_1^{1/2}V_{n+1,m}&=Z_{n+1,m}Z_1^{-1/2}\\
    &=(I_d-Z_{n+1,m}\hat{Z}_{n+1,m})^{-1}Z_1^{1/2}
\end{split}
\end{equation}
as well as
\begin{equation}
\begin{split}
    V_{n+1,m}Z_1^{1/2}&=Z_1^{-1/2}Z_{n+1,m}\\
    &=Z_1^{1/2}(I_d-\hat{Z}_{n+1,m}Z_{n+1,m})^{-1}
\end{split}
\end{equation}
and similarly for $Z_2^{1/2}$ and $V_{n,m+1}$. Inserting these identities in \eqref{eq:matrix-log-gamma-SP2-intermediate}, this leads to

\begin{equation}
\begin{split}
   \hat{Z}_{n,m}
   &= \int_0^\infty \rmd t\, e^{-tZ_1^{1/2}}[\hat{Z}_{n+1,m}(I_d-Z_{n+1,m}\hat{Z}_{n+1,m})^{-1}Z_1^{1/2}+Z_1^{1/2}(I_d-\hat{Z}_{n+1,m}Z_{n+1,m})^{-1}\hat{Z}_{n+1,m}] e^{-tZ_1^{1/2}}\\
   &+ \int_0^\infty \rmd t\, e^{-tZ_2^{1/2}}[\hat{Z}_{n,m+1}(I_d-Z_{n,m+1}\hat{Z}_{n,m+1})^{-1}Z_2^{1/2}+Z_2^{1/2}(I_d-\hat{Z}_{n,m+1}Z_{n,m+1})^{-1}\hat{Z}_{n,m+1}] e^{-tZ_2^{1/2}}\\
   &= -\int_0^\infty \rmd t\, \p_t(e^{-tZ_1^{1/2}}\hat{Z}_{n+1,m}(I_d-Z_{n+1,m}\hat{Z}_{n+1,m})^{-1} e^{-tZ_1^{1/2}})\\
    &-\int_0^\infty \rmd t\, \p_t(e^{-tZ_2^{1/2}}\hat{Z}_{n,m+1}(I_d-Z_{n,m+1}\hat{Z}_{n,m+1})^{-1} e^{-tZ_2^{1/2}})\\
   &= \hat{Z}_{n+1,m}(I_d-Z_{n+1,m}\hat{Z}_{n+1,m})^{-1} +\hat{Z}_{n,m+1}(I_d-Z_{n,m+1}\hat{Z}_{n,m+1})^{-1} \\
   \end{split}
\end{equation}
where in presence of a source $J$ the term $J_{n,m}$ should be added to the r.h.s. To symmetrize the equation, we introduce a modified response field $Y_{n,m}$ which reads
\begin{equation}
\begin{split}
    Y_{n,m}= \hat{Z}_{n,m}(I_d-Z_{n,m}\hat{Z}_{n,m})^{-1} &\Longleftrightarrow \hat{Z}_{n,m}=(I_d+Y_{n,m}Z_{n,m})^{-1}Y_{n,m} \\
    &\Longleftrightarrow (I_d+Y_{n,m}Z_{n,m})(I_d-\hat{Z}_{n,m}Z_{n,m})=I_d
    \end{split}
\end{equation}
To summarise we have obtained that solving the WNT of the \LGPol by the saddle point method is equivalent to solving the following discrete nonlinear matrix system (Eqs.~\eqref{eq:WNT-matrix-LG} in the main text)
\begin{equation} \label{systemLG} 
    \begin{split}
           Z_{n,m}(I_d+Y_{n,m}Z_{n,m})^{-1}&=Z_{n-1,m}  + Z_{n,m-1}     \\
        (I_d+Y_{n,m}Z_{n,m})^{-1}Y_{n,m}&=Y_{n+1,m}  + Y_{n,m+1} 
    \end{split}
\end{equation}
where in presence of a source $J$ the term $J_{n,m}$ should be added to the r.h.s. of the second equation. \\

To show that this system is integrable we exhibit an explicit Lax pair 
so that the nonlinear system \eqref{systemLG} can be equivalently rewritten as a linear system
$\vec v_{n+1,m}=L_{n,m} \vec v_{n,m}$ and $\vec v_{n,m+1}=U_{n,m} \vec v_{n.m}$,
where $\Vec{v}$ is a two component pair of $d \times d$-matrices.
The explicit $2d \times 2d$ Lax matrices  are presented in the main text in Eqs.~\eqref{eq:LaxPair-matrix-LG-1} and \eqref{eq:LaxPair-matrix-LG-2} (which have been normalised so that $\Det L_{n,m}=-1$ and $\Det U_{n,m}=1$ and are thus independent of the spectral parameter $\lambda$). Let us rewrite them here in a more compact form
\be 
L_{n,m} =  \left(
\begin{array}{cc}
 \frac{I_d}{\lambda } & \frac{Z_{n,m-1}}{\lambda }
   \\
 -\frac{Y_{n,m}}{\lambda } & -\frac{Y_{n,m} 
   Z_{n,m-1}}{\lambda }-\lambda I_d  \\
\end{array}
\right)
\quad , \quad 
U_{n,m} = \left(
\begin{array}{cc}
 \frac{I_d}{\sqrt{\lambda ^2+1}} &
   -\frac{Z_{n-1,m}}{\sqrt{\lambda ^2+1}} \\
 \frac{Y_{n,m}}{\sqrt{\lambda ^2+1}} &
   \sqrt{\lambda ^2+1}I_d-\frac{Y_{n,m}
   Z_{n-1,m}}{\sqrt{\lambda ^2+1}} \\
\end{array}
\right)
\ee

To be able to rewrite the nonlinear system \eqref{systemLG} as a linear system one needs that the Lax matrices 
satisfy the compatibility condition $L_{n,m+1} U_{n,m} - U_{n+1,m} L_{n,m}= 0$ (obtained by writing $\vec v_{n+1,m+1}$
in two different ways). Let us now show
explicitly that if $Z,Y$ satisfy the system \eqref{systemLG} then this condition holds.\\

Let us define the $2 \times 2$ matrix $M = \lambda \sqrt{1+ \lambda^2} (L_{n,m+1} U_{n,m} - U_{n+1,m} L_{n,m})$.
It is equal to 
\bea 
&& M = \left(
\begin{array}{cc}
 I_d &  Z_{n,m}
   \\
 - Y_{n,m+1} & - Y_{n,m+1} Z_{n,m}-\lambda^2 I_d  \\
\end{array}
\right)  \left(
\begin{array}{cc}
 I_d &
   - Z_{n-1,m} \\
 Y_{n,m}  &
   (1+\lambda^2)I_d - Y_{n,m}
   Z_{n-1,m}  \\
\end{array}
\right) \\
&& - \left(
\begin{array}{cc}
 I_d &
   - Z_{n,m} \\
 Y_{n+1,m}  &
   (1+\lambda^2)I_d - Y_{n+1,m}
   Z_{n,m}  \\
\end{array}
\right) 
\left(
\begin{array}{cc}
 I_d &  Z_{n,m-1}
   \\
 - Y_{n,m} & - Y_{n,m} Z_{n,m-1}-\lambda^2 I_d  \\
\end{array}
\right) 
\eea 
Let us perform the block products being careful with the ordering of the matrices. One finds that $M(1,1)=0$ since each term is equal to $I_d + Z_{n,m} Y_{n,m}$.
Consider now the element $M(1,2)$. It is equal to 
\be 
M(1,2) = Z_{n,m} - (I_d + Z_{n,m} Y_{n,m}) (Z_{n-1,m}  + Z_{n,m-1}) 
\ee 
This vanishes from the first equation in \eqref{systemLG} using that $Z_{n,m}(I_d+Y_{n,m}Z_{n,m})^{-1} = (I_d+Z_{n,m} Y_{n,m})^{-1} Z_{n,m}$
as can be checked by expanding the inverse in series on both sides. 
Next one has 
\be 
M(2,1) = Y_{n,m} - (Y_{n,m+1} + Y_{n+1,m} ) (I_d + Z_{n,m} Y_{n,m}) 
\ee 
Again it vanishes from the second equation in \eqref{systemLG} using that  
$(I_d+Y_{n,m}Z_{n,m})^{-1}Y_{n,m}= Y_{n,m} (I_d+Z_{n,m} Y_{n,m})^{-1}$.
The term $M(2,2)$ has a part proportional to $\lambda^2$ which reads
\be 
M(2,2)|_{\lambda^2} = Y_{n,m} (Z_{n-1,m} + Z_{n,m-1}) -  (Y_{n+1,m} + Y_{n,m+1}) Z_{n,m} 
\ee 
If we use \eqref{systemLG} we obtain
\be 
M(2,2)|_{\lambda^2} = Y_{n,m} Z_{n,m} (I_d+ Y_{n,m} Z_{n,m})^{-1} - (I_d+ Y_{n,m} Z_{n,m})^{-1} Y_{n,m} Z_{n,m} = 0 
\ee  
Finally the remain part of $M(2,2)$ reads
\be \label{rest} 
M(2,2)|_{\lambda=0} = - Y_{n,m+1} Z_{n,m} + Y_{n,m+1} (I_d + Z_{n,m} Y_{n,m}) Z_{n-1,m} 
+ Y_{n,m} Z_{n,m-1} - Y_{n+1,m} (I_d + Z_{n,m} Y_{n,m}) Z_{n,m-1} 
\ee 
Let us rewrite the first equation in \eqref{systemLG} as $(I_d+Z_{n,m} Y_{n,m})^{-1} Z_{n,m} = Z_{n-1,m}  + Z_{n,m-1}$,
equivalently as $(I_d+Z_{n,m} Y_{n,m}) Z_{n-1,m} = Z_{n,m} -  (I_d+Z_{n,m} Y_{n,m}) Z_{n,m-1} $ and
substitute in the second term in \eqref{rest}. One obtains
\bea \label{rest2} 
&& M(2,2)|_{\lambda=0} \\
&& = - Y_{n,m+1} Z_{n,m} + Y_{n,m+1} Z_{n,m} 
- Y_{n,m+1} (I_d+Z_{n,m} Y_{n,m}) Z_{n,m-1} 
+ Y_{n,m} Z_{n,m-1} + Y_{n+1,m} (I_d + Z_{n,m} Y_{n,m}) Z_{n,m-1} \nn \\
&& = - (Y_{n,m+1} + Y_{n+1,m} ) (I_d + Z_{n,m} Y_{n,m}) Z_{n,m-1} + Y_{n,m} Z_{n,m-1} = 0 
\eea   
where in the last equation we have used the second equation
in \eqref{systemLG} in the equivalent form $ Y_{n,m} (I_d + Z_{n,m} Y_{n,m})^{-1}= Y_{n,m+1} + Y_{n+1,m}   $.\\

Having shown integrability it becomes possible to compute some large deviation rate functions. For instance consider the observable $\overline{e^{- \alpha \Tr[B Z_{N-1,M-1}]}}$ (which corresponds to choosing a source $J_{n,m}= - B \delta_{n,N-1} \delta_{m,M-1}$) for the point-to-point polymer, i.e., with initial data \eqref{PPinitial}. Since the field $\hat Z$ vanishes at infinity, it is  sufficient/equivalent to solve the
system \eqref{systemLG} without the source, i.e., setting $J=0$, and with the same initial data on $Z$, but with the final boundary condition on the response field 
\be 
\hat Z_{N-1,M-1}= - B \quad , \quad \hat Z_{n \geq N,m}=0, \, \forall m\geq 0, \quad \hat Z_{n,m \geq M}=0, \, \forall n\geq 0
\ee 
This can be done by studying the scattering problem for the linear system, as done here for the MSHE,
but is left for future studies.

 \subsection{FD transformation of the matrix Log-Gamma}
 
We will obtain below the invariant measure of the matrix Log-Gamma polymer by constructing a FD symmetry of the field theoretical action
 \be
 \label{app:eq:action-log-gamma}
 \begin{split}
& \hat{S}_0[Z,\hat{Z},V] = \sum_{(n,m)\in \mathcal{C}}  \Tr \left[ \hat{Z}_{n,m} \left(Z_{n,m}-
 ( Z_{n-1,m}  + Z_{n,m-1} )^{1/2} V_{n,m}  ( Z_{n-1,m}  + Z_{n,m-1} )^{1/2} \right)\right] \\
 & +  \alpha \log \det V_{n,m} + 
  \Tr \, V_{n,m}^{-1} 
 \end{split} 
\ee 

We suppose that the original dynamics is studied in the rectangle $\mathcal{C}$, see Fig.~\ref{fig:loggamma-polymer-lightcone-wnt}. The recursion
is thus applied in that rectangle, with fixed initial conditions $Z_i$ ($\hat{Z}_{n,m}$ is set to zero outside the region $\mathcal{C}$). The FD symmetry is defined using the following inversion of the field $Z_{n,m}$ and the affine transformation of the response field.

\be
\begin{split}
\label{eq:change-var-log-gamma-FDT}
& Z_{n,m} = Q_{n'-1,m'-1}^{-1} \quad , \quad n'=N-n \quad , \quad m'=M-m \\
&  Z_{n,m}^{1/2}\hat{Z}_{n,m} Z_{n,m}^{1/2} + \tilde{V}_{n,m}^{-1} = Q_{n',m'}^{1/2}\hat{Q}_{n',m'} Q_{n',m'}^{1/2} + \tilde{W}_{n',m'}^{-1}\\
& (n,m)\in \mathcal{C} \Leftrightarrow (n',m')\in \mathcal{C}
\end{split}
\ee  
This defines a map $(Z,\hat{Z}) \to (Q,\hat{Q})$ where all indices of all fields belong to ${\cal C}$.
One can think of $Q_{n',m'}$ as the process reversed in space and time from $Z_{n,m}$. \\

We need to relate the randomness of the initial process $V_{n,m}$ with the noise of the reversed process which we define as $W_{n',m'}$. The factors $\tilde{V}$ and $\tilde{W}$ defined in the affine transformation of the response field will subsequently be related to the noises through a similarity transformation $\tilde{V}=CVC^{-1}$ and $\tilde{W}=DW D^{-1}$ (see below). This similarity transformation preserves the determinant and the trace of the two noises. We further proceed by replacing the change of variable \eqref{eq:change-var-log-gamma-FDT} into the original action \eqref{app:eq:action-log-gamma}, we obtain
\be
\begin{split}
& \hat{S}_0[Z,\hat{Z},V] = \sum_{(n,m)\in \mathcal{C}}  \Tr \left[ Z_{n,m}^{1/2}\hat{Z}_{n,m}Z_{n,m}^{1/2} \left(I_d-
 Z_{n,m}^{-1/2}( Z_{n-1,m}  + Z_{n,m-1} )^{1/2} V_{n,m}  ( Z_{n-1,m}  + Z_{n,m-1} )^{1/2}Z_{n,m}^{-1/2} \right)\right] \\
 & + \alpha  \log \det V_{n,m} + 
  \Tr \, V_{n,m}^{-1} \\
   &= \sum_{(n',m')\in \mathcal{C}}  \Tr \bigg[ (Q_{n',m'}^{1/2}\hat{Q}_{n',m'} Q_{n',m'}^{1/2} +  \tilde W_{n',m'}^{-1}-\tilde V_{n,m}^{-1}  )\\
   &\left(I_d-
 Q_{n'-1,m'-1}^{1/2}( Q_{n',m'-1}^{-1}  + Q_{n'-1,m'}^{-1} )^{1/2} V_{n,m}  ( Q_{n',m'-1}^{-1}  + Q_{n'-1,m'}^{-1} )^{1/2}Q_{n'-1,m'-1}^{1/2} \right)\bigg] \\
 & + \alpha  \log \det V_{n,m} + 
  \Tr \, V_{n,m}^{-1} \\
\end{split}
\ee 
We identify the terms corresponding to the action in the novel variables and obtain

\be
\begin{split}
& \hat{S}_0[Z,\hat{Z},V]    = \hat{S}_0[Q,\hat{Q},W]+\sum_{(n',m')\in \mathcal{C}}  \Tr \bigg[ Q_{n',m'}^{1/2}\hat{Q}_{n',m'} Q_{n',m'}^{1/2}\\
   &\big(
 Q_{n',m'}^{-1/2}( Q_{n'-1,m'}  + Q_{n',m'-1} )^{1/2} W_{n',m'}  ( Q_{n'-1,m'}  + Q_{n',m'-1} )^{1/2}Q_{n',m'}^{-1/2} \\
 &-
 Q_{n'-1,m'-1}^{1/2}( Q_{n',m'-1}^{-1}  + Q_{n'-1,m'}^{-1} )^{1/2} V_{n,m}  ( Q_{n',m'-1}^{-1}  + Q_{n'-1,m'}^{-1} )^{1/2}Q_{n'-1,m'-1}^{1/2} \big) \\
 &- (\tilde{W}_{n',m'}^{-1}-\tilde{V}_{n,m}^{-1})\left(
 Q_{n'-1,m'-1}^{1/2}( Q_{n',m'-1}^{-1}  + Q_{n'-1,m'}^{-1} )^{1/2} V_{n,m}  ( Q_{n',m'-1}^{-1}  + Q_{n'-1,m'}^{-1} )^{1/2}Q_{n'-1,m'-1}^{1/2} \right)\bigg] \\
 & + \alpha  \log \det V_{n,m} - \alpha \log \det W_{n',m'} \\
\end{split}
\ee 

The relation between the two noises is found by cancelling the term proportional to the response field $\hat{Q}$. 
\begin{equation}
\label{eq:fdt-loggamma-noise}
\begin{split}
     &Q_{n',m'}^{-1/2}( Q_{n'-1,m'}  + Q_{n',m'-1} )^{1/2} W_{n',m'}  ( Q_{n'-1,m'}  + Q_{n',m'-1} )^{1/2}Q_{n',m'}^{-1/2} =\\
 &
 Q_{n'-1,m'-1}^{1/2}( Q_{n',m'-1}^{-1}  + Q_{n'-1,m'}^{-1} )^{1/2} V_{n,m}  ( Q_{n',m'-1}^{-1}  + Q_{n'-1,m'}^{-1} )^{1/2}Q_{n'-1,m'-1}^{1/2} 
 \end{split}
\end{equation}

This identity allows to relate the two log-Determinants as
\begin{equation}
\begin{split}
  \mathcal{I}_1&=\alpha  \log \det V_{n,m} - \alpha\log \det W_{n',m'} \\
  &= \alpha\log \det [Q_{n',m'}^{-1}( Q_{n'-1,m'}  + Q_{n',m'-1} )]  - \alpha\log \det [Q_{n'-1,m'-1}( Q_{n',m'-1}^{-1}  + Q_{n'-1,m'}^{-1} )]\\ 
    &= \alpha\log \Det  \left[\frac{Q_{n'-1,m'} Q_{n',m'-1}}{Q_{n'-1,m'-1} Q_{n',m'}} \right]
\end{split}
\end{equation}
where we have used the algebraic identity $(A+B) (A^{-1} + B^{-1})^{-1} = (I + A B^{-1}) B (I + A B^{-1})^{-1} A$, implying that $\Det \big[ (A+B) (A^{-1} + B^{-1})^{-1} \big] = \Det A \, \Det B$. These terms are telescopic and upon summation over $n'$ and $m'$ will generate boundary terms (see below).  The last term to treat is then 
\begin{equation}
\begin{split}
  \mathcal{I}_2=  - \Tr (\tilde{W}_{n',m'}^{-1}-\tilde{V}_{n,m}^{-1})\left(
 Q_{n'-1,m'-1}^{1/2}( Q_{n',m'-1}^{-1}  + Q_{n'-1,m'}^{-1} )^{1/2} V_{n,m}  ( Q_{n',m'-1}^{-1}  + Q_{n'-1,m'}^{-1} )^{1/2}Q_{n'-1,m'-1}^{1/2} \right)
\end{split}
\end{equation}
Writing the relation between the noises \eqref{eq:fdt-loggamma-noise} as $DWD^\intercal = CVC^\intercal$, this reads
\begin{equation}
\begin{split}
 \mathcal{I}_2=   \Tr[\tilde{V}_{n,m}^{-1}C_{n,m}V_{n,m}C_{n,m}^{\intercal}]- \Tr [\tilde{W}_{n',m'}^{-1}D_{n',m'}W_{n',m'}D_{n',m'}^{\intercal}]
\end{split}
\end{equation}
We now write explicitly the similarity transformation for the noises as
\begin{equation}
\label{eq:log-gamma-similarity-transformation-noise-fd}
    \tilde{W}_{n',m'}=D_{n',m'}W_{n',m'}D^{-1}_{n',m'}, \quad \tilde{V}_{n',m'}=C_{n',m'}V_{n',m'}C^{-1}_{n',m'}
\end{equation}
and thus obtain the remaining term
\begin{equation}
\begin{split}
 \mathcal{I}_2&=    \Tr [C_{n',m'}C_{n',m'}^{\intercal}] -  \Tr[D_{n,m}D_{n,m}^{\intercal}] \\
 &=\Tr[ Q_{n'-1,m'-1}( Q_{n',m'-1}^{-1}  + Q_{n'-1,m'}^{-1} ) ] - \Tr[ Q_{n',m'}^{-1}( Q_{n'-1,m'}  + Q_{n',m'-1} )] 
\end{split}
\end{equation}
These terms are also telescopic and upon summation over $n'$ and $m'$ will generate boundary terms. As a summary, we have obtained that
\begin{equation}
    \begin{split}
& \hat{S}_0[Z,\hat{Z},V]    = \hat{S}_0[Q,\hat{Q},W]+\sum_{(n',m')\in \mathcal{C}} \mathcal{I}_1+ \mathcal{I}_2  
    \end{split}
\end{equation}
The addition term  $\sum_{(n',m')\in \mathcal{C}} \mathcal{I}_1+ \mathcal{I}_2$ is a telescopic sum, 
and we now show that it can be expressed from consecutive ratios of partition sums over
the four boundaries in Fig.~\ref{fig:loggamma-polymer-lightcone-wnt}. \\

The sum over $\mathcal{I}_1$ involves solely the four corners in Fig.~\ref{fig:loggamma-polymer-lightcone-wnt} upon simplification. For any $\kappa \in ]\frac{d-1}{2\alpha}, 1-\frac{d-1}{2\alpha}[$  (which imposes further that $\alpha>d-1$, this restriction being necessary to normalise the measure over the ratios, see below) we have
\begin{equation}
\begin{split}
    \sum_{(n',m')\in \mathcal{C}}\mathcal{I}_1=& \alpha (-\log \Det Q_{0,0}-\log \Det Q_{N-1,M-1}+\log \Det Q_{0,M-1}+\log \Det Q_{N-1,0})\\
    =&\alpha \big(\kappa(\log \Det Q_{N-1,0}-\log \Det Q_{0,0})+(1-\kappa)(\log \Det Q_{0,M-1}-\log \Det Q_{0,0})\\
    &+\kappa(\log \Det Q_{0,M-1}-\log \Det Q_{N-1,M-1})+(1-\kappa)(\log \Det Q_{N-1,0}-\log \Det Q_{N-1,M-1})\big)\\
    =&\alpha \kappa  \sum_{n'=1}^{N-1} \big(\log \Det \frac{Q_{n',0}}{Q_{n'-1,0}}- \log \Det \frac{Q_{n',M-1}}{Q_{n'-1,M-1}}\big)+\alpha (1-\kappa)\sum_{m'=1}^{M-1} \big(\log \Det \frac{Q_{0,m'}}{Q_{0,m'-1}}-\log \Det \frac{Q_{N-1,m'}}{Q_{N-1,m'-1}} \big)\\
\end{split}
\end{equation}
We use the short-hand notation $AB^{-1}:=\frac{A}{B}$ when there is no ambiguity. The sum over $\mathcal{I}_2$ involves all consecutive ratios on the boundary of the domain, i.e.,
\begin{equation}
    \begin{split}
      \sum_{(n',m')\in \mathcal{C}}\mathcal{I}_2=& \sum_{n'=1}^{N-1} \Tr\bigg[  \frac{Q_{n'-1,0}}{ Q_{n',0}} -\frac{Q_{n'-1,M-1}}{ Q_{n',M-1}}  \bigg]+\sum_{m'=1}^{M-1} \Tr \bigg[ \frac{Q_{0,m'-1}}{ Q_{0,m'}} -\frac{Q_{N-1,m'-1}}{ Q_{N-1,m'}} \bigg]
    \end{split}
\end{equation}

leading to our final result
\begin{equation}
\label{eq:FDT-matrix-log-gamma-action}
\begin{split}
    &\hat{S}_0[Z,\hat{Z},V]  = \hat{S}_0[Q,\hat{Q},W]\\
    &+\sum_{n'=1}^{N-1} \Tr\bigg[  \frac{Q_{n'-1,0}}{ Q_{n',0}} -\frac{Q_{n'-1,M-1}}{ Q_{n',M-1}}  \bigg]+\alpha \kappa  \sum_{n'=1}^{N-1} \big(\log \Det \frac{Q_{n',0}}{Q_{n'-1,0}}- \log \Det \frac{Q_{n',M-1}}{Q_{n'-1,M-1}}\big)\\
    &+\sum_{m'=1}^{M-1} \Tr \bigg[ \frac{Q_{0,m'-1}}{ Q_{0,m'}} -\frac{Q_{N-1,m'-1}}{ Q_{N-1,m'}} \bigg]+\alpha (1-\kappa)\sum_{m'=1}^{M-1} \big(\log \Det \frac{Q_{0,m'}}{Q_{0,m'-1}}-\log \Det \frac{Q_{N-1,m'}}{Q_{N-1,m'-1}} \big)
    \end{split}
\end{equation}

Thus, upon the FD transformation, the action is preserved up to boundary terms. We can rewrite this FD relation by expressing the boundary terms in terms of $Z$ rather than $Q$.

\begin{equation}
\label{eq:FDT-matrix-log-gamma-action-with-Z}
\begin{split}
    &\hat{S}_0[Z,\hat{Z},V]  = \hat{S}_0[Q,\hat{Q},W]\\
    &+\sum_{n=1}^{N-1} \Tr\bigg[  \frac{Z_{n-1,M-1}}{ Z_{n,M-1}} -\frac{Z_{n-1,0}}{ Z_{n,0}}  \bigg]-\alpha \kappa  \sum_{n=1}^{N-1} \big(\log \Det  \frac{Z_{n-1,M-1}}{ Z_{n,M-1}}- \log \Det\frac{Z_{n-1,0}}{ Z_{n,0}}\big)\\
    &+\sum_{m=1}^{M-1} \Tr \bigg[ \frac{Z_{N-1,m-1}}{ Z_{N-1,m}} -\frac{Z_{0,m-1}}{ Z_{0,m}} \bigg]-\alpha (1-\kappa)\sum_{m=1}^{M-1} \big(\log \Det \frac{Z_{N-1,m-1}}{ Z_{N-1,m}} -\log \Det \frac{Z_{0,m-1}}{ Z_{0,m}} \big)
    \end{split}
\end{equation}

\subsection{Conventions and identities for the measures on the partition function ratios}

Until now there was no ordering ambiguity for the expression of the partition function ratios, since they appear only in traces and determinants.
However below we will need to express probability measures on such ratios and an ambiguity will appear. In this subsection, as a preliminary step, we define the two possible ordering for such ratios, and we explain their properties.  Introducing the notations for the symmetrised ratios of two matrices $A,B \in {\cal P}_d$
\begin{equation}
\label{eq:definition-convention-ratios}
   r\left(\frac{A}{B}\right)=B^{-1/2} A B^{-1/2}, \quad  \tilde{r}\left(\frac{A}{B}\right)=A^{1/2} B^{-1} A^{1/2}
\end{equation}
One can show the following identities on the product measures, using the relations 
\eqref{eq:measure-psd},  \eqref{detinverse} and\eqref{detg}
\be \label{AB}
\mu(\rmd B)  \mu(\rmd r\left(\frac{A}{B}\right)) = \mu (\rmd B)  \mu(\rmd A ) \quad , \quad 
\mu(\rmd A)  \mu(\rmd \tilde{r}\left(\frac{A}{B}\right)) = \mu(\rmd A ) \mu (\rmd B^{-1}) =  \mu(\rmd A ) \mu (\rmd B) 
\ee  

This allows to define a reference measure on the initial condition $Z_i$ defined in \eqref{eq:log-gamma-initial-values} which will be useful in the following as a building block of the invariant measure.
Applying iteratively \eqref{AB} this reference measure takes the four equivalent forms either (i) in terms of the partition functions only 
(ii) in terms of the consecutive ratios and one chosen "zero mode" (also called reference point)

\begin{equation}
\label{eq:log-gamma-jacobian-ratio-partitionfunc-finalline-identity-1}
    \begin{split}
           &\mu(\rmd Z_{0,0})\prod_{n=1}^{N-1}  \mu(\rmd Z_{n,0}) \prod_{m=1}^{M-1} \mu(\rmd  Z_{0,m})\\
           &=\mu(\rmd Z_{0,0}) \prod_{n=1}^{N-1}\mu(\rmd r\left(\frac{Z_{n,0}}{Z_{n-1,0}}\right))  \prod_{m=1}^{M-1} \mu (\rmd r\left(\frac{Z_{0,m}}{Z_{0,m-1}}\right)) \\
           &=\mu(\rmd Z_{N-1,0})\prod_{n=1}^{N-1}\mu(\rmd \tilde{r}\left(\frac{Z_{n,0}}{Z_{n-1,0}}\right))\prod_{m=1}^{M-1} \mu (\rmd r\left(\frac{Z_{0,m}}{Z_{0,m-1}}\right))  \\
           &=\mu(\rmd Z_{0,M-1}) \prod_{m=1}^{M-1} \mu (\rmd \tilde{r}\left(\frac{Z_{0,m}}{Z_{0,m-1}}\right))\prod_{n=1}^{N-1}\mu(\rmd r\left(\frac{Z_{n,0}}{Z_{n-1,0}}\right)) \\
    \end{split}
\end{equation}

Moving around the reference point $Z_{0,0}$ to the corners $Z_{N-1,0}$ or $Z_{0,M-1}$ has the effect to transform $r$ into $\tilde{r}$ in the measure, which amounts to change the orientation of the symmetrized ratio. This is particularly important to study different paths of the matrix Log-Gamma polymer, either a departing path from the origin $n=0,m=0$ or a down-right path. These distinctions between $r$ and $\tilde r$ are of course irrelevant for the scalar Log-Gamma polymer since
all the variables commute for $d=1$.\\

Applying again iteratively \eqref{AB}, we also obtain a reference
measure on the final condition $Z'_f$ on the final condition $Z_f'$, defined in \eqref{eq:log-gamma-final-values} 

\begin{equation}
\label{eq:log-gamma-jacobian-ratio-partitionfunc-finalline-identity-2}
\begin{split}
  & \mu(\rmd Z_{N-1,M-1})   \prod_{n=0}^{N-2} \mu(\rmd Z_{n,M-1}) \prod_{m=0}^{M-2} \mu(\rmd Z_{N-1,m})  \\
  &= \mu( \rmd Z_{N-1,M-1} ) \prod_{n=1}^{N-1} \mu(\rmd  \tilde{r}\left(\frac{Z_{n,M-1}}{Z_{n-1,M-1}}\right))\prod_{m=1}^{M-1} \mu(\rmd  \tilde{r}\left(\frac{Z_{N-1,m}}{Z_{N-1,m-1}}\right))\\
          &=  \mu( \rmd Z_{N-1,0} ) \prod_{m=1}^{M-1} \mu (\rmd r\left(\frac{Z_{N-1,m}}{Z_{N-1,m-1}}\right))\prod_{n=1}^{N-1} \mu (\rmd \tilde{r}\left( \frac{Z_{n,M-1}}{Z_{n-1,M-1}}\right))\\
          &=  \mu( \rmd Z_{0,M-1} ) \prod_{n=1}^{N-1} \mu (\rmd r\left( \frac{Z_{n,M-1}}{Z_{n-1,M-1}}\right))\prod_{m=1}^{M-1} \mu (\rmd \tilde{r}\left(\frac{Z_{N-1,m}}{Z_{N-1,m-1}}\right))
    \end{split}
\end{equation}

 \subsection{Invariant measure of the matrix Log-Gamma: calculation on four sites}
 \label{sec:invariant-log-gamma-4sites}

\begin{figure}[h!]
\begin{tikzpicture}[scale=2.6]

% Drawing grid
\draw[step=2cm,gray,very thin] (0,0) rectangle (1,1);

% Nodes
\node[above] at (0,1) {$Z_{0,1}$};
\node[below] at (1,0) {$Z_{1,0}$};
\node[below] at (0,0) {$Z_{0,0}$};
\node[above right] at (1,1) {$Z_{1,1}, \, \tilde{Z}_{1,1}, \, V_{1,1}$};

% Dashed lines with labels and big arrowheads
\draw[->, dashed, line width=0.6mm, >=latex'] (0,1) -- (1,1);
\draw[->, dashed, line width=0.6mm, >=latex'] (1,0) --   (1,1) node[regular polygon, regular polygon sides=4, fill, minimum size=7pt, inner sep=0pt, outer sep=0pt] {};

\end{tikzpicture} 
\caption{Recursion of the matrix Log-Gamma polymer on four sites. }
\label{fig:FDT_for_the_log_gamma_on_4_sites}
\end{figure}
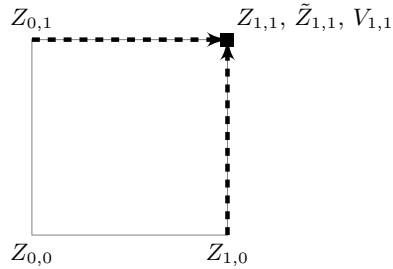

As a warm-up we now obtain the invariant measure from the FD transformation for $M=N=2$ which corresponds to a set of four sites as depicted in Fig.~\ref{fig:FDT_for_the_log_gamma_on_4_sites}. On four sites the "initial condition"
is $Z_i=(Z_{0,0},Z_{1,0},Z_{0,1})$, the terminal condition is $Z_f=(Z_{1,1})$
and the final condition is $Z'_f=(Z_{1,1},Z_{0,1},Z_{1,0})$. The final point $Z_{1,1}$ is determined by the recursion 
$Z_{1,1}=
 ( Z_{0,1}  + Z_{1,0} )^{1/2} V_{1,1}  ( Z_{0,1}  + Z_{1,0} )^{1/2}$. The action simply reads

 \be
 \begin{split}
& \hat{S}_0[Z,\hat{Z},V] =   \Tr \left[ \hat{Z}_{1,1} \left(Z_{1,1}-
 ( Z_{0,1}  + Z_{1,0} )^{1/2} V_{1,1}  ( Z_{0,1}  + Z_{1,0} )^{1/2} \right)\right]  +  \alpha \log \det V_{1,1} + \Tr \, V_{1,1}^{-1} 
 \end{split} 
\ee 
Using Eq.~\eqref{eq:change-var-log-gamma-FDT}, the FD transformation on four sites read
\begin{equation}
    \begin{split}
         &Z_{0,0} =Q_{1,1}^{-1}, \quad  Z_{1,0}=Q_{0,1}^{-1}, \quad   Z_{0,1}=Q_{1,0}^{-1} , \quad  Z_{1,1} = Q_{0,0}^{-1}   \, .
    \end{split}
\end{equation}
From Eq.~\eqref{eq:fdt-loggamma-noise}, the noises are related as
\begin{equation}
\begin{split}
     &Q_{1,1}^{-1/2}( Q_{0,1}  + Q_{1,0} )^{1/2} W_{1,1}  ( Q_{0,1}  + Q_{1,0} )^{1/2}Q_{1,1}^{-1/2} =
 Q_{0,0}^{1/2}( Q_{1,0}^{-1}  + Q_{0,1}^{-1} )^{1/2} V_{1,1}  ( Q_{1,0}^{-1}  + Q_{0,1}^{-1} )^{1/2}Q_{0,0}^{1/2} 
 \end{split}
\end{equation}

and from Eq.~\eqref{eq:change-var-log-gamma-FDT} the response fields are related as 

\begin{equation}
    Z_{1,1}^{1/2}\hat{Z}_{1,1} Z_{1,1}^{1/2} + \tilde{V}_{1,1}^{-1} = Q_{1,1}^{1/2}\hat{Q}_{1,1} Q_{1,1}^{1/2} + \tilde{W}_{1,1}^{-1}
\end{equation}
where the noises $\tilde{V}$ and $\tilde{W}$ are defined as similarity transformations of the original noise in Eq.~\eqref{eq:log-gamma-similarity-transformation-noise-fd}. The FD symmetry on the action \eqref{eq:FDT-matrix-log-gamma-action} read here as follows

\begin{equation}
\label{eq:app-log-gamma-fd-4sites-stationary-measure}
\begin{split}
    \hat{S}_0[Z,\hat{Z},V]  = \hat{S}_0[Q,\hat{Q},W]&+ \Tr\bigg[  \frac{Q_{0,0}}{ Q_{1,0}} -\frac{Q_{0,1}}{ Q_{1,1}}  \bigg]+\alpha \kappa   \big(\log \Det \frac{Q_{1,0}}{Q_{0,0}}- \log \Det \frac{Q_{1,1}}{Q_{0,1}}\big)\\
    &+\Tr \bigg[ \frac{Q_{0,0}}{ Q_{0,1}} -\frac{Q_{1,0}}{ Q_{1,1}} \bigg]+\alpha (1-\kappa) \big(\log \Det \frac{Q_{0,1}}{Q_{0,0}}-\log \Det \frac{Q_{1,1}}{Q_{1,0}} \big)
    \end{split}
\end{equation}

Taking the exponential of minus \eqref{eq:app-log-gamma-fd-4sites-stationary-measure} on both sides, integrating over $V_{1,1}$ and $\hat{Z}_{1,1}$ and inserting the observable $\mathcal{O}(Z'_f)$, the FD relation \eqref{eq:app-log-gamma-fd-4sites-stationary-measure} implies
\begin{equation}
\label{eq:MSR-4-sites-fd-transformation}
    \begin{split}
        &\iint \mu(\rmd  V_{1,1}) \rmd \hat{Z}_{1,1}e^{-\hat{S}_0[Z,\hat{Z},V]}\mathcal{O}(Z'_f)|  \Det \frac{Z_{0,1}}{Z_{0,0}}|^{-\alpha (1-\kappa)}|\Det \frac{Z_{1,0}}{Z_{0,0}}|^{-\alpha \kappa}e^{-\Tr\big[  \frac{Z_{0,0}}{ Z_{1,0}}  +\frac{Z_{0,0}}{ Z_{0,1}} \big]} =|\Det Z_{1,1}|^{-\frac{d+1}{2}}|\Det Z_{0,0}|^{-\frac{d+1}{2}}\\
        &  \times \iint \mu(\rmd W_{1,1}) d\hat{Q}_{1,1}e^{-\hat{S}_0[Q,\hat{Q},W]} \mathcal{O}(Z'_f) |\Det \frac{Z_{1,1}}{Z_{0,1}}|^{-\alpha \kappa} |  \Det \frac{Z_{1,1}}{Z_{1,0}}|^{-\alpha (1-\kappa)}e^{-\Tr\big[  \frac{Z_{0,1}}{ Z_{1,1}} + \frac{Z_{1,0}}{ Z_{1,1}}  \big]} \\
    \end{split}
\end{equation}
We have used that since all the variables $Z$ are fixed and since the measure over the noise is $GL_d$ invariant, one has $\mu (\rmd V_{1,1})=\mu(\rmd  W_{1,1})$. Furthermore, we have used that the only non-trivial Jacobian comes from the response field
\begin{equation}
    \rmd \hat{Z}_{1,1}=\rmd\hat{Q}_{1,1} |\Det Q_{1,1}|^{\frac{d+1}{2}}|\Det Q_{0,0}|^{\frac{d+1}{2}}=\rmd\hat{Q}_{1,1} |\Det Z_{0,0}|^{-\frac{d+1}{2}}|\Det Z_{1,1}|^{-\frac{d+1}{2}}
\end{equation}
 which comes from Eq.~\eqref{eq:change-var-log-gamma-FDT} together with \eqref{detg}.\\

We now multiply both terms of \eqref{eq:MSR-4-sites-fd-transformation} by the measure

\begin{equation}
\label{eq:ratio-jacobian-4sites-log-gamma}
   \rmd Z_{1,1} \mu(\rmd Z_{0,0}) \mu (\rmd Z_{1,0}) \mu (\rmd Z_{0,1}) \, .
\end{equation}
and integrate over all the variables $Z$. We call the resulting equation $LHS=RHS$ 
then study separately the right side $RHS$ and then the left hand side $LHS$.

\paragraph{Right hand side RHS.} As a first step, we will use the normalisation of the right hand side of \eqref{eq:MSR-4-sites-fd-transformation} with respect to the final field $Q_{1,1}$ (i.e., the standard MSR integral normalisation discussed below \eqref{eq:app-log-gamma-action-three-fields})
\be \label{norm2} 
1 = \int \rmd Q_{1,1} \iint \mu(\rmd  W_{1,1}) \rmd\hat{Q}_{1,1}e^{-\hat{S}_0[Q,\hat{Q},W]} = 
\int \frac{\rmd Z_{0,0}}{ |\det Z_{0,0}|^{d+1} }\iint \mu(\rmd W_{1,1}) \rmd\hat{Q}_{1,1}e^{-\hat{S}_0[Q,\hat{Q},W]} 
\ee 
Note that r.h.s. of \eqref{norm2} contains the Jacobian from $Q_{1,1}$ to $Z_{0,0}$ which is obtained from \eqref{detinverse}.
We then obtain

\begin{equation}
\label{eq:log-gamma-invariant-measure-rhs}
    \begin{split}
        &RHS= \iiint \mathcal{O}(Z'_f)\mu(\rmd Z_{1,1})\mu(\rmd Z_{1,0} )\mu(\rmd Z_{0,1})|\Det \frac{Z_{1,1}}{Z_{0,1}}|^{-\alpha \kappa} |  \Det \frac{Z_{1,1}}{Z_{1,0}}|^{-\alpha (1-\kappa)}e^{-\Tr\big[  \frac{Z_{0,1}}{ Z_{1,1}} + \frac{Z_{1,0}}{ Z_{1,1}}  \big]} \\
    \end{split}
\end{equation}

\paragraph{Left hand side LHS.}
From Eqs.~\eqref{eq:average-observable-msr-log-gamma} and \eqref{eq:MSR-log-gamma-expectation-value}, LHS is a conditional expectation of the final observables with respect to the initial values

\begin{equation}
\label{eq:log-gamma-invariant-measure-lhs}
\begin{split}
  LHS=&\iiint \mathbb{E}[ \mathcal{O} (Z_f') | Z_i ]
\mu(\rmd Z_{0,0}) \mu(\rmd Z_{1,0})\mu(\rmd Z_{0,1})  |  \Det \frac{Z_{0,1}}{Z_{0,0}}|^{-\alpha (1-\kappa)}|\Det \frac{Z_{1,0}}{Z_{0,0}}|^{-\alpha \kappa}e^{-\Tr\big[  \frac{Z_{0,0}}{ Z_{1,0}}  +\frac{Z_{0,0}}{ Z_{0,1}} \big]}   
\end{split}
\end{equation}

Overall, equating \eqref{eq:log-gamma-invariant-measure-rhs} and \eqref{eq:log-gamma-invariant-measure-lhs}, we obtain our final result

\begin{equation}
\begin{split}
  &\iiint \mathbb{E}[ \mathcal{O} ( Z_f') | Z_i]
\mu(\rmd Z_{0,0}) \mu(\rmd Z_{1,0})\mu(\rmd Z_{0,1})   |  \Det \frac{Z_{0,1}}{Z_{0,0}}|^{-\alpha (1-\kappa)}|\Det \frac{Z_{1,0}}{Z_{0,0}}|^{-\alpha \kappa}e^{-\Tr\big[  \frac{Z_{0,0}}{ Z_{1,0}}  +\frac{Z_{0,0}}{ Z_{0,1}} \big]} \\
&=\iiint \mathcal{O}(Z'_f)\mu(\rmd Z_{1,1})\mu(\rmd Z_{1,0} )\mu(\rmd Z_{0,1})|\Det \frac{Z_{1,1}}{Z_{0,1}}|^{-\alpha \kappa} |  \Det \frac{Z_{1,1}}{Z_{1,0}}|^{-\alpha (1-\kappa)}e^{-\Tr\big[  \frac{Z_{0,1}}{ Z_{1,1}} + \frac{Z_{1,0}}{ Z_{1,1}}  \big]}
\end{split}
\end{equation}
which show that the measure on the triplet $Z_{0,0},Z_{1,0},Z_{0,1}$ in the l.h.s 
is transported by the dynamics (i.e., the log gamma recursion) onto the measure on the triplet $Z_{1,1},Z_{1,0},Z_{0,1}$. 
It is now possible to use the relations \eqref{eq:log-gamma-jacobian-ratio-partitionfunc-finalline-identity-1} and \eqref{eq:log-gamma-jacobian-ratio-partitionfunc-finalline-identity-2} to express these measures more conveniently
on the partition sum ratios, which shows explicitly that these ratios are independent inverse Wishart distributed. To this aim we fix the zero mode to be $Z_{0,1}$ for both initial and final sets, and the initial measure reads
\be 
\mu(\rmd Z_{0,1}) \rmd P^{iW}_{\alpha(1-\kappa)}[\tilde{r}\left(\frac{Z_{0,1}}{Z_{0,0}} \right)]\rmd P^{iW}_{\alpha \kappa}[r\left(\frac{Z_{1,0}}{Z_{0,0}} \right)] 
\ee 
while the final measure reads
\be 
\mu(\rmd Z_{0,1}) \rmd P^{iW}_{\alpha (1-\kappa)}[\tilde{r}\left(\frac{Z_{1,1}}{Z_{1,0}} \right)] \rmd P^{iW}_{\alpha \kappa}[r\left(\frac{Z_{1,1}}{Z_{0,1}} \right)]
\ee 
One sees that now these define the invariant measure on the ratios.

\begin{enumerate}
    \item All ratios are sampled independently.
    \item The ratios along the horizontal axis defined with the matrix ordering convention $r$, see \eqref{eq:definition-convention-ratios}, 
    are sampled from the Inverse Wishart distribution with parameter $\kappa \alpha$.
    \item The ratios along the vertical axis defined with the matrix ordering convention $\tilde{r}$ are sampled from the Inverse Wishart distribution with parameter and $(1-\kappa)\alpha$.
    \item The zero mode $Z_{0,1}$ is uniformly distributed over $\mathcal{P}_d$.
    \item We have the constraint $\kappa \in ]\frac{d-1}{2\alpha}, 1-\frac{d-1}{2\alpha}[$ --  where $\alpha > d-1$ so that the interval is not empty -- the constraint coming from the normalisation condition of the inverse Wishart distributions on the ratios.
\end{enumerate}

To obtain the invariant measure over a lattice composed of more than four sites, one propagates the invariant cell by cell until covering the macroscopic lattice described in Fig.~\ref{fig:loggamma-polymer-lightcone-wnt}. In the next Section, we will carry the calculation explicitly for the invariant measure on a lattice of arbitrary size.

\subsection{Invariant measure of the matrix Log-Gamma: calculation on an arbitrary rectangle}
To investigate the invariant measure on a lattice of arbitrary size, i.e., $\mathcal{C}$ (see Fig.~\ref{fig:loggamma-polymer-lightcone-wnt}), one needs to take into account the Jacobian of the transformation
\begin{equation}
    (Z,\hat{Z},V) \to (Q,\hat{Q},W)
\end{equation}
In a first stage, we will fix the initial values $Z_i$ as well as the terminal values $Z_f$ \eqref{eq:log-gamma-terminal-values} but we will integrate over all the other variables. Therefore one needs to separate the integration over $\mathcal{C}_{bulk}$ (see definition in Fig.~\ref{fig:loggamma-polymer-lightcone-wnt})
and over the indices in $Z_f$, equivalently $\mathcal{C} \setminus \mathcal{C}_{bulk}$. We calculate in the following two sections the Jacobians respectively in the bulk and on the terminal line.

\subsubsection{Jacobian of the FD transformation in the bulk of the lattice $\mathcal{C}_{bulk}$}

We now show that in the bulk $\mathcal{C}_{bulk}$ the Jacobian over the partition function, the response field and the noise reads
\be 
\label{eq:invariant-log-gamma-jacobian1}
\begin{split}
&{\cal D}_\mu V {\cal D} Z {\cal D} \hat Z =  {\cal D}_\mu W {\cal D} Q {\cal D} \hat Q \times J_{bulk} \\
&J_{bulk} =\prod_{n=1}^{N-2}\prod_{m=1}^{M-2}\frac{|\Det Z_{n,m}|^{\frac{d+1}{2}}}{|\Det Z_{n-1,m-1}|^{\frac{d+1}{2}}}
\end{split}
\ee 
Since  $Q$ depends only on $(Z)$, and $W$ depends only on $(Z,V )$ and $\hat{Q}$ depends on $(Z,V,\hat{Z})$, the Jacobian matrix of the change of variable is triangular and thus we only have to consider the diagonal contributions.\\

The transformation of the noise \eqref{eq:fdt-loggamma-noise} relates $V_{n,m}$ and $W_{n',m'}$ through an action of the group $GL_d$ that leaves the measure $\mathcal{D}_\mu$ invariant. Thus the Jacobian arising from the transformation of the noise is equal to unity. Therefore the Jacobian $J_{bulk}$ is 

\begin{equation}
    J_{bulk}= |\Det \frac{\delta Z_{n,m}}{\delta Q_{n'',m''}}| \times| \Det  \frac{\delta \hat{Z}_{n,m}}{\delta \hat{Q}_{n'',m''}}| 
\end{equation}

Using \eqref{eq:change-var-log-gamma-FDT} together with
\eqref{detinverse}, the first factor reads

\begin{equation}
   |\Det \frac{\delta Z_{n,m}}{\delta Q_{n'',m''}}|= \prod_{n=1}^{N-2}\prod_{m=1}^{M-2}|\Det Z_{n,m}|^{d+1}
\end{equation}
Using \eqref{eq:change-var-log-gamma-FDT} together with
\eqref{detg}, the second factor reads 
\begin{equation}
\label{eq:invariant-log-gamma-jacobian2-2}
   | \Det  \frac{\delta \hat{Z}_{n,m}}{\delta \hat{Q}_{n'',m''}}| = \prod_{n=1}^{N-2}\prod_{m=1}^{M-2} |\Det Z_{n,m}|^{-\frac{d+1}{2}}|\Det Z_{n-1,m-1}|^{-\frac{d+1}{2}}
\end{equation}
The product of these factors finally yields \eqref{eq:invariant-log-gamma-jacobian1}.

\subsubsection{Jacobian on the terminal line of the lattice $\mathcal{C} \setminus \mathcal{C}_{bulk}$}
On the terminal line $Z_f$ \eqref{eq:log-gamma-terminal-values}, the partition function is fixed while the response field is not. Therefore there is a Jacobian for the response field for all indices in the set $Z_f$. Note that this excludes the two corners, see Fig.~\ref{fig:total-jacobian-fd-log-gamma}. Using \eqref{eq:change-var-log-gamma-FDT} together with \eqref{detg}, this Jacobian is

\begin{equation}
\label{eq:invariant-log-gamma-jacobian2}
\begin{split}
    J_{final}&=| \Det  \frac{\delta \hat{Z}_{n,m}}{\delta \hat{Q}_{n'',m''}}|_{Z_{n,m}\in Z_f}\\
    &=\prod_{n=1}^{N-1} |\Det Z_{n-1,M-2}|^{-\frac{d+1}{2}} |\Det Z_{n,M-1}|^{-\frac{d+1}{2}} \prod_{m=1}^{M-1}|\Det Z_{N-2,m-1}|^{-\frac{d+1}{2}} |\Det Z_{N-1,m}|^{-\frac{d+1}{2}} \\
    &\times |\Det Z_{N-1,M-1}|^{\frac{d+1}{2}}|\Det Z_{N-2,M-2}|^{\frac{d+1}{2}}\\
\end{split}
\end{equation}

\begin{figure}[t!]
    \centering  \includegraphics[width=0.4\linewidth]{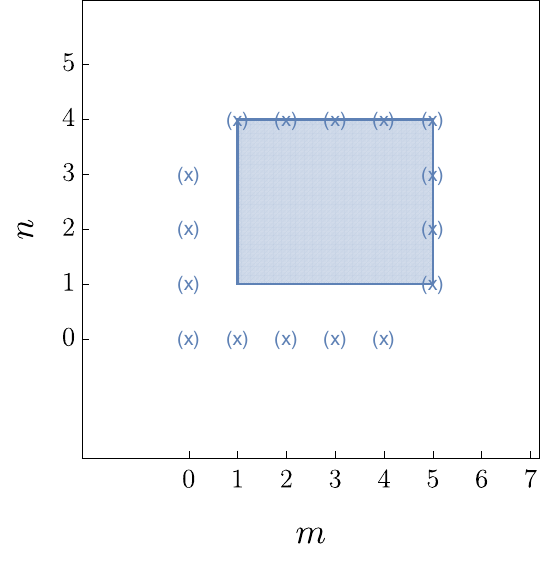}
    \caption{We represent for $N=5$ and $M=6$ the coordinates points that appear in the combination of the Jacobian of the FD transformation in the bulk and the Jacobian on the terminal line, i.e., \eqref{eq:invariant-measure-log-gamma-jacobian3}. The rectangle filled describes the region where the recursion of the Log-Gamma polymer is applied.}
    \label{fig:total-jacobian-fd-log-gamma}
\end{figure}

\begin{remark}
Note that the product of \eqref{eq:invariant-log-gamma-jacobian2} with \eqref{eq:invariant-log-gamma-jacobian2-2} reads $ \prod_{n=1}^{N-1}\prod_{m=1}^{M-1} |\Det Z_{n,m}|^{-\frac{d+1}{2}}|\Det Z_{n-1,m-1}|^{-\frac{d+1}{2}}$.
\end{remark}

\subsubsection{Total Jacobian of the FD transformation}
The product of the Jacobians from \eqref{eq:invariant-log-gamma-jacobian1} and \eqref{eq:invariant-log-gamma-jacobian2} is quite simple and reads

\begin{equation}
\label{eq:invariant-measure-log-gamma-jacobian3}
 J_{bulk} J_{final} =  \prod_{n=0}^{N-2} |\Det  Z_{n,0}|^{-\frac{d+1}{2}} \prod_{m=1}^{M-2} |\Det Z_{0,m}|^{-\frac{d+1}{2}} \prod_{n=1}^{N-1}|\Det Z_{n,M-1}|^{-\frac{d+1}{2}} 
   \prod_{m=1}^{M-2}|\Det Z_{N-1,m}|^{-\frac{d+1}{2}} 
\end{equation}
We have represented in Fig.~\ref{fig:total-jacobian-fd-log-gamma} the coordinates of the partition functions present in the Jacobian by the markers $(\times)$ for $N=5$ and $M=6$.

\subsubsection{Invariant measure}

In the case of a lattice of arbitrary size, we need to consider the following normalisation for the path integral with the $Q$ variables
\begin{equation}
\label{eq:invariant-log-gamma-normalisation-end-point}
    1=\int \prod_{n=1}^{N-1}\rmd Q_{n,M-1} \prod_{m=1}^{M-2}\rmd Q_{N-1,m} \iint \mathcal{D}_\mu W\mathcal{D}\hat{Q} \int_{\mathcal{C}_{bulk}} \mathcal{D}Q e^{-  \hat{S}_0[Q,\hat{Q},W]}
\end{equation}
where the indices of the terminal line have been explicitly written. Using \eqref{eq:change-var-log-gamma-FDT} together with \eqref{detinverse}, the integrals over the terminal line can be rewritten with the $Z$ variables as
\begin{equation}
\label{eq:invariant-log-gamma-normalisation-end-point-2}
    \int \prod_{n=1}^{N-1}\rmd Q_{n,M-1} \prod_{m=1}^{M-2}\rmd Q_{N-1,m} = \int \prod_{n=0}^{N-2} \frac{\rmd Z_{n,0}}{|\Det  Z_{n,0}|^{d+1}} \prod_{m=1}^{M-2} \frac{\rmd Z_{0,m}}{|\Det Z_{0,m}|^{d+1}}
\end{equation}

\begin{remark}
From the total Jacobian \eqref{eq:invariant-measure-log-gamma-jacobian3}, we can extract the factor
\begin{equation}
\label{eq:log-gamma-fdt-jacobian-remark}
    \prod_{n=0}^{N-2}|\Det  Z_{n,0}|^{-\frac{d+1}{2}} \prod_{m=1}^{M-2} |\Det Z_{0,m}|^{-\frac{d+1}{2}} 
\end{equation}
which is half of the powers we need to make use of the normalisation of the MSR integral.
\end{remark}

The strategy to obtain the invariant measure is to first (1) take the exponential of minus \eqref{eq:FDT-matrix-log-gamma-action} on both sides, and (2) integrate over all noises $V$ and all response fields $\hat{Z}$ with indices in $\mathcal{C}$, (3) integrate over the fields $Z$ in the bulk $\mathcal{C}_{bulk}$ and (4) insert the observable $\mathcal{O}(Z_f')$ over the final values defined in \eqref{eq:log-gamma-final-values}. This leads, upon reorganisation and inserting the Jacobian of the FD relation \eqref{eq:invariant-measure-log-gamma-jacobian3}, to

\begin{equation}
\label{eq:invariant-measure-log-gamma-fdt-after-jacobian}
\begin{split}
    &\iint \mathcal{D}_\mu V {\cal D}\hat Z \int_{\mathcal{C}_{bulk}} {\cal D} Z e^{-  \hat{S}_0[Z,\hat{Z},V]} \mathcal{O}(Z_f')\times \prod_{m=1}^{M-1} |\Det \frac{Z_{0,m}}{Z_{0,m-1}} |^{-\alpha (1-\kappa)} e^{- \Tr \big[ \frac{Z_{0,m-1}}{ Z_{0,m}} \big]} \prod_{n=1}^{N-1} |\Det  \frac{Z_{n,0}}{Z_{n-1,0}}|^{-\alpha \kappa }e^{-\Tr\big[  \frac{Z_{n-1,0}}{ Z_{n,0}}   \big]}\\
    &= \iint \mathcal{D}_\mu W {\cal D}\hat{Q} \int_{\mathcal{C}_{bulk}} {\cal D} Q e^{-  \hat{S}_0[Q,\hat{Q},W]} \mathcal{O}(Z_f')\times \prod_{m=1}^{M-1} |\Det \frac{Z_{N-1,m}}{Z_{N-1,m-1}} |^{-\alpha (1-\kappa)} e^{- \Tr \big[ \frac{Z_{N-1,m-1}}{ Z_{N-1,m}} \big]} \\
    &\times \prod_{n=1}^{N-1} |\Det  \frac{Z_{n,M-1}}{Z_{n-1,M-1}}|^{-\alpha \kappa }e^{-\Tr\big[  \frac{Z_{n-1,M-1}}{ Z_{n,M-1}}   \big]} \\
    &\times    \prod_{n=0}^{N-2} |\Det  Z_{n,0}|^{-\frac{d+1}{2}} \prod_{m=1}^{M-2} |\Det Z_{0,m}|^{-\frac{d+1}{2}} \prod_{n=1}^{N-1}|\Det Z_{n,M-1}|^{-\frac{d+1}{2}} 
   \prod_{m=1}^{M-2}|\Det Z_{N-1,m}|^{-\frac{d+1}{2}} \\
    \end{split}
\end{equation}

One then multiplies both sides of the Eq.~\eqref{eq:invariant-measure-log-gamma-fdt-after-jacobian} by several measures, namely 

\begin{equation}
   \underbrace{ \prod_{n=0}^{N-1} \mu(\rmd Z_{n,0} )\prod_{m=1}^{M-1} \mu(\rmd Z_{0,m} )}_{\text{initial values}} \underbrace{\prod_{n=1}^{N-1}\rmd Z_{n,M-1}\prod_{m=1}^{M-2}\rmd Z_{N-1,m}}_{\text{terminal values}}
\end{equation}

After these multiplications we integrate over all the variables $Z$. We call the resulting equation $LHS=RHS$ and then study separately the right side RHS and then the left side LHS.
\\

\paragraph{Right hand side RHS.}

We use the normalisation condition \eqref{eq:invariant-log-gamma-normalisation-end-point}-\eqref{eq:invariant-log-gamma-normalisation-end-point-2} for the RHS. The resulting equation thus reads

\begin{equation}
\label{eq:log-gamma-invariant-measure-rhs-general}
\begin{split}
     \text{RHS} = \iiint &\mathcal{O}(Z_f')\prod_{n=0}^{N-1}\mu(\rmd Z_{n,M-1})\prod_{m=0}^{M-2}\mu(\rmd Z_{N-1,m})   \\
    &\times \prod_{m=1}^{M-1} |\Det \frac{Z_{N-1,m}}{Z_{N-1,m-1}} |^{-\alpha (1-\kappa)} e^{- \Tr \big[ \frac{Z_{N-1,m-1}}{ Z_{N-1,m}} \big]}\prod_{n=1}^{N-1} |\Det  \frac{Z_{n,M-1}}{Z_{n-1,M-1}}|^{-\alpha \kappa }e^{-\Tr\big[  \frac{Z_{n-1,M-1}}{ Z_{n,M-1}}   \big]}
    \end{split}
\end{equation}

\paragraph{Left hand side LHS.}
From Eqs.~\eqref{eq:average-observable-msr-log-gamma} and \eqref{eq:MSR-log-gamma-expectation-value}, the l.h.s is a conditional expectation of the final observables with respect to the initial set

\begin{equation}
\label{eq:log-gamma-invariant-measure-lhs-general}
\begin{split}
  \text{LHS} =\iiint &\mathbb{E}[ \mathcal{O}( Z_f') |Z_i]\prod_{n=0}^{N-1} \mu(\rmd Z_{n,0} )\prod_{m=1}^{M-1} \mu(\rmd Z_{0,m} ) \\
&\times\prod_{m=1}^{M-1} |\Det \frac{Z_{0,m}}{Z_{0,m-1}} |^{-\alpha (1-\kappa)} e^{- \Tr \big[ \frac{Z_{0,m-1}}{ Z_{0,m}} \big]}  \prod_{n=1}^{N-1} |\Det  \frac{Z_{n,0}}{Z_{n-1,0}}|^{-\alpha \kappa }e^{-\Tr\big[  \frac{Z_{n-1,0}}{ Z_{n,0}}   \big]}
\end{split}
\end{equation}

Equating Eqs.~\eqref{eq:log-gamma-invariant-measure-rhs-general} and \eqref{eq:log-gamma-invariant-measure-lhs-general} 
shows that the measure on the initial values $Z_i$ in LHS 
is transported by the dynamics (i.e., the log gamma recursion) onto the measure on the final values $Z_f'$ in RHS. 
It is now again possible to use the relations \eqref{eq:log-gamma-jacobian-ratio-partitionfunc-finalline-identity-1} and \eqref{eq:log-gamma-jacobian-ratio-partitionfunc-finalline-identity-2} to express these measures more conveniently
in terms of the partition sum ratios, which shows explicitly that these ratios are independent inverse Wishart distributed. To this aim we fix the zero mode to be $Z_{0,M-1}$ for both initial and final sets, and the initial measure reads
\be 
\mu(\rmd Z_{0,M-1}) \prod_{m=1}^{M-1}\rmd P^{iW}_{\alpha(1-\kappa)}[\tilde{r}\left(\frac{Z_{0,m}}{Z_{0,m-1}} \right)]\prod_{n=1}^{N-1} \rmd P^{iW}_{\alpha \kappa}[r\left(\frac{Z_{n,0}}{Z_{n-1,0}} \right)] 
\ee 
while the final measure reads
\be 
\mu(\rmd Z_{0,M-1}) \prod_{m=1}^{M-1}\rmd P^{iW}_{\alpha (1-\kappa)}[\tilde{r}\left(\frac{Z_{N-1,m}}{Z_{N-1,m-1}} \right)] \prod_{n=1}^{N-1}\rmd P^{iW}_{\alpha \kappa}[r\left(\frac{Z_{n,M-1}}{Z_{n-1,M-1}} \right)]
\ee 

One sees that now these define the invariant measure on the ratios.

\begin{enumerate}
    \item All ratios are sampled independently.
    \item The ratios along the horizontal axis defined with the matrix ordering convention $r$, see \eqref{eq:definition-convention-ratios}, 
    are sampled from the Inverse Wishart distribution with parameter $\kappa \alpha$.
    \item The ratios along the vertical axis defined with the matrix ordering convention $\tilde{r}$ are sampled from the Inverse Wishart distribution with parameter and $(1-\kappa)\alpha$.
    \item The zero mode $Z_{0,M-1}$ is uniformly distributed over $\mathcal{P}_d$.
    \item We have the constraint $\kappa \in ]\frac{d-1}{2\alpha}, 1-\frac{d-1}{2\alpha}[$ --  where $\alpha > d-1$ so that the interval is not empty -- the constraint coming from the normalisation condition of the inverse Wishart distributions on the ratios.
\end{enumerate}

\section{Field theory of the matrix O'Connell-Yor polymer} A matrix version of the Yor-O'Connell polymer was first introduced in Ref.~\cite{matrixOYpolymer}. Here, we consider a noise matrix $W_n(t)$ which is a $d\times d$ real symmetric matrix with correlator 
\begin{equation}
    \overline{[W_n(t)]_{ij}[W_{n'}(t)]_{kl}} = g (\delta_{ik}\delta_{jl}+\delta_{il}\delta_{jk})\delta_{nn'}\delta(t-t')
\end{equation}
and its probability measure is the path integral measure
\begin{equation}
    \mathcal{D}W \exp \left(-\frac{1}{4g}\sum_n \int \rmd t \, \Tr W^2 \right), \quad \mathcal{D}W=\prod_{n,t}\rmd W_n(t) \,  \quad \rmd W_n(t)=\prod_{i \leq j}\rmd [W_n(t)]_{ij}
\end{equation}
Consider now the evolution for the matrix field $Z=Z_n(t) \in \mathcal{P}_d$ described 
by the stochastic equation 

\be 
\label{eq:matrix-YO-process}
 \p_t Z_{n} = Z_{n-1} - Z_{n} +
 Z_{n}^{1/2} W_{n}(t) Z_{n}^{1/2}
\ee 
This equation should be complemented with an "initial condition". The paradigmatic example of initial condition is the droplet initial condition defined as
\begin{equation}
    Z_n(t_i)=\delta_{n,1} \,  I_d , \quad Z_0(t)=0, \; \forall t \geq t_i
\end{equation}
In that case $Z_{N-1}(t_f)$ gives the point-to-point matrix OY partition function, from $(0,t_i)$ to $(N-1,t_f)$.  Note again that ${\rm Tr}[ Z_n(t)]$ has an interpretation as a sum of ordered matrix products over semi-discrete
polymer paths, as for $d=1$. \\

The deterministic part of Eq.~\eqref{eq:matrix-YO-process} can be solved by convoluting the initial condition with the Poisson kernel $P(n,t)=e^{-t}t^n/n!$ and thus preserves the positivity of the initial condition. The stochastic part of Eq.~\eqref{eq:matrix-YO-process}  is identical to the zero-dimensional evolution \eqref{eq:EvolutionLambdai} which also preserves positivity \cite{GautieMatrixKesten}. Thus by a similar argument as for the MSHE, we expect \eqref{eq:EvolutionLambdai} to preserve positivity.
\\

A more general class of initial condition specifies the set of initial values $Z=Z_i$
\begin{equation}
\label{eq:YO-initial-values}
    \begin{split}
 Z_i = \{ Z_0(t) , t \in [t_i,t_f] \quad \text{and} \quad Z_1(t_i), \dots, Z_{N-1}(t_i) \} 
    \end{split}
\end{equation}

One can then ask about the set of terminal values ${\cal Z}=Z_f$
\begin{equation}
\label{eq:YO-terminal-values}
    \begin{split}
          Z_f = \{ Z_{N-1}(t) , t \in [t_i^+,t_f] \quad \text{and} \quad Z_1(t_f), \dots, Z_{N-2}(t_f) \} 
    \end{split}
\end{equation}
We additionally define the set of final values which includes the "corners" (by analogy with the log Gamma polymer)
\begin{equation}
\label{eq:YO-final-values}
    \begin{split}
          Z_f' = \{ Z_{N-1}(t) , t \in [t_i,t_f] \quad \text{and} \quad Z_0(t_f), \dots, Z_{N-2}(t_f) \} 
    \end{split}
\end{equation}

This relation between initial set, and terminal/final set will be useful to discuss the stationary OY polymer below. \\

Using the MSR method, the expectation value over the noise of any functional $\mathcal{O}[\mathcal{Z}]$ of the field ${\cal Z}$ 
with a given initial condition $Z_i$
can be written as a path integral
\begin{equation}
   \overline{\mathcal{O}[\mathcal{Z}]}=\iiint \mathcal{D}W\mathcal{D}{\hat{Z}}\mathcal{D}Z \, \mathcal{O}[Z] e^{- S_0[Z,\hat{Z},W]}
\end{equation}
where the action involves the symmetric response matrix field $\hat{Z}_n(t)$
\begin{equation}
\begin{split}
    S_0[Z,\hat{Z},W]&=\sum_n \int_t \rmd t \, \Tr \big[  \hat{Z}_n (\p_t Z_n-Z_{n-1}+Z_n-Z_n^{1/2} W_n Z_n^{1/2})+ \frac{1}{4g} W_n^2 \big] \\
\end{split}
\end{equation}
The integration is performed over the noise $W$, over the response field $\hat{Z}$ and over the partition function $Z$ on all space-time indices where the stochastic evolution \eqref{eq:matrix-YO-process} takes place. There is no integration over the initial condition $Z_i$ so that the expectation is a conditional one. The integration measure over each of theses three fields is Lebesgue as defined in Section~\ref{subsec:definition-measures}. 
The MSR integral is normalised to unity, consistent with $\overline{1}=1$ (upon choosing $\mathcal{O}[\cdot]=1$). The integration over the response field enforces the stochastic dynamic \eqref{eq:matrix-YO-process}.\\

Equivalently, to emphasize that the expectation over the noise of the functional $\mathcal{O}[{\cal Z}]$ is conditioned on the initial set of values, we introduce the notation
\begin{equation}
\label{eq:MSR-YO-expectation-value}
    \overline{\mathcal{O}[\mathcal{Z}]}:=\E[\mathcal{O}(Z)|Z_i]
\end{equation}

It is possible to integrate explicitly over the noise $\mathcal{W}$ and obtain a representation of the MSR path integral solely in function of $Z$ and $\hat{Z}$, 
\begin{equation}
    \overline{\mathcal{O}[\mathcal{Z}]}=\iint \mathcal{D}{\hat{Z}}\mathcal{D}Z \, \mathcal{O}[Z] e^{- S[Z,\hat{Z}]}
\end{equation}
where the reduced action reads
\begin{equation}
    S[Z,\hat{Z}]=\sum_n \int \rmd t \, \Tr\big[  \hat{Z}_n (\p_t Z_n-Z_{n-1}+Z_n)-g \hat{Z}_nZ_n\hat{Z}_nZ_n \big]
\end{equation}

\subsection{Weak-noise and saddle point of the matrix O'Connell-Yor polymer}
The weak-noise theory for the scalar version of the OY polymer was derived and studied in \cite{krajenbrink2023weak}.
Here we investigate the weak-noise theory of the matrix generalisation of the OY polymer. To this
aim we first replace $g \to \varepsilon g$ and consider the observable 
\be 
\mathcal{O}[\mathcal{Z}] = \exp\left(  \frac{1}{\varepsilon} \int \rmd t  \sum_n  \, \Tr \left[J_n(t) Z_n(t) \right] \right)
\ee 
where $J_n(t)$ is a matrix source field (vanishing at $t=\infty$). This allows us to obtain that 
\begin{equation}
    \overline{e^{  \frac{1}{\varepsilon} \int \rmd t  \sum_n  \, \Tr \left[J_n(t) Z_n(t) \right] }} = \iint \mathcal{D}{\hat{Z}}\mathcal{D}Z \, e^{- S[Z,\hat{Z}]+ \frac{1}{\varepsilon} \int \rmd t  \sum_n   \Tr \left[J_n(t) Z_n(t) \right] } 
\end{equation}
Upon rescaling of the response field $\hat{Z} \to \hat{Z}/\varepsilon$, we obtain the modified action
\begin{equation}
    \overline{e^{  \frac{1}{\varepsilon} \int \rmd t  \sum_n  \, \Tr \left[J_n(t) Z_n(t) \right] }} = \iint \mathcal{D}{\hat{Z}}\mathcal{D}Z \,  e^{-\frac{1}{\varepsilon} S_J[Z,\hat{Z}]} 
\end{equation}
with
\begin{equation} \label{actionOY}
    S_J[Z,\hat{Z}] = \sum_n \int \rmd t \, \Tr\big[  \hat{Z}_n (\p_t Z_n-Z_{n-1}+Z_n)-g \hat{Z}_nZ_n\hat{Z}_nZ_n - J_n Z_n \big]
\end{equation}
In the limit $\varepsilon \ll 1$, the path integral is dominated by its saddle point. Taking derivatives with respect to the fields $\{Z,\hat{Z} \}$ and after some algebraic manipulations, we obtain the following matrix equations
\begin{equation} \label{systemOYmat}
    \begin{split}
        \p_t Z_n &= Z_{n-1}-Z_n +2g Z_n \hat{Z}_n Z_n \\
        -\p_t \hat{Z}_n &=\hat{Z}_{n+1}-\hat{Z}_n+2g \hat{Z}_n Z_n \hat{Z}_n+J_n  \\
    \end{split}
\end{equation}

This system turns out to be integrable. The two Lax matrices are given in the main text in 
\eqref{eq:supp-mat-factorisation-lax-Ln} and in \eqref{UOY}, which we rewrite here in the form
\bea 
L_n = \left(
\begin{array}{cc}
 \frac{1}{\lambda } & \frac{Z_n}{\lambda } \\
 -\frac{2 g \hat Z_n}{\lambda } & \lambda -\frac{2 g \hat Z_n
   Z_n}{\lambda } \\
\end{array}
\right) \quad , \quad     
U_{n}=
\begin{pmatrix}
\frac{\lambda^2-1}{2}I_d  &   -Z_{n-1} \\ 
&\\
2g\hat{Z}_n &  \frac{1-\lambda^2}{2}I_d
\end{pmatrix}
\eea 
Let us now check explicitly that the system \eqref{systemOYmat} implies that the 
compatibility equation $\p_t L_n=U_{n+1}L_n -L_n U_n$ holds. 
Let us write
\bea 
&& U_{n+1}L_n -L_n U_n \\
&& = \begin{pmatrix}
\frac{\lambda^2-1}{2}I_d  &   -Z_{n} \\ 
&\\
2g\hat{Z}_{n+1} &  \frac{1-\lambda^2}{2}I_d
\end{pmatrix} \left(
\begin{array}{cc}
 \frac{1}{\lambda } & \frac{Z_n}{\lambda } \\
 -\frac{2 g \hat Z_n}{\lambda } & \lambda -\frac{2 g \hat Z_n
   Z_n}{\lambda } \\
\end{array}
\right) - 
\left(
\begin{array}{cc}
 \frac{1}{\lambda } & \frac{Z_n}{\lambda } \\
 -\frac{2 g \hat Z_n}{\lambda } & \lambda -\frac{2 g \hat Z_n
   Z_n}{\lambda } \\
\end{array}
\right) \begin{pmatrix}
\frac{\lambda^2-1}{2}I_d  &   -Z_{n-1} \\ 
&\\
2g\hat{Z}_n &  \frac{1-\lambda^2}{2}I_d
\end{pmatrix} \nn 
\eea 
and perform the block products taking care of the ordering of the matrices. 
The element $(1,1)$ vanishes. The element $(2,1)$ is
\be 
\frac{2 g}{\lambda} ( \hat Z_{n+1} - \hat Z_n + 2 g \hat Z_n Z_n \hat Z_n ) 
\ee 
which according to \eqref{systemOYmat} coincides with $- \frac{2 g}{\lambda} \partial_t \hat Z_n$. 
The element $(1,2)$ is
\be 
\frac{1}{\lambda} (Z_{n-1} - Z_n + 2 g Z_n \hat Z_n Z_n ) 
\ee 
which coincides with $\frac{1}{\lambda} \partial_t Z_n$. Finally the $(2,2)$ element simplifies into
\be 
\frac{2 g}{\lambda} (\hat Z_{n+1} Z_n - \hat Z_n Z_{n-1} )
\ee 
which as one can check using \eqref{systemOYmat} coincides with $- \frac{2 g}{\lambda} ((\partial_t \hat Z_n ) Z_n + \hat Z_n \partial_t Z_n)$.

\subsection{FD transformation for the matrix O'Connell-Yor polymer}
We will obtain below the invariant measure of the matrix O'Connell-Yor polymer through a FD symmetry from the field theory of the model. 
We will use the previous Section on the log-Gamma polymer as a guide for the various steps of the calculation. 
In particular, to deal with the (rather subtle) Jacobians of the FD transformation associated to the initial and terminal
conditions in a similar way as was done for the
log-Gamma, we will need to introduce in some steps a time regularization. \\

To simplify the calculation, we first get rid of the diagonal contribution to the action by proceeding to the change of variable
    \begin{equation}
        Z_n \to Z_n e^{-\gamma t}, \quad \hat{Z}_n \to \hat{Z}_n e^{\gamma t}
    \end{equation}
The new action upon this change reads
\begin{equation}
\label{app:eq:action-YO-2}
\begin{split}
    S_0[Z,\hat{Z},W]&=\sum_{n=1}^{N-1} \int_{t_i^+}^{t_f} \rmd t \, \Tr \big[  \hat{Z}_n (\p_t Z_n-Z_{n-1}+(1-\gamma)Z_n-Z_n^{1/2} W_n Z_n^{1/2})+ \frac{1}{4g} W_n^2 \big] \\
\end{split}
\end{equation}
We will subsequently choose $\gamma=1$. From now on we suppose that the original dynamics is studied in the rectangle defined by $n \in [1,N-1]$ and $t \in [t_i^+, t_f]$ where $t_i^+ = t_i+\epsilon$ with $\epsilon=0^+$. Subsequently we will need $t_f^-=t_f-\epsilon$. The matrix OY recursion is applied in this region, with the fixed "initial conditions"
\begin{equation}
 Z_i = \{ Z_0(t) , t \in [t_i,t_f^-] \quad \text{and} \quad Z_1(t_i), \dots, Z_{N-1}(t_i) \}   
\end{equation}
(one may as well consider $\hat{Z}_n$ to be set to zero outside this region). The lower bound of the time integral of \eqref{app:eq:action-YO-2} arises from the time regularization. \\

The FD symmetry is defined by the following inversion of the fields from the set
$\{ Z_0(t),\dots,Z_{N-1}(t)\}$ to the set $\{ Q_0(\tau),\dots,Q_{N-1}(\tau)\}$ together with 
the affine transformation of the response field from the set $\{ \hat Z_1(t),\dots,\hat Z_{N-1}(t) \}$ to the set $\{ \hat Q_1(\tau),\dots,\hat Q_{N-1}(\tau)\}$ 
\begin{equation}
\label{eq:change-var-YO-FDT}
\begin{split}
    &Z_n(t)=Q_{n'-1}^{-1}(\tau^-=\tau-\epsilon), \quad n'=N-n, \; \tau=\tau(t) = t_f+t_i^+-t   \quad \quad \text{for} ~~  n =0,\dots,N-1 \\
    &Z_n(t)^{1/2}\hat{Z}_n(t)Z_n(t)^{1/2} = Q_{n'}(\tau)^{1/2}\hat{Q}_{n'}(\tau)Q_{n'}(\tau)^{1/2}+\frac{1}{2g}\alpha_{n'} \quad \quad \text{for} ~~ n =1,\dots,N-1 \\
    &W_n(t) = \hat{W}_{n'}(\tau)+\alpha_{n'}\\
    & n\in [1,N-1] \Leftrightarrow n' \in [1,N-1] , \quad n\in [0,N-1] \Leftrightarrow n' \in [1,N] 
\end{split}
\end{equation}
where the $\alpha_n$ are defined below in \eqref{eq:fdt-yo-alpha}. The time transformation is chosen so that
\begin{equation}
\label{eq:change-var-YO-FDT-indices}
    \tau([t_i^+,t_f])=[t_i^+,t_f]\, , \quad Z_n(t_i) = Q_{n'-1}^{-1}(t_f) \, , \quad  Z_n(t_f) = Q_{n'-1}^{-1}(t_i), \quad \tau^-([t_i,t_f])=[t_i,t_f]
\end{equation}
which preserves the integral of integration and maps the initial condition to the final condition.\\

The space and time coordinates are reversed for the process obtained through the map $(Z,\hat{Z})\to (Q,\hat{Q})$. We now need to relate the randomness of the initial process $W_n$ with the noise of the reversed process which we define as $\hat{W}_{n'}$, and we need to determine the factor $\alpha_{n'}$ included in the affine transformation of the response field. We further proceed by performing the change of variable \eqref{eq:change-var-YO-FDT} into the action \eqref{app:eq:action-YO-2}, 
using the identity $\int_{t_i^+}^{t_f} \rmd tA(t) = \int^{\tau(t_i^+)=t_f}_{\tau(t_f)=t_i^+} ds A(t(s))$, and 
we obtain 

\begin{equation}
\begin{split}
    &S_0[Z,\hat{Z},W]    =\sum_{n=1}^{N-1} \int_{t_i^+}^{t_f}  \rmd t \, \Tr \big[  Z_n^{1/2}\hat{Z}_nZ_n^{1/2} (Z_n^{-1/2}\p_t Z_nZ_n^{-1/2}-Z_n^{-1/2}Z_{n-1}Z_n^{-1/2}- W_n )+ \frac{1}{4g} W_n^2 \big] \\
  &=\sum_{n'=1}^{N-1} \int_{t_i^+}^{t_f} \rmd \tau \, \Tr \big[  (Q_{n'}^{1/2}\hat{Q}_{n'}Q_{n'}^{1/2}+\frac{1}{2g}\alpha_{n'}) (Q_{n'-1}^{-1/2}\p_\tau Q_{n'-1} Q_{n'-1}^{-1/2}|_{\tau^-}-Q_{n'-1}^{1/2}Q_{n'}^{-1}Q_{n'-1}^{1/2}|_{\tau^-}- \hat{W}_{n'}-\alpha_{n'} )+ \frac{1}{4g} (\hat{W}_{n'}+\alpha_{n'})^2 \big] \\
\end{split}
\end{equation} 
where we used the notation $Q|_{\tau^-}=Q(\tau^-)=Q(\tau-\epsilon)$ arising from the map $Z \to Q$ and we note that
some terms are evaluated at $\tau$ and others at $\tau^-$.\\ 

We identify the terms corresponding to the action in the novel variables (and thus add and subtract the missing terms) to obtain
\begin{equation}
    \begin{split}
        &S_0[Z,\hat{Z},W]  =\hat{S}_0[Q,\hat{Q},\hat{W}]+\sum_{n'=1}^{N-1} \int_{t_i^+}^{t_f}  \rmd \tau \, \Tr \big[  Q_{n'}^{1/2}\hat{Q}_{n'}Q_{n'}^{1/2} \bigg( Q_{n'-1}^{-1/2}\p_\tau Q_{n'-1} Q_{n'-1}^{-1/2}|_{\tau^-}-Q_{n'}^{-1/2}\p_\tau Q_{n'} Q_{n'}^{-1/2}\\
        &+Q_{n'}^{-1/2}Q_{n'-1} Q_{n'}^{-1/2}-Q_{n'-1}^{1/2}Q_{n'}^{-1}Q_{n'-1}^{1/2}|_{\tau^-}-\alpha_{n'} \bigg)+ \frac{1}{2g}\alpha_{n'} (Q_{n'-1}^{-1/2}\p_\tau Q_{n'-1} Q_{n'-1}^{-1/2}|_{\tau^-}-Q_{n'-1}^{1/2}Q_{n'}^{-1}Q_{n'-1}^{1/2}|_{\tau^-} )- \frac{1}{4g}\alpha_{n'}^2\big] \\
    \end{split}
\end{equation} 

The coefficient $\alpha_{n'}$ is found by cancelling the term proportional to the response field $\hat{Q}_{n'}$. 
\begin{equation}
\label{eq:fdt-yo-alpha}
    \alpha_{n'}= Q_{n'-1}^{-1/2}\p_\tau Q_{n'-1} Q_{n'-1}^{-1/2}|_{\tau^-}-Q_{n'}^{-1/2}\p_\tau Q_{n'} Q_{n'}^{-1/2} +Q_{n'}^{-1/2}Q_{n'-1} Q_{n'}^{-1/2}-Q_{n'-1}^{1/2}Q_{n'}^{-1}Q_{n'-1}^{1/2}|_{\tau^-}
\end{equation}
 Upon the replacement of the expression of $\alpha_{n'}$, we finally obtain that the difference of two actions  solely generates boundary terms. 

\be
\begin{split}
S_0[Z,\hat{Z},W]    =& \hat{S}_0[Q,\hat{Q},\hat{W}]+\\
&    \frac{1}{4g}\sum_{n'=1}^{N-1} \int_{t_i^+}^{t_f}  \rmd \tau \, \Tr \big[  (Q_{n'-1}^{-1}\p_\tau Q_{n'-1})^2|_{\tau^-}-    (Q_{n'}^{-1}\p_\tau Q_{n'})^2 +(Q_{n'-1} Q_{n'}^{-1})^2|_{\tau^-}-(Q_{n'-1} Q_{n'}^{-1})^2\\
    &-2 (Q_{n'}^{-1}\p_\tau Q_{n'-1})|_{\tau^-}-2(\p_\tau(Q_{n'}^{-1}) Q_{n'-1}) \big]\\
\underset{\epsilon\to 0}{=} & \hat{S}_0[Q,\hat{Q},\hat{W}]+\frac{1}{4g} \int_{t_i}^{t_f} \rmd \tau \, \Tr \big[ (Q_{0}^{-1} \partial_\tau Q_{0})^2 - (Q_{N-1}^{-1} \partial_\tau Q_{N-1})^2\big] 
- \frac{1}{2g}\sum_{n'=1}^{N-1} \Tr[Q_{n'}^{-1} Q_{n'-1}]_{\tau=t_i}^{t_f}  \\
\end{split}
\ee
In the limit $\epsilon \to 0$ several sums become telescopic in the second line and the two terms in the third line combine into a total time derivative.\\

There is an additional intrinsic degree of freedom in this identity. Let $\kappa > g(d-1)$ (this restriction being necessary for normalisation, see below),  we have  

\be
\begin{split}
&S_0[Z,\hat{Z},W]  =\hat{S}_0[Q,\hat{Q},\hat{W}]+\frac{1}{4g} \int_{t_i}^{t_f} \rmd \tau \, \Tr \big[ (Q_{0}^{-1} \partial_\tau Q_{0})^2 - (Q_{N-1}^{-1} \partial_\tau Q_{N-1})^2\big] \\
&- \frac{1}{2g}\sum_{n'=1}^{N-1} \Tr[Q_{n'}^{-1} Q_{n'-1}]_{\tau=t_i}^{t_f}+\frac{\kappa}{2g}\sum_{n'=1}^{N-1}\left[\log \Det[Q_{n'}^{-1} Q_{n'-1}]\right]_{\tau=t_i}^{t_f} -\frac{\kappa}{2g} \left[\log \Det Q_0 - \log \Det Q_{N-1}\right]_{\tau=t_i}^{t_f}\\
\end{split}
\ee

We rewrite the last terms as 
\begin{equation}
    \begin{split}
     \frac{\kappa}{2g} \left[\log \Det Q_0 - \log \Det Q_{N-1}\right]_{\tau=t_i}^{t_f} &=\frac{\kappa}{2g}   \int_{t_i}^{t_f} \rmd \tau \, \p_\tau \log \Det Q_0- \p_\tau \log \Det Q_{N-1}\\
     &=\frac{\kappa}{2g}    \int_{t_i}^{t_f} \rmd \tau \,  \Tr  [Q_0^{-1}\p_\tau Q_0] - \Tr [Q_{N-1}^{-1}\p_\tau Q_{N-1}]
    \end{split}
\end{equation}
Therefore 
\be
\label{eq:app-FDT-YO-transformation-action}
\begin{split}
S_0[Z,\hat{Z},W]  =&\hat{S}_0[Q,\hat{Q},\hat{W}]+\frac{1}{4g} \int_{t_i}^{t_f} \rmd \tau \, \Tr \big[ (Q_{0}^{-1} \partial_\tau Q_{0}-\kappa)^2 - (Q_{N-1}^{-1} \partial_\tau Q_{N-1}-\kappa)^2\big] \\
&- \frac{1}{2g}\sum_{n'=1}^{N-1} \Tr[Q_{n'}^{-1} Q_{n'-1}]_{\tau=t_i}^{t_f}+\frac{\kappa}{2g}\left[\log \Det[Q_{n'}^{-1} Q_{n'-1}]\right]_{\tau=t_i}^{t_f} 
\end{split}
\ee
which we also rewrite in terms of the partition function $Z_n(t)$ as 
\be
\label{eq:app-FDT-YO-transformation-action-withW}
\begin{split}
S_0[Z,\hat{Z},W]  =&\hat{S}_0[Q,\hat{Q},\hat{W}]+\frac{1}{4g} \int_{t_i}^{t_f} \rmd t \, \Tr \big[ (Z_{N-1}^{-1} \partial_t Z_{N-1}-\kappa)^2 - (Z_{0}^{-1} \partial_t Z_{0}-\kappa)^2\big] \\
&+ \frac{1}{2g}\sum_{n=1}^{N-1} \Tr[Z_{n}^{-1} Z_{n-1}]_{t=t_i}^{t_f}-\frac{\kappa}{2g}\left[\log \Det[Z_{n}^{-1} Z_{n-1}]\right]_{t=t_i}^{t_f} 
\end{split}
\ee

Thus, upon the FD transformation, the action is preserved up to boundary terms in space and time, obtained
here explicitly. As we will see below, the regularisation involving $\epsilon=0^+$ is important for the integration measures in the path integral but not in the action \eqref{eq:app-FDT-YO-transformation-action-withW}.

\subsection{Conventions and identities for the measures on the partition function ratios and log-derivative}\label{subsec:convention-log-derivative}

Until now there was no ordering ambiguity for the expression of the partition function ratios and log-derivative, since they appear only in traces and determinants.
However below we will need to express probability measures on such ratios and an ordering ambiguity will appear. This is quite analogous to
what was done for the log gamma polymer which we will use as a guide. In this section we use the same 
notation for matrix ratios as defined in \eqref{eq:definition-convention-ratios}.\\

For the log-derivative of the partition function, we will use the convention
\begin{equation}
    Z^{-1/2}_n(t)\p_t Z_n(t) Z^{-1/2}_n(t)
\end{equation}
which is the continuum limit of either $r$ or $\tilde{r}$, the two limits being identical (the continuum limit is performed at the vicinity of the identity matrix). From the definition of the recursion of the OY polymer \eqref{eq:matrix-YO-process}, this log-derivative defines a symmetric matrix. Taking the continuum limit of \eqref{AB}, one obtains the identity
\begin{equation}
\label{eq:AB-YO}
    \mu(\rmd B(t))\mu(\rmd B(t+0^+)) = \mu(\rmd B(t)) \rmd (B^{-1/2}\p_t B B^{-1/2})
\end{equation}
The last measure on the log-derivative is the Lebesgue measure as in the continuum limit the $\mathcal{P}_d$ invariant measure on the ratios converges at the vicinity of the identity to the Lebesgue measure as the determinant factor converges to unity.\\

By iterating the identity \eqref{eq:AB-YO} over a time interval $[t_i,t_f]$, we obtain that
\begin{equation}
\label{eq:AB-YO-2}
\begin{split}
    \prod_{t=t_i}^{t_f} \mu(\rmd B(t)) &=  \mu(\rmd B(t_i)) \prod_{t=t_i}^{t_f} \rmd (B^{-1/2}\p_t B B^{-1/2})
\end{split}
\end{equation}

This allows to define a reference measure on the initial condition $Z_i$ defined in \eqref{eq:YO-initial-values} which will be useful in the following as a building block of the invariant measure.
Applying iteratively \eqref{AB} and \eqref{eq:AB-YO} this reference measure takes the four equivalent forms either (i) in terms of the partition functions only 
(ii) in terms of the consecutive ratios as well as log-derivatives, and one chosen "zero mode" (also called reference point)

\begin{equation}
\label{eq:YO-jacobian-ratio-partitionfunc-finalline-identity-1}
    \begin{split}
           &\prod_{n=1}^{N-1}  \mu(\rmd Z_{n}(t_i)) \prod_{t=t_i}^{t_f} \mu(\rmd  Z_{0}(t))\\
           &=\mu(\rmd Z_{0}(t_i)) \prod_{n=1}^{N-1}\mu(\rmd r\left(\frac{Z_{n}(t_i)}{Z_{n-1}(t_i)}\right))  \prod_{t=t_i}^{t_f} \rmd (Z_0^{-1/2}\p_t Z_0 Z_0^{-1/2}) \\
           &=\mu(\rmd Z_{N-1}(t_i))\prod_{n=1}^{N-1}\mu(\rmd \tilde{r}\left(\frac{Z_{n}(t_i)}{Z_{n-1}(t_i)}\right))\prod_{t=t_i}^{t_f} \rmd (Z_0^{-1/2}\p_t Z_0 Z_0^{-1/2})   \\
           &=\mu(\rmd Z_{0}(t_f)) \prod_{n=1}^{N-1}\mu(\rmd r\left(\frac{Z_{n}(t_i)}{Z_{n-1}(t_i)}\right))  \prod_{t=t_i}^{t_f} \rmd (Z_0^{-1/2}\p_t Z_0 Z_0^{-1/2}) \\
    \end{split}
\end{equation}

Moving around the reference point $Z_{0}(t_i)$ to the corners $Z_{N-1}(t_i)$ or $Z_{0}(t_f)$ has the effect to transform $r$ into $\tilde{r}$ in the measure, which amounts to change the matrix ordering within the symmetrized ratio. This is particularly important to study different paths of the matrix OY polymer, either a departing path from the origin $(n=0,t=t_i)$ or a down-right path. These distinctions between $r$ and $\tilde r$ are of course irrelevant for the scalar OY polymer since all the variables commute for $d=1$.\\

Applying again iteratively \eqref{AB} and \eqref{eq:AB-YO}, we also obtain a reference
measure on the final condition $Z'_f$ on the final condition $Z_f'$, defined in \eqref{eq:log-gamma-final-values}

\begin{equation}
\label{eq:YO-jacobian-ratio-partitionfunc-finalline-identity-2}
\begin{split}
  &    \prod_{n=0}^{N-2} \mu(\rmd Z_{n}(t_f)) \prod_{t=t_i}^{t_f} \mu(\rmd Z_{N-1}(t))  \\
  &= \mu( \rmd Z_{N-1}(t_f) ) \prod_{n=1}^{N-1} \mu(\rmd  \tilde{r}\left(\frac{Z_{n}(t_f)}{Z_{n-1}(t_f)}\right))\prod_{t=t_i}^{t_f} \rmd(  Z_{N-1}^{-1/2}\p_t Z_{N-1} Z_{N-1}^{-1/2})\\
          &=  \mu( \rmd Z_{0}(t_f) )\prod_{n=1}^{N-1} \mu (\rmd r\left( \frac{Z_{n}(t_f)}{Z_{n-1}(t_f)}\right))\prod_{t=t_i}^{t_f} \rmd(  Z_{N-1}^{-1/2}\p_t Z_{N-1} Z_{N-1}^{-1/2})\\
          &=  \mu( \rmd Z_{N-1}(t_i) )\prod_{n=1}^{N-1} \mu (\rmd \tilde{r}\left( \frac{Z_{n}(t_f)}{Z_{n-1}(t_f)}\right))\prod_{t=t_i}^{t_f} \rmd(  Z_{N-1}^{-1/2}\p_t Z_{N-1} Z_{N-1}^{-1/2})
    \end{split}
\end{equation}

\subsection{Invariant measure of the matrix O'Connell-Yor polymer}
 We will use the FD symmetry to obtain the invariant measure of the OY polymer. One needs to take into account the Jacobian

\begin{equation}
    (Z,\hat{Z},W) \to (Q,\hat{Q},\hat{W})
\end{equation}

\subsubsection{Jacobian of the FD transformation}

Everywhere below, we will distinguish three regions for the space time indices of the fields (by analogy with the log-Gamma polymer)
and in agreement with the definitions given above \eqref{eq:YO-initial-values} and \eqref{eq:YO-terminal-values} 
\begin{enumerate}
    \item The bulk of the dynamics $(n,t)\in [1,N-2]\times [t_i^+,t_f^-]$
    \item The terminal boundary $(n,t)=\{N-1\}\times [t_i^+,t_f]$ and $ (n,t)=[1,N-2] \times \{t_f\} $
    \item The initial boundary $(n,t)=\{ 0 \} \times [t_i,t_f]$ and $(n,t)=[1,N-1]\times \{t_i\}$
\end{enumerate}

The FD transformation induces a Jacobian from the map of the fields $(Z,\hat{Z},W)\to (Q,\hat{Q},\hat{W})$. Since this map is triangular, its total Jacobian is the product of its diagonal elements. As the FD transformation over the noise is additive \eqref{eq:change-var-YO-FDT}, the associated Jacobian is equal to unity.\\

     The response field $\hat{Z}$ have space-time indices  in the bulk and on the terminal boundary, and so does 
     $\hat{Q}$, see \eqref{eq:change-var-YO-FDT-indices} and \eqref{eq:change-var-YO-FDT}. We denote the associated Jacobian $|\frac{\mathcal{D}\hat{Z}}{\mathcal{D}\hat{Q}}|_{\rm bulk+terminal}$. From the FD transformation \eqref{eq:change-var-YO-FDT} and \eqref{detg}, the Jacobian reads

\begin{equation}
\begin{split}
    |\frac{\mathcal{D}\hat{Z}}{\mathcal{D}\hat{Q}}|_{\rm bulk+terminal} &=  \prod_{n=1}^{N-1} \prod_{t=t_i^+}^{t_f} |\det Z_{n}(t)|^{-\frac{d+1}{2}}   |\det Q_{n'}(\tau)|^{\frac{d+1}{2}} \\
    &=  \prod_{n=1}^{N-1} \prod_{t=t_i^+}^{t_f} |\det Z_{n}(t)|^{-\frac{d+1}{2}}  \prod_{t=t_i}^{t_f^-} |\det Z_{n-1}(t)|^{-\frac{d+1}{2}} 
\end{split}
\end{equation}
where we recall on the first line $n'=N-n$, $\tau(t)=t_f+t_i^+-t$ as well as
\begin{equation}
\begin{split}
    Q^{-1}_{n'}(\tau(t))&=Z_{N-n'-1}(t_i+t_f-\tau(t))=Z_{N-n'-1}(t_i+t_f-(t_f+t_i^+-t ))=Z_{N-n'-1}(t^-)\\
    \end{split}
\end{equation}

We consider the Jacobian of the partition function solely in the bulk as the initial and terminal boundaries will be handled separately. We will denote this Jacobian $|\frac{\mathcal{D}Z}{\mathcal{D}Q}|_{\rm bulk }$. From \eqref{detinverse} and \eqref{eq:change-var-YO-FDT}, it reads

\begin{equation}
\begin{split}
   | \frac{\mathcal{D}Z}{\mathcal{D}Q}|_{\rm bulk } &= \prod_{n=1}^{N-2}\prod_{t=t_i^+}^{t_f^-}|\Det Z_n(t) |^{d+1} 
    \end{split}
\end{equation}

The product of the two Jacobians read
\begin{equation}
\label{eq:YO-total-jacobian-MSR-invariant}
\begin{split}
    J_{total}&=|\frac{\mathcal{D}\hat{Z}}{\mathcal{D}\hat{Q}}|_{\rm bulk+terminal}|\frac{\mathcal{D}Z}{\mathcal{D}Q}|_{\rm bulk }  \\
    &= \prod_{n=1}^{N-2}|\Det Z_n(t_i)|^{-\frac{d+1}{2}} |\Det Z_n(t_f)|^{-\frac{d+1}{2}} \prod_{t=t_i}^{t_f^-}|\Det Z_0(t)|^{-\frac{d+1}{2}} \prod_{t=t_i^+}^{t_f}|\Det Z_{N-1}(t)|^{-\frac{d+1}{2}}
    \end{split}
\end{equation}
which is the analog of the result obtained for the Log-Gamma in \eqref{eq:invariant-measure-log-gamma-jacobian3}. The space index and time variables arising in \eqref{eq:YO-total-jacobian-MSR-invariant} correspond to the initial and terminal boundaries without the corners $(n=0,t=t_f)$ and $(n=N-1,t=t_i)$.

\subsubsection{Application of the FD symmetry to the MSR path integral}

We now exponentiate the FD relation \eqref{eq:app-FDT-YO-transformation-action-withW}
and rearrange the left and right hand sides. Next, we 
integrate over all noises $W_n(t)$ and response fields $\hat Z_n(t)$ with  $(n,t) \in [1,N-1]\times [t_i^+,t_f]$ (in the bulk and terminal boundary) as well as over all the variables $Z_n(t)$ with indices $(n,t)$ in the bulk. This integration is performed on both sides of the equation and thus requires the Jacobian \eqref{eq:YO-total-jacobian-MSR-invariant} on the r.h.s. One obtains after insertion of an observable over the final fields $\mathcal{O}(Z_f')$
\be
\label{eq:YO-MSR-invariant-measure}
\begin{split}
& \int_{bulk} \mathcal{D}Z\iint  \mathcal{D} W \mathcal{D}\hat{Z} e^{-S_0[Z,\hat{Z},W]} \mathcal{O}(Z_f')\\
&\times  e^{- \frac{1}{4g} \int_{t_i}^{t_f} \rmd t \, \Tr \big[  (Z_{0}^{-1} \partial_t Z_{0}-\kappa)^2\big]}\prod_{n=1}^{N-1} |\Det [ Z_{n}^{-1}(t_i)Z_{n-1}(t_i)]|^{\frac{\kappa}{2g}} e^{-\frac{1}{2g}\Tr[ Z_{n}^{-1}(t_i)Z_{n-1}(t_i)]}
\\
& = \int_{bulk} \mathcal{D}Q\iint \mathcal{D} \hat{W} \mathcal{D}\hat{Q} e^{-\hat{S}_0[Q,\hat{Q},\hat{W}]}\mathcal{O}(Z_f')J_{total} \\
& \times 
e^{- \frac{1}{4g} \int_{t_i}^{t_f} \rmd t \, \Tr \big[  (Z_{N-1}^{-1} \partial_t Z_{N-1}-\kappa)^2\big]} \prod_{n=1}^{N-1} |\Det [ Z_{n}^{-1}(t_f)Z_{n-1}(t_f)]|^{\frac{\kappa}{2g}} e^{-\frac{1}{2g}\Tr[ Z_{n}^{-1}(t_f)Z_{n-1}(t_f)]}  
\end{split}
\ee  
At this stage the initial $Z_i$ and terminal $Z_f$ fields are fixed (they have not yet been integrated over). We now multiply both sides of \eqref{eq:YO-MSR-invariant-measure} by the following measure over the initial and terminal fields
\begin{equation}
\label{eq:YO-MSR-invariant-measure-integration-partition-func}
    \underbrace{\prod_{t=t_i}^{t_f}\mu(\rmd Z_{0}(t))\prod_{n=1}^{N-1} \mu(\rmd Z_{n}(t_i))}_{\text{initial values}} \times \underbrace{\prod_{t=t_i^+}^{t_f}\rmd Z_{N-1}(t)  \prod_{n=1}^{N-2} \rmd Z_{n}(t_f) }_{\text{terminal values}}
\end{equation}
and integrate over all $Z_n(t)$ on the initial and terminal conditions, obtaining schematically the new equation

\begin{equation}
\label{eq:YO-MSR-invariant-measure-integrated}
    \int_{initial}\int_{terminal} \eqref {eq:YO-MSR-invariant-measure-integration-partition-func} \times \, l.h.s \eqref{eq:YO-MSR-invariant-measure} =\int_{initial}\int_{terminal} \eqref {eq:YO-MSR-invariant-measure-integration-partition-func} \times \, r.h.s \eqref{eq:YO-MSR-invariant-measure} 
\end{equation}
We now treat separately the l.h.s and r.h.s of \eqref{eq:YO-MSR-invariant-measure-integrated}. In the right hand side, we will use the normalisation of the MSR path integral whereas in the left hand side we will interpret the path integral as an expectation of the observable $\mathcal{O}(Z_f')$ conditioned on the initial values from the set $Z_i^{(r)}$.

\subsubsection{Right hand side of \eqref{eq:YO-MSR-invariant-measure-integrated} }

The measure \eqref{eq:YO-MSR-invariant-measure-integration-partition-func} multiplied by the Jacobian of the map $(Z,\hat{Z},W) \to (Q,\hat{Q},\hat{W})$ \eqref{eq:YO-total-jacobian-MSR-invariant} provides an overall factor equal to
\begin{equation}
\begin{split}
     &\prod_{n=1}^{N-2}|\Det Z_n(t_i)|^{-\frac{d+1}{2}} |\Det Z_n(t_f)|^{-\frac{d+1}{2}} \prod_{t=t_i}^{t_f^-}|\Det Z_0(t)|^{-\frac{d+1}{2}} \prod_{t=t_i^+}^{t_f}|\Det Z_{N-1}(t)|^{-\frac{d+1}{2}} \\
     &\times \prod_{t=t_i}^{t_f}\frac{\rmd Z_{0}(t)}{|\Det Z_{0}(t)|^{\frac{d+1}{2}}}  \prod_{n=1}^{N-1} \frac{\rmd Z_{n}(t_i)}{|\Det Z_{n}(t_i)|^{\frac{d+1}{2}}} \prod_{t=t_i^+}^{t_f}\rmd Z_{N-1}(t)  \prod_{n=1}^{N-2} \rmd Z_{n}(t_f) \\
     &=\prod_{n=0}^{N-2}\frac{\rmd Z_{n}(t_i)}{|\Det Z_{n}(t_i)|^{d+1 }}\prod_{t=t_i^+}^{t_f^-} \frac{\rmd Z_{0}(t) }{|\Det Z_{0}(t)|^{d+1}}\prod_{t=t_i}^{t_f}\frac{\rmd Z_{N-1}(t)}{|\Det Z_{N-1}(t)|^{\frac{d+1}{2}}}  \prod_{n=0}^{N-2} \frac{\rmd Z_{n}(t_f)}{|\Det Z_n(t_f)|^{\frac{d+1}{2}} } 
     \end{split}
\end{equation}

The integration over the initial fields in the right hand side of \eqref{eq:YO-MSR-invariant-measure} can be done explicitly through the normalisation of the MSR path integral. Indeed, the MSR normalisation reads, where $Q_0(\tau)$, $\tau \in [t_i^+, t_f]$ $\{ Q_n(\tau=t_i)\}_{n=0,N-1}$ are fixed
\begin{equation}
    \iint\prod_{n=1}^{N-1}\rmd Q_{n}(t_f) \prod_{\tau=t_i^+}^{t_f^-} \rmd Q_{N-1}(\tau) 
   \int_{\rm bulk} \mathcal{D}Q \iint\mathcal{D}\hat{W} \mathcal{D}\hat{Q} e^{-S_0[Q,\hat{Q},\hat{W}]}=1 
\end{equation}
This implies using the FD transformation 
\begin{equation}
    \iint\prod_{n=0}^{N-2}\frac{\rmd Z_{n}(t_i)}{|\Det Z_{n}(t_i)|^{d+1 }}\prod_{t=t_i^+}^{t_f^-} \frac{\rmd Z_{0}(t) }{|\Det Z_{0}(t)|^{d+1}}
    \int_{\rm bulk} \mathcal{D}Q \iint\mathcal{D}\hat{W} \mathcal{D}\hat{Q}  e^{-S_0[Q,\hat{Q},\hat{W}]}=1 
\end{equation}
Using this identity, 
the right hand side \eqref{eq:YO-MSR-invariant-measure-integrated} becomes
\begin{equation} \label{eq:YO-MSR-invariant-rhs-final}
\begin{split}
 r.h.s \, \eqref{eq:YO-MSR-invariant-measure-integrated}&=   \iint \prod_{t=t_i}^{t_f}\mu(\rmd Z_{N-1}(t))\prod_{n=0}^{N-2}\mu(\rmd Z_{n}(t_f)) \mathcal{O}(Z_f') \\
 &e^{- \frac{1}{4g} \int_{t_i}^{t_f} \rmd t \, \Tr \big[  (Z_{N-1}^{-1} \partial_t Z_{N-1}-\kappa)^2\big]} \prod_{n=1}^{N-1} |\Det [ Z_{n}^{-1}(t_f)Z_{n-1}(t_f)]|^{\frac{\kappa}{2g}} e^{-\frac{1}{2g}\Tr[ Z_{n}^{-1}(t_f)Z_{n-1}(t_f)]}  
 \end{split}
\end{equation}

\subsubsection{Left hand side 
of \eqref{eq:YO-MSR-invariant-measure-integrated}}

We now consider the left hand side of \eqref{eq:YO-MSR-invariant-measure} after multiplication by 
\eqref{eq:YO-MSR-invariant-measure-integration-partition-func} and integration

\begin{equation}
\label{eq:YO-MSR-invariant-lhs-final}
    \begin{split}
        &l.h.s \, \eqref{eq:YO-MSR-invariant-measure-integrated}= \int \prod_{t=t_i}^{t_f}\mu(\rmd Z_{0}(t)) \prod_{n=1}^{N-1} \mu(\rmd Z_{n}(t_i)) \prod_{t=t_i^+}^{t_f}\rmd Z_{N-1}(t)  \prod_{n=1}^{N-2} \rmd Z_{n}(t_f)  \\
        &\times\int_{bulk} \mathcal{D}Z \iint  \mathcal{D} W \mathcal{D}\hat{Z} e^{-S_0[Z,\hat{Z},W]} \mathcal{O}(Z_f')
e^{- \frac{1}{4g} \int_{t_i}^{t_f} \rmd t \, \Tr \big[  (Z_{0}^{-1} \partial_t Z_{0}-\kappa)^2\big]}\prod_{n=1}^{N-1} |\Det [ Z_{n}^{-1}(t_i)Z_{n-1}(t_i)]|^{\frac{\kappa}{2g}} e^{-\frac{1}{2g}\Tr[ Z_{n}^{-1}(t_i)Z_{n-1}(t_i)]}\\
&= \iint \prod_{t=t_i}^{t_f}\mu(\rmd Z_{0}(t))\prod_{n=1}^{N-1} \mu(\rmd Z_{n}(t_i))\E[\mathcal{O}(Z_f')|Z_i] \\
& \times e^{- \frac{1}{4g} \int_{t_i}^{t_f} \rmd t \, \Tr \big[  (Z_{0}^{-1} \partial_t Z_{0}-\kappa)^2\big]}\prod_{n=1}^{N-1} |\Det [ Z_{n}^{-1}(t_i)Z_{n-1}(t_i)]|^{\frac{\kappa}{2g}} e^{-\frac{1}{2g}\Tr[ Z_{n}^{-1}(t_i)Z_{n-1}(t_i)]}
    \end{split}
\end{equation}
where $\E[\mathcal{O}(Z_f')|Z_i] $ denotes the average over the noise of the functional of the final values of the stochastic process
for fixed values of the initial condition $Z_i$, see \eqref{eq:MSR-YO-expectation-value}. \\

\subsubsection{Summary and invariant measure}

Equating Eqs.~\eqref{eq:YO-MSR-invariant-rhs-final} and \eqref{eq:YO-MSR-invariant-lhs-final} shows how the measure 
on the initial values $Z_i$ in \eqref{eq:YO-MSR-invariant-rhs-final}
is transported by the OY polymer dynamics onto the measure on the final values $Z_f'$ in \eqref{eq:YO-MSR-invariant-lhs-final}.
We recall that $Z_i$ is defined in \eqref{eq:YO-initial-values}
and $Z'_f$ in \eqref{eq:YO-final-values}.

We now use the relations \eqref{eq:YO-jacobian-ratio-partitionfunc-finalline-identity-1} and \eqref{eq:YO-jacobian-ratio-partitionfunc-finalline-identity-2} to express these measures more conveniently
in terms of (i) partition sum ratios (discrete part) and (ii) log derivatives (continuum part).
This shows explicitly that (i) the ratios are independent inverse Wishart distributed and 
(ii) the continuum part is geometric matrix Brownian motion. More precisely, here we choose
$Z_{0}(t_i)$ as the zero mode for the initial condition, and $Z_{0}(t_f)$ as the zero mode for the
final condition. Then the initial measure reads
\be 
\label{eq:invariant-measure-YO-compressed-1}
\mu(\rmd Z_{0}(t_i)) \prod_{n=1}^{N-1}\rmd P^{iW}_{\kappa/(2g),g}[ \,  r\left(\frac{Z_{n}(t_i)}{Z_{n-1}(t_i)} \right)]\prod_{t=t_i}^{t_f}  \rmd (Z_0^{-1/2}\p_t Z_0 Z_0^{-1/2}) e^{- \frac{1}{4g} \int_{t_i}^{t_f} \rmd t \, \Tr \big[  (Z_{0}^{-1} \partial_t Z_{0}-\kappa)^2\big]}
\ee 
while the final measure reads
\be 
\mu(\rmd Z_{0}(t_f)) \prod_{n=1}^{N-1}\rmd P^{iW}_{\kappa/(2g),g}[ \, r\left(\frac{Z_{n}(t_f)}{Z_{n-1}(t_f)} \right)] \prod_{t=t_i}^{t_f}\rmd(  Z_{N-1}^{-1/2}\p_t Z_{N-1} Z_{N-1}^{-1/2}) e^{- \frac{1}{4g} \int_{t_i}^{t_f} \rmd t \, \Tr \big[  (Z_{N-1}^{-1} \partial_t Z_{N-1}-\kappa)^2\big]}
\ee 
where we recall the definition of the inverse Wishart measure from \eqref{eq:def-inverse-wishart-pdf}.
Note that these measures can be equivalently rewritten using the identity \eqref{eq:AB-YO-2}.\\

One sees that now these define the invariant measure on the ratios and log-derivatives.

\begin{enumerate}
    \item All ratios and log-derivatives are sampled independently.
    \item The initial line $Z_{0}(t)$ and the final line $Z_{N-1}(t)$, $t \in [t_i,t_f]$ are matrix geometric Brownian motions with drift $\kappa \times I_d$
    and diffusion coefficient $2 g$.
    \item The ratios along the vertical (space) axis defined with the matrix ordering convention $r$, see \eqref{eq:definition-convention-ratios}, are sampled from the Inverse Wishart distribution with parameter $(\kappa /(2g),g)$.
    \item The initial zero mode $Z_{0}(t_i)$ and the final zero mode $Z_{0}(t_f)$ are uniformly distributed over $\mathcal{P}_d$. Note that they are respectively initial and final conditions of the geometric Brownian motions $Z_0(t)$.
    \item We have the constraint $\kappa/(2g)> \frac{d-1}{2} $ -- the constraint coming from the normalisation condition of the inverse Wishart distribution.
\end{enumerate}

\begin{remark}
    The initial measure \eqref{eq:invariant-measure-YO-compressed-1} could also be re-written by replacing $\mu(\rmd Z_{0}(t_i)) $ by $\mu(\rmd Z_{0}(t_f)) $ so that the initial and final measure share a common point. The interpretation of this replacement would amount to sample uniformly either the initial point or of the final point of the geometric Brownian motion.
\end{remark}

\begin{remark}
In Ref.~\cite{matrixOYpolymer} a version of the matrix OY polymer was constructed and studied using
Markov generators. In \cite[Section~3]{matrixOYpolymer} an invariant measure was obtained 
(see Eq.~(3.4) and Theorem 2 for $N=2$, and generalisation for any $N$ below) by probabilistic methods (different from the one used here).
\end{remark}

\section{Field theory of the matrix strict-weak polymer}
\label{sec:strict-weak-app}

The strict-weak polymer is defined by the following recursion the partition function  $Z_{n,t} \in {\mathcal P}_d$
\begin{equation}
    \label{eq:app-strict-weak-def}Z_{n,t+1}=Z_{n,t}^{1/2}Y_{n,t}Z_{n,t}^{1/2} +Z_{n-1,t} \, .
\end{equation}
 The randomness $Y_{n,t}$ is a Wishart random matrix. Its probability distribution function reads

\bea 
\rmd P^W_\alpha[ Y] =\frac{1}{\Gamma_d(\alpha) } |\det Y|^{\alpha}  e^{-  \Tr \, Y } \mu(dY) 
\eea 
with $\alpha > \frac{d-1}{2}$ and where $\mu$ is the natural measure on ${\cal P}_d$ given in Eq.~\eqref{eq:measure-psd} and the normalisation is the same as \eqref{eq:normalisation-inverse-wishart}.\\

The recursion is applied for $1 \leq n \leq N-1$, $1 \leq t \leq T-1$ with the additional constraints $ n \leq t$ and $n-(N-1)\geq t-(T-1)$,
as depicted in Fig.~\ref{fig:strict-weakpolymer-lightcone-wnt}. The grey band in the figure correspond the "initial data" which is then
propagated up to the point $(T-1,N-1)$. The initial data corresponding to the point-to-point 
strict-weak polymer have a localised initial condition given by
\begin{equation}
    \begin{cases}
       Z_{n,n-1}=0, \quad & 2 \leq n \leq N-1 \\ 
       Z_{1,0}= I_d \\
       Z_{0,t}=0 
    \end{cases}
\end{equation}
Below we will also study the stationary initial condition. \\

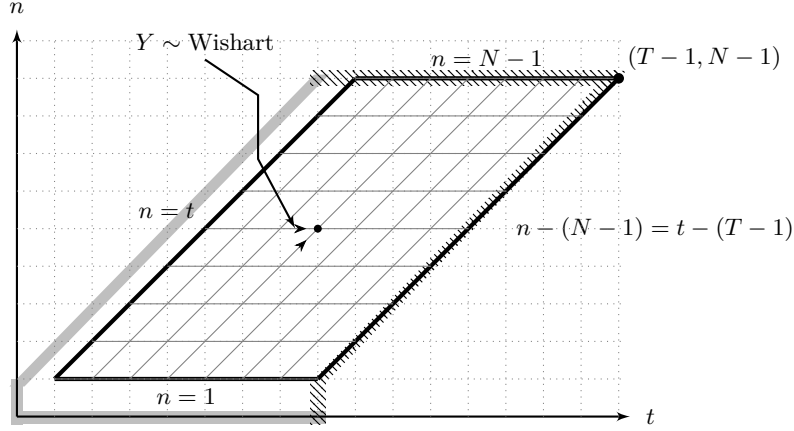
\begin{figure}[t!]
\begin{tikzpicture}[scale=0.5]

\fill[gray!50] (0,-0.2) rectangle (8.2,0.13);
\fill[gray!50] (-0.15,-0.2) rectangle (0.15,1);
% \fill[gray!50] (-0.2,-0.2) rectangle (0.2,1);
\fill[gray!50] (1-1.2,1-0.2) -- (7.9,9.1) -- (8.3,9.1) -- (1-0.8,1-0.2) -- cycle;

\fill[pattern=north west lines, pattern color=black] (7.8,8.8) rectangle (16,9.2);
\fill[pattern=north west lines, pattern color=black] (15.8,9.2) -- (16.2,9.2) -- (8.3,1) -- (7.9,1) -- cycle;
\fill[pattern=north west lines, pattern color=black]  (7.8,-0.2) rectangle (8.2,1);

% 
%% axis
\draw[->, thick,>=latex'] (0, 0) -- (0, 10.3);
\draw[->, thick,>=latex'] (0,0)--( 16.3, 0);

% %% light cone for z
% \draw[ultra thick] (0,1) -- (8,9);
% \draw[ultra thick] (0,1) -- (7,1);

% %% light cone for ztilde
% \draw[ultra thick] (8,9) -- (15,9);
% \draw[ultra thick] (7,1) -- (15,9);

%% Light cone for z

% \draw[ultra thick] (1,1) -- (1,2);
\draw[ultra thick] (1,1) -- (9,9) node[midway, above left] {$n=t$};
\draw[ultra thick] (1,1) -- (8,1) node[midway, below] {$n=1$};

%% Light cone for ztilde
\draw[ultra thick] (9,9) -- (16,9) node[midway, above] {$n=N-1$};
\draw[ultra thick] (8,1) -- (16,9) node[midway, right, xshift=0.5cm] {$n-(N-1)=t-(T-1)$};

\foreach \k in {1,2, ..., 16}
	{\draw[gray, dotted] (\k, 0) -- (\k, 10.1);}
\foreach \k in {1,2, ..., 10}
	{\draw[gray, dotted] (0,\k) -- (16.1, \k);}

\clip (-2, -1) rectangle (20.5, 11.3);

\foreach \k in { 1, 2, ..., 9}
	{\draw[gray] (\k,\k) -- (\k+7, \k);}
\foreach \k in{ 2, 3, ..., 7}
	{\draw[gray] (\k, 1) -- (\k+8,9);}

\draw(16,9) node[anchor=south west]{$(T-1, N-1)$};
\fill (16,9) circle(4pt);
% \draw (0,1) node[anchor=east]{$(0,1)$};

\node[above] at (0,10.5) {$n$};
\node[right] at (16.5,0) {$t$};

 \draw[black, thick, decorate, decoration={zigzag, segment length=30mm, amplitude=4mm, post length=6mm}, ->,>=latex'] (5,9.5) node[above] {$Y\sim \mathrm{Wishart}$} -- (7.4,5);

\draw[-Stealth] (7.5,4.5) -- (7.7,4.7);
\draw[-Stealth] (7.5,5) -- (7.7,5);
\fill (8,5) circle(3pt);

\end{tikzpicture}
\caption{The black thick lines define the region - a parallelogram $\mathcal{C}$ - where the recursion of the Strict-Weak polymer is applied. The gray band corresponds to the initial values of the partition function $Z_i$ (this defines the minimal boundary conditions necessary to determine the value of the partition function at position $(N-1,T-1)$, see also Eq.~\eqref{eq:SW1-initial-values}) and the band filled with diagonal lines corresponds to the final values of the partition function $Z'_f$ (see Eq.~\eqref{eq:SW1-final-values}). The two arrowheads describe how the recursion propagates on the lattice. Note that the convention on the figure  for the axes $(n,t)$ have been changed to $(t,n)$ to follow Ref.~\cite{corwin2015strict}.}
\label{fig:strict-weakpolymer-lightcone-wnt}
\end{figure}

The expectation value over the noise of any functional ${\cal O}[Z]$ of the space time matrix field $Z= Z_{n,t}$ can be written as a path integral (up to normalisation) involving a response matrix field $\hat{Z}$
\be 
\label{eq:average-observable-msr-strictweak}
\overline{  {\cal O}[Z] } = \iiint \mathcal{D}_\mu Y {\cal D}\hat Z  {\cal D} Z \, {\cal O}[Z] \, 
e^{-  \hat{S}_0[Z,\hat{Z},Y] }
\ee 
where $\mathcal{D}_\mu Y=\prod_{n,t}\mu(dY_{n,t})  $ where the action reads
\be
\label{eq:app-strict-weak-action-three-fields}
 \hat{S}_0[Z,\hat{Z},Y] = \sum_{n,t}  \Tr \left[ \hat{Z}_{n,t} \left(Z_{n,t+1}-Z_{n,t}^{1/2}Y_{n,t}Z_{n,t}^{1/2} -Z_{n-1,t} \right)\right]  - \alpha  \log \det Y_{n,t} + 
  \Tr \, Y_{n,t}
\ee
The MSR integral which appears in the r.h.s of \eqref{eq:average-observable-msr-strictweak} is normalised
to unity, as a consequence of the identity $
\iint \rmd Z \rmd \hat Z e^{ {\rm Tr} \hat Z (Z - Z_0)}  = 1 $, consistent with $\E_V[1] =1$
(upon choosing $\mathcal{O}[\cdot]=1$).\\

The noise can be fully integrated out using the expression for the characteristic function of a Wishart matrix which reads
\begin{equation}
 \E[e^{- \Tr[XY]}]=\Det(I+X)^{-\alpha}   
\end{equation}
This then yields $\overline{  {\cal O}[Z] } = \iint   {\cal D}\hat Z  {\cal D} Z \, {\cal O}[Z] \, 
e^{-  S[Z,\hat{Z}] }$ with the following action
\begin{equation}
\label{eq:app-strict-weak-action-two-fields}
    S[Z,\hat{Z}]= \sum_{n,t}  \Tr \left[\hat{Z}_{n,t}  (Z_{n,t+1}-Z_{n-1,t} )\right] + \alpha \log \Det (I_d- Z_{n,t}\hat{Z}_{n,t} )
\end{equation}

Similarly to the case of the log Gamma polymer, one can represent the expectation value over any observable of the recursion \eqref{eq:app-strict-weak-def} using a path integral formula analog to the one in \eqref{eq:average-observable-msr-log-gamma}. To treat more general initial/final conditions, the evolution of the partition function starts from a list of "initial values"
\begin{equation}
\label{eq:SW1-initial-values}
    \begin{split}
 Z_i &= ( Z_{0,0}, Z_{0,1}\dots, Z_{0,T-N+1} ; Z_{1,0}, \dots , Z_{N-1,N-2} )\in \mathcal{P}_d^{N+T-1} 
    \end{split}
\end{equation}
and ends with a list of a "terminal values"
\begin{equation}
\label{eq:SW1-terminal-values}
    \begin{split}
         Z_f &= ( Z_{N-1,N-1}, \dots Z_{N-1,T-1} ;  Z_{1,T-N+1} \dots Z_{N-2,T-2} )\in \mathcal{P}_d^{N+T-3} 
    \end{split}
\end{equation}
Note that the two corners $Z_{N-1,N-2}$ and $Z_{0,T-N-1}$ are not included in $Z_f$, since the recursion is not applied at these points. 
Below, we will also need the list of "final values" which are the terminal values complemented with the corners, 
\begin{equation}
\label{eq:SW1-final-values}
    \begin{split}
         Z'_f &= ( Z_{N-1,N-2}, \dots Z_{N-1,T-1} ; Z_{0,T-N+1}, Z_{1,T-N+1} \dots Z_{N-2,T-2}  )\in \mathcal{P}_d^{N+M-1} 
    \end{split}
\end{equation}

\subsection{Weak-noise limit and saddle point of the matrix Strict-Weak polymer}
To investigate the weak-noise theory of the matrix Strict-Weak polymer, one needs to first proceed to the change of variable
\begin{equation}
\label{eq:wnt-strict-weak-app}
    Y_{n,t} \to \alpha Y_{n,t} , \quad 
Z_{n,t} \to \alpha^{t-n} Z_{n,t}  \quad , \quad 
\hat{Z}_{n,t} \to  \alpha^{n-t} \hat{Z}_{n,t} 
\end{equation}
This ensures that the whole action \eqref{eq:app-strict-weak-action-three-fields} is now proportional to $\alpha$ which we consider large, i.e., $\alpha \gg 1$, to obtain the saddle point equations. We start with the saddle point equation obtained by differentiating $\hat S_0[Z,\hat Z, Y]$ in
\eqref{eq:app-strict-weak-action-three-fields} w.r.t the response field, which yields

\be 
\label{eq:WNT-strict-weak-SP1-intermediate}
Z_{n,t+1}=Z_{n,t}^{1/2}Y_{n,t}Z_{n,t}^{1/2} +Z_{n-1,t}
\ee 
The saddle point equation obtained by differentiating $\hat S_0[Z,\hat Z, Y]$ w.r.t
the noise reads 
\begin{equation}
\label{eq:WNT-strict-weak-optimal-noise}
    I_d-Y_{n,t}^{-1}=Z_{n,t}^{1/2}\hat{Z}_{n,t}Z_{n,t}^{1/2} \quad \Longleftrightarrow   \quad   Y_{n,t}=(I_d-Z_{n,t}^{1/2}\hat{Z}_{n,t}Z_{n,t}^{1/2})^{-1}
\end{equation}
Injecting the expression of $Y_{n,t}$ obtained in \eqref{eq:WNT-strict-weak-optimal-noise} into \eqref{eq:WNT-strict-weak-SP1-intermediate}, we obtain the first equation of the saddle point system
\begin{equation}
\begin{split}
Z_{n,t+1}&=Z_{n,t}^{1/2}(I_d-Z_{n,t}^{1/2}\hat{Z}_{n,t}Z_{n,t}^{1/2})^{-1}Z_{n,t}^{1/2} +Z_{n-1,t}\\
&=Z_{n,t}(I_d-\hat{Z}_{n,t}Z_{n,t})^{-1}+Z_{n-1,t}\\
\end{split}
\end{equation}

The last saddle point equation is obtained by differentiating $\hat S_0[Z,\hat Z, Y]$ with respect to the field $Z_{n,t}$ using the expansion \eqref{eq:saddle-point-sqrt}. It reads
\begin{equation}
\begin{split}
  &\hat{Z}_{n,t-1}-\hat{Z}_{n+1,t}=\int_0^{\infty} \rmd u \, e^{-uZ_{n,t}^{1/2}}[\hat{Z}_{n,t}\sqrt{Z}_{n,t}Y_{n,t}+Y_{n,t}\sqrt{Z}_{n,t}\hat{Z}_{n,t}]  e^{-uZ_{n,t}^{1/2}}\\
  &=\int_0^{\infty} \rmd u \, e^{-uZ_{n,t}^{1/2}}[\hat{Z}_{n,t}Z_{n,t}^{1/2}(I_d-Z_{n,t}^{1/2}\hat{Z}_{n,t}Z_{n,t}^{1/2})^{-1}+(I_d-Z_{n,t}^{1/2}\hat{Z}_{n,t}Z_{n,t}^{1/2})^{-1}Z_{n,t}^{1/2}\hat{Z}_{n,t}]  e^{-u Z_{n,t}^{1/2}}\\
  &=\int_0^{\infty} \rmd u \, e^{-u Z_{n,t}^{1/2}}[\hat{Z}_{n,t}(I_d-Z_{n,t}\hat{Z}_{n,t})^{-1}Z_{n,t}^{1/2}+Z_{n,t}^{1/2}(I_d-\hat{Z}_{n,t}Z_{n,t})^{-1}\hat{Z}_{n,t}]  e^{-uZ_{n,t}^{1/2}}\\
  &=-\int_0^{\infty} \rmd u \, \p_u \left( e^{-uZ_{n,t}^{1/2}}\hat{Z}_{n,t}(I_d-Z_{n,t}\hat{Z}_{n,t})^{-1}  e^{-uZ_{n,t}^{1/2}}\right)\\
  &=\hat{Z}_{n,t}(I_d-Z_{n,t}\hat{Z}_{n,t})^{-1}
   \end{split}
\end{equation}
where from the first line to the second line we have replaced the expression for the optimal noise \eqref{eq:WNT-strict-weak-optimal-noise}. To summarise we have obtained the WNT equations~\eqref{eq:WNT-matrix-SW} of the matrix strict-weak polymer, i.e.,
\begin{equation}
    \begin{split}
Z_{n,t+1}&=Z_{n,t}(I_d-\hat{Z}_{n,t}Z_{n,t})^{-1}+Z_{n-1,t}\\
\hat{Z}_{n,t-1}&=\hat{Z}_{n,t}(I_d-Z_{n,t}\hat{Z}_{n,t})^{-1}+\hat{Z}_{n+1,t}
    \end{split}
\end{equation}

We have found this system to be integrable and have obtained its Lax pair, which displayed in Eqs.~\eqref{eq:LaxPair-matrix-SW-1}--\eqref{eq:LaxPair-matrix-SW-2}.
As for the log Gamma polymer, even for the scalar case $d=1$, there are original results. 

\begin{remark}
  Alternatively, proceeding to the rescaling \eqref{eq:wnt-strict-weak-app} and taking the limit $\alpha \gg 1$, one can take a saddle point with respect to the two fields on the action \eqref{eq:app-strict-weak-action-two-fields} and obtains the WNT equations \eqref{eq:WNT-matrix-SW}.  
\end{remark}

\subsection{FD transformation of the matrix strict-weak polymer }
We obtain the invariant measure of the matrix strict-weak polymer through a FD symmetry of the field theoretical action 
\be
\label{app:eq:action-strict-weak}
 \hat{S}_0[Z,\hat{Z},Y] =  \sum_{(n,t)\in \mathcal{C}}   \Tr \left[ \hat{Z}_{n,t-1} \left(Z_{n,t}-Z_{n,t-1}^{1/2}Y_{n,t-1}Z_{n,t-1}^{1/2} -Z_{n-1,t-1} \right)\right]  - \alpha \log \det Y_{n,t-1} + 
  \Tr \, Y_{n,t-1}
\ee

Let us now suppose that the original dynamics is studied in the parallelogram $\mathcal{C}$ (depicted in Fig.~\ref{fig:strict-weakpolymer-lightcone-wnt}) so that $\sum_{(n,t) \in \mathcal{C}} = \sum_{n=1}^{N-1} \sum_{t=1}^{T-1}$ with the additional constraints $n\leq t$ and $n-(N-1)\geq t-(T-1)$. The recursion is thus applied in that parallelogram, with fixed initial conditions $Z_i$ (one may as well consider $\hat{Z}_{n,t-1}$ to be set to zero outside that region). The FDT symmetry is found using the following inversion of the field $Z_{n,t}$ and the affine transformation of the response field.
\be
\label{eq:change-var-strict-weak-FDT}
\begin{split}
& Z_{n,t} = Q_{n'-1,t'-1}^{-1} \quad , \quad n'=N-n \quad , \quad t'=T-t \\
&  (Z_{n,t} -Z_{n-1,t-1} )^{1/2}\hat{Z}_{n,t-1}  (Z_{n,t} -Z_{n-1,t-1} )^{1/2}  = (Q_{n',t'} -Q_{n'-1,t'-1} )^{1/2}\hat{Q}_{n',t'-1}  (Q_{n',t'} -Q_{n'-1,t'-1} )^{1/2} + \alpha_{n',t'}\\
& (n,t)\in \mathcal{C} \Leftrightarrow (n',t')\in \mathcal{C}
\end{split}
\ee

The space and time coordinates are reversed for the symmetrised process obtained through the map $(Z,\hat{Z})\mapsto (Q,\hat{Q})$. In particular we obtain that the summation over parallelogram $\mathcal{C}$ is preserved by the transformation, i.e., $\sum_{(n,t)\in \mathcal{C}} \Leftrightarrow \sum_{(n',t')\in \mathcal{C}}$ which will be implicit in what follows.\\

We now need to relate the randomness of the initial process $Y_{n,t}$ with the noise of the reversed process which we define as $W_{n',t'}$ and we need to determine the factor $\alpha_{n',t'}$ included in the affine transformation of the response field. We further proceed by replacing the change of variable \eqref{eq:change-var-strict-weak-FDT} into the original action \eqref{app:eq:action-strict-weak}, we obtain 
\be
\begin{split}
 &\hat{S}_0[Z,\hat{Z},Y] =  \sum_{(n,t)\in \mathcal{C}}   \Tr \bigg[ (Z_{n,t} -Z_{n-1,t-1} )^{1/2}\hat{Z}_{n,t-1} (Z_{n,t} -Z_{n-1,t-1} )^{1/2}\\
 &\left(I_d-(Z_{n,t} -Z_{n-1,t-1} )^{-1/2}Z_{n,t-1}^{1/2}Y_{n,t-1}Z_{n,t-1}^{1/2}(Z_{n,t} -Z_{n-1,t-1} )^{-1/2}  \right)\bigg]  - \alpha \log \det Y_{n,t-1} + 
  \Tr \, Y_{n,t-1}\\
  &=  \sum_{(n',t')\in \mathcal{C}}   \Tr \bigg[((Q_{n',t'} -Q_{n'-1,t'-1} )^{1/2}\hat{Q}_{n',t'-1}  (Q_{n',t'} -Q_{n'-1,t'-1} )^{1/2} + \alpha_{n',t'})\\
 &\left(I_d-(Q_{n'-1,t'-1}^{-1} -Q_{n',t'}^{-1} )^{-1/2}Q_{n'-1,t'}^{-1/2}Y_{n,t-1}Q_{n'-1,t'}^{-1/2}(Q_{n'-1,t'-1}^{-1} -Q_{n',t'}^{-1} )^{-1/2}  \right)\bigg]  - \alpha \log \det Y_{n,t-1} + 
  \Tr \, Y_{n,t-1}\\
  \end{split}
\ee
We identify the terms corresponding to the action in the novel variables and obtain
\be
\begin{split}
 &\hat{S}_0[Z,\hat{Z},Y]   = S[Q,\hat{Q},W]+ \sum_{(n',t')\in \mathcal{C}}   \Tr \bigg[(Q_{n',t'} -Q_{n'-1,t'-1} )^{1/2}\hat{Q}_{n',t'-1}  (Q_{n',t'} -Q_{n'-1,t'-1} )^{1/2}\\
  &\bigg((Q_{n',t'} -Q_{n'-1,t'-1} )^{-1/2}Q_{n',t'-1}^{1/2}W_{n',t'-1}Q_{n',t'-1}^{1/2}(Q_{n',t'} -Q_{n'-1,t'-1} )^{-1/2}\\
  &-(Q_{n'-1,t'-1}^{-1} -Q_{n',t'}^{-1} )^{-1/2}Q_{n'-1,t}^{-1/2}Y_{n,t-1}Q_{n'-1,t}^{-1/2}(Q_{n'-1,t'-1}^{-1} -Q_{n',t'}^{-1} )^{-1/2}  \bigg)\\
 &+\alpha_{n',t'}\left(I_d-(Q_{n'-1,t'-1}^{-1} -Q_{n',t'}^{-1} )^{-1/2}Q_{n'-1,t'}^{-1/2}Y_{n,t-1}Q_{n'-1,t'}^{-1/2}(Q_{n'-1,t'-1}^{-1} -Q_{n',t'}^{-1} )^{-1/2}  \right)\bigg]\\
 &- \alpha \log \det Y_{n,t-1}+\alpha \log \det W_{n',t'-1} + 
  \Tr \, Y_{n,t-1}- 
  \Tr \, W_{n',t'-1}\\
  \end{split}
\ee
The relation between the two noises  $W_{n',t'-1}$ and $Y_{n,t-1}$ is found by cancelling the term proportional to the response field $\hat{Q}$. 
\begin{equation}
\label{eq:fdt-strictweak-noise}
    \begin{split}
        &(Q_{n',t'} -Q_{n'-1,t'-1} )^{-1/2}Q_{n',t'-1}^{1/2}W_{n',t'-1}Q_{n',t'-1}^{1/2}(Q_{n',t'} -Q_{n'-1,t'-1} )^{-1/2}=\\
        &(Q_{n'-1,t'-1}^{-1} -Q_{n',t'}^{-1} )^{-1/2}Q_{n'-1,t'}^{-1/2}Y_{n,t-1}Q_{n'-1,t'}^{-1/2}(Q_{n'-1,t'-1}^{-1} -Q_{n',t'}^{-1} )^{-1/2}
    \end{split}
\end{equation}
This identity allows to relate the two log-Determinants as
\begin{equation}
    \begin{split}
        \mathcal{T}_1&:= -\alpha  \log \det Y_{n,t-1}+\alpha \log \det W_{n',t'-1} \\
        &= \alpha \log \Det (Q_{n'-1,t'-1}^{-1} -Q_{n',t'}^{-1} )^{-1}  Q_{n'-1,t'}^{-1}(Q_{n',t'} -Q_{n'-1,t'-1} )^{-1}  Q_{n',t'-1}^{-1} \\
        &=\alpha \log \Det \frac{Q_{n',t'}Q_{n'-1,t'-1}}{Q_{n'-1,t'}Q_{n',t'-1}}
    \end{split}
\end{equation}
where we have used

\begin{equation}
\label{eq_det-identity-2}
    \Det \left[ (A-B) (B^{-1}-A^{-1})^{-1} \right] = 
\Det \left[ (A B^{-1}-I ) B (A B^{-1}-I)^{-1} A \right] = 
\Det A \, \Det B 
\end{equation}
These terms are telescopic and upon summation over $(n',t')$ will generate boundary terms. The last term to treat is then
\begin{equation}
    \begin{split}
        \mathcal{T}_2 &=\Tr \left[ \alpha_{n',t'}\left(I_d-(Q_{n'-1,t'-1}^{-1} -Q_{n',t'}^{-1} )^{-1/2}Q_{n'-1,t'}^{-1/2}Y_{n,t-1}Q_{n'-1,t'}^{-1/2}(Q_{n'-1,t'-1}^{-1} -Q_{n',t'}^{-1} )^{-1/2}  \right)\right] \\
   &+\Tr Y_{n,t-1}-\Tr W_{n',t'-1}
    \end{split}
\end{equation}

Writing the relation between the noises \eqref{eq:fdt-strictweak-noise} as $AWA^\intercal = BYB^\intercal$, we choose
\begin{equation}
\label{eq:fd-strict-weak-auxiliary-noise}
    \alpha_{n',t'}=(B^\intercal)^{-1}_{n',t'}B^{-1}_{n',t'}-(A^\intercal)^{-1}_{n',t'}A^{-1}_{n',t'}
\end{equation}
The remaining term is
\begin{equation}
    \begin{split}
        \mathcal{T}_2&=\Tr [\alpha_{n',t'}]\\
        &= \Tr [(B^\intercal)^{-1}_{n',t'}B^{-1}_{n',t'}]-\Tr [(A^\intercal)^{-1}_{n',t'}A^{-1}_{n',t'}]\\
        &=\Tr [Q_{n'-1,t'}(Q_{n'-1,t'-1}^{-1}-Q_{n',t'}^{-1})]-\Tr [Q_{n',t'-1}^{-1}(Q_{n',t'}-Q_{n'-1,t'-1})]
    \end{split}
\end{equation}

These terms are also telescopic and upon summation over $n'$ and $t'$ will generate boundary terms. As a summary we have obtained that 
\begin{equation}
    \begin{split}
        &\hat{S}_0[Z,\hat{Z},Y]   = S_0[Q,\hat{Q},W]+ \sum_{(n',t')\in \mathcal{C}}\mathcal{T}_1+\mathcal{T}_2
    \end{split}
\end{equation}
Since the sum $\sum_{(n',t')\in \mathcal{C}}\mathcal{T}_1+\mathcal{T}_2$ is telescopic, we need to properly consider the range of summation to define its boundary.

\begin{figure}[t!]
    \centering
    \includegraphics[scale=0.7]{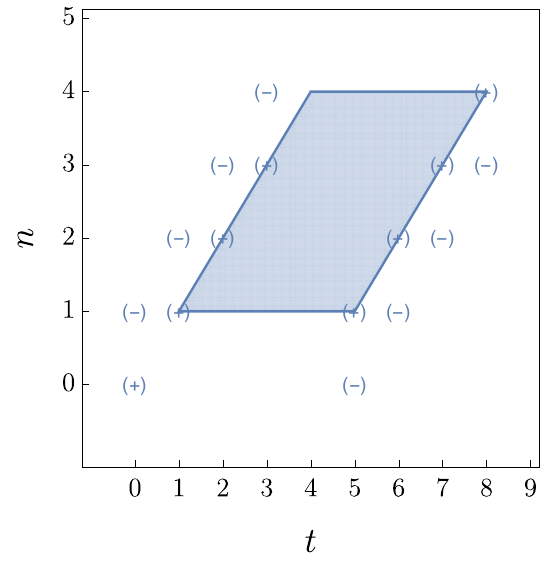}
    \caption{Summation of the terms in $\mathcal{T}_1$ for $N=5$ and $T=9$. The markers $(+)$ or $(-)$ indicate the sign of $\log \Det Q_{n,t}$ at the corresponding site in the term $\mathcal{T}_1$.}
    \label{fig:boundary-terms-strict-weak}
\end{figure}

The sum over $\mathcal{T}_1$ involves solely bands on the left and right boundaries as depicted in Fig.~\ref{fig:boundary-terms-strict-weak}. 

\begin{equation}
\begin{split}
    &\sum_{(n',t')\in \mathcal{C}}\mathcal{T}_1     = \alpha \kappa \sum_{n'=1}^{N-1}\big( \log \Det \frac{Q_{n'-1,n'-1}}{Q_{n',n'-1}}-\log \Det \frac{Q_{n'-1,T-N+n'}}{Q_{n',T-N+n'}}\big) \\
    &+\alpha (1-\kappa)\sum_{n'=1}^{N-2}\big( \log \det \frac{Q_{n',n'}}{Q_{n',n'-1}}-\log \det \frac{Q_{n',T-N+n'+1}}{Q_{n',T-N+n'}}\big)+\alpha (1-\kappa) \big(\log \det \frac{Q_{0,0}}{Q_{0,T-N+1}}-\log \det \frac{Q_{N-1,N-2}}{Q_{N-1,T-1}}\big)
\end{split}
\end{equation}
The first line gives a weight $\alpha \kappa$ to all vertical increments, the second line gives a weight $\alpha (1-\kappa)$ to the horizontal increments and the third line describes the horizontal increments of the bottom line $n=0$ and the top line $n=N-1$. Although the above expression is valid for any $\kappa$, we will restrict to $\kappa <-\frac{d-1}{2\alpha} $ for normalisation considerations, see below. Note that these increments can be rewritten as a telescopic sum to involve the whole boundary, i.e.,
\begin{equation}
    \begin{split}
        \log \det \frac{Q_{0,0}}{Q_{0,T-N+1}}&= \sum_{t'=1}^{T-N+1}\log \det  \frac{Q_{0,t'-1}}{Q_{0,t'}}, \quad 
        \log \det \frac{Q_{N-1,N-2}}{Q_{N-1,T-1}}= \sum_{t'=1}^{T-N+1}\log \det  \frac{Q_{N-1,N-3+t'}}{Q_{N-1,N-2+t'}}\\
    \end{split}
\end{equation}

The sum over $\mathcal{T}_2$ involves all consecutive ratios on the boundary of the domain, i.e.,
\begin{equation}
    \begin{split}
       \sum_{(n',t')\in \mathcal{C}}\mathcal{T}_2=& \sum_{n'=1}^{N-1}\Tr [\frac{Q_{n'-1,n'-1}}{Q_{n',n'-1}}-\frac{Q_{n'-1,T-N+n'}}{Q_{n',T-N+n'}}]-\sum_{n'=1}^{N-2}\Tr [\frac{Q_{n',n'}}{Q_{n',n'-1}}-\frac{Q_{n',T-N+1+n'}}{Q_{n',T-N+n'}}]\\
       &+\sum_{t'=1}^{T-N+1}\Tr[\frac{Q_{0,t'}}{Q_{0,t'-1}}-\frac{Q_{N-1,N-2+t'}}{Q_{N-1,N-3+t'}}]
    \end{split}
\end{equation}

Thus 
\begin{equation}
\label{eq:FDT-matrix-strict-weak-action}
    \begin{split}
        &\hat{S}_0[Z,\hat{Z},Y]   = S_0[Q,\hat{Q},W]\\
       &+\sum_{n'=1}^{N-1}\Tr [\frac{Q_{n'-1,n'-1}}{Q_{n',n'-1}}-\frac{Q_{n'-1,T-N+n'}}{Q_{n',T-N+n'}}]+ \alpha \kappa \sum_{n'=1}^{N-1}\big( \log \Det \frac{Q_{n'-1,n'-1}}{Q_{n',n'-1}}-\log \Det \frac{Q_{n'-1,T-N+n'}}{Q_{n',T-N+n'}}\big)  \\
       &-\sum_{n'=1}^{N-2}\Tr [\frac{Q_{n',n'}}{Q_{n',n'-1}}-\frac{Q_{n',T-N+1+n'}}{Q_{n',T-N+n'}}]+\alpha (1-\kappa)\sum_{n'=1}^{N-2}\big( \log \det \frac{Q_{n',n'}}{Q_{n',n'-1}}-\log \det \frac{Q_{n',T-N+n'+1}}{Q_{n',T-N+n'}}\big)\\
       &+\sum_{t'=1}^{T-N+1}\Tr[\frac{Q_{0,t'}}{Q_{0,t'-1}}-\frac{Q_{N-1,N-2+t'}}{Q_{N-1,N-3+t'}}] +\alpha (1-\kappa)\sum_{t'=1}^{T-N+1}\big( \log \det  \frac{Q_{0,t'-1}}{Q_{0,t'}}-\log \det  \frac{Q_{N-1,N-3+t'}}{Q_{N-1,N-2+t'}} \big) 
    \end{split}
\end{equation}

Thus, upon the FD transformation, the action is preserved up to boundary terms. 

\subsection{Invariant measure of the matrix strict-weak: calculation on four sites} \label{app:subsec:invariant-strict-weak}

We now obtain the invariant measure from the FD transformation over four sites depicted in Fig.~\ref{fig:FDT_for_the_strict_weak_on_4_sites}.

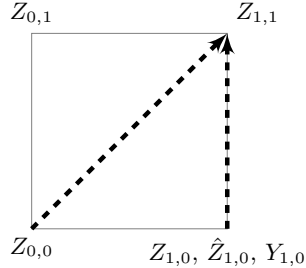
\begin{figure}[h!]
\begin{tikzpicture}[scale=2.6]

% Drawing grid
\draw[step=2cm,gray,very thin] (0,0) rectangle (1,1);

% Nodes
\node[above] at (0,1) {$Z_{0,1}$};
\node[below] at (1,0) {$Z_{1,0}, \, \hat{Z}_{1,0}, \, Y_{1,0} $};
\node[below] at (0,0) {$Z_{0,0}$};
\node[above right] at (1,1) {$Z_{1,1} \, $};

% Dashed lines with labels and big arrowheads
\draw[->, dashed, line width=0.6mm, >=latex'] (0,0) -- (1,1);
\draw[->, dashed, line width=0.6mm, >=latex'] (1,0) --  (1,1) ;

\end{tikzpicture} 
\caption{Recursion of the matrix strict-weak polymer on four sites.}
\label{fig:FDT_for_the_strict_weak_on_4_sites}
\end{figure}

On four sites the "initial condition"
is $Z_i=(Z_{0,0},Z_{1,0},Z_{0,1})$, the terminal condition is $Z_f=(Z_{1,1})$ and the final condition is $Z'_f=(Z_{1,1},Z_{0,1},Z_{1,0})$. The final point $Z_{1,1}$ is determined by the recursion  $Z_{1,1}=Z_{1,0}^{1/2}Y_{1,0}Z_{1,0}^{1/2} +Z_{0,0} $. The action simply reads
 \be
 \hat{S}_0[Z,\hat{Z},Y] =  \Tr \left[ \hat{Z}_{1,0} \left(Z_{1,1}-Z_{1,0}^{1/2}Y_{1,0}Z_{1,0}^{1/2} -Z_{0,0} \right)\right]  - \alpha  \log \det Y_{1,0} +  \Tr \, Y_{1,0}
\ee
Using \eqref{eq:change-var-strict-weak-FDT}, The FD transformation on four sites read
\begin{equation}
    \begin{split}
         &Z_{0,0} =Q_{1,1}^{-1}, \quad  Z_{1,0}=Q_{0,1}^{-1}, \quad   Z_{0,1}=Q_{1,0}^{-1} , \quad  Z_{1,1} = Q_{0,0}^{-1}   \, .
    \end{split}
\end{equation}
From Eq.~\eqref{eq:fdt-strictweak-noise}, the noises are related as
\begin{equation}
\label{eq:fdt-strictweak-noise-4sites-1}
    \begin{split}
        &(Q_{1,1} -Q_{0,0} )^{-1/2}Q_{1,0}^{1/2}W_{1,0}Q_{1,0}^{1/2}(Q_{1,1} -Q_{0,0} )^{-1/2}=
        (Q_{0,0}^{-1} -Q_{1,1}^{-1} )^{-1/2}Q_{0,1}^{-1/2}Y_{1,0}Q_{0,1}^{-1/2}(Q_{0,0}^{-1} -Q_{1,1}^{-1} )^{-1/2}
    \end{split}
\end{equation}
and from \eqref{eq:change-var-strict-weak-FDT} the response fields are related as 
\begin{equation}
    (Z_{1,1} -Z_{0,0} )^{1/2}\hat{Z}_{1,0}  (Z_{1,1} -Z_{0,0} )^{1/2}  = (Q_{1,1} -Q_{0,0} )^{1/2}\hat{Q}_{1,0}  (Q_{1,1} -Q_{0,0} )^{1/2} + \alpha_{1,1}\\
\end{equation}
where the expression of $\alpha_{1,1}$ is given in \eqref{eq:fd-strict-weak-auxiliary-noise}. The FD symmetry on the action \eqref{eq:FDT-matrix-strict-weak-action} reads here as follows

\begin{equation}
\label{eq:app-strict-weak-fd-4sites-stationary-measure}
    \begin{split}
        &\hat{S}_0[Z,\hat{Z},Y]   = S_0[Q,\hat{Q},W]+\Tr [\frac{Q_{0,0}}{Q_{1,0}}-\frac{Q_{0,1}}{Q_{1,1}}]+ \alpha \kappa\big( \log \Det \frac{Q_{0,0}}{Q_{1,0}}-\log \Det \frac{Q_{0,1}}{Q_{1,1}}\big)  \\
       &+\Tr[\frac{Q_{0,1}}{Q_{0,0}}-\frac{Q_{1,1}}{Q_{1,0}}] +\alpha (1-\kappa)\big( \log \det  \frac{Q_{0,0}}{Q_{0,1}}-\log \det  \frac{Q_{1,0}}{Q_{1,1}} \big) \\
       &  =S_0[Q,\hat{Q},W]+ \Tr [\frac{Z_{0,1}}{Z_{1,1}}-\frac{Z_{0,0}}{Z_{1,0}}]+ \alpha \kappa\big( \log \Det \frac{Z_{0,1}}{Z_{1,1}} -\log \Det \frac{Z_{0,0}}{Z_{1,0}} \big)  \\
       &  + \Tr[\frac{Z_{1,1}}{Z_{1,0}}-\frac{Z_{0,1}}{Z_{0,0}}] +\alpha (1-\kappa)\big( \log \det  \frac{Z_{1,0}}{Z_{1,1}} -\log \det \frac{Z_{0,0}}{Z_{0,1}} \big) 
    \end{split}
\end{equation}

Taking the exponential of minus \eqref{eq:app-strict-weak-fd-4sites-stationary-measure} on both sides, integrating over $Y_{1,0}$ and $\hat{Z}_{1,0}$ and inserting the observable $\mathcal{O}(Z'_f)$, the FD relation \eqref{eq:app-strict-weak-fd-4sites-stationary-measure} implies

\begin{equation}
\label{eq:MSR-4-sites-fd-transformation-strictweak}
    \begin{split}
        &\iint \mu(\rmd Y_{1,0}) \rmd \hat{Z}_{1,0}e^{-\hat{S}_0[Z,\hat{Z},V]} \mathcal{O}(Z'_f)e^{-\Tr [\frac{Z_{0,0}}{Z_{1,0}}+\frac{Z_{0,1}}{Z_{0,0}}]} |\Det \frac{Z_{0,0}}{Z_{1,0}}|^{-\alpha \kappa} |\Det  \frac{Z_{0,0}}{Z_{0,1}}|^{-\alpha (1-\kappa)}=|\Det Z_{1,1}|^{-\frac{d+1}{2}}|\Det Z_{0,0}|^{-\frac{d+1}{2}}\\
        &\times \iint \mu(\rmd  W_{1,0}) \rmd \hat{Q}_{1,0}e^{-\hat{S}_0[Q,\hat{Q},W]}\mathcal{O}(Z'_f) e^{-\Tr[\frac{Z_{0,1}}{Z_{1,1}}+\frac{Z_{1,1}}{Z_{1,0}}] } |\Det \frac{Z_{0,1}}{Z_{1,1}}|^{-\alpha \kappa }|\Det\frac{Z_{1,0}}{Z_{1,1}}|^{-\alpha (1-\kappa) } \\
    \end{split}
\end{equation}

We now use the FD symmetry. First of all, since all $Z$ are fixed and since the measure over the noise of $GL_d$ invariant, one has $\mu (\rmd Y_{1,0})=\mu (\rmd W_{1,0})$ as from \eqref{eq:fdt-strictweak-noise-4sites-1} the noises are related by a $GL_d$ transformation involving solely $Z$. Furthermore, we have used that the only non-trivial Jacobian comes from the response field
\begin{equation}
    \rmd \hat{Z}_{1,0}=\rmd \hat{Q}_{1,0} \Det |Q_{1,1}|^{\frac{d+1}{2}}|\Det Q_{0,0}|^{\frac{d+1}{2}}=\mathcal{D}\hat{Q}_{1,0} |\Det Z_{0,0}|^{-\frac{d+1}{2}} \Det |Z_{1,1}|^{-\frac{d+1}{2}}
\end{equation}
which can be derived using the identities \eqref{detg} and \eqref{eq_det-identity-2}.\\

We now multiply both terms of \eqref{eq:MSR-4-sites-fd-transformation-strictweak} by the measure

\begin{equation}
\label{eq:ratio-jacobian-4sites-strict-weak}
   \rmd Z_{1,1} \mu(\rmd Z_{0,0}) \mu (\rmd Z_{1,0}) \mu (\rmd Z_{0,1}) \, .
\end{equation}
and integrate over all the variables $Z$. We call the resulting equation $LHS=RHS$ 
then study separately the right side $RHS$ and then the left hand side $LHS$.\\

\paragraph{Right hand side RHS.} As a first step, we will use the normalisation of the right hand side of \eqref{eq:MSR-4-sites-fd-transformation-strictweak} with respect to the final field $Q_{1,1}$ (i.e., the standard MSR integral normalisation discussed below \eqref{eq:app-log-gamma-action-three-fields})
\be \label{norm2-sw} 
1 = \int \rmd Q_{1,1} \iint \mu(\rmd  W_{1,0}) \rmd\hat{Q}_{1,0}e^{-\hat{S}_0[Q,\hat{Q},W]} = 
\int \frac{\rmd Z_{0,0}}{ |\det Z_{0,0}|^{d+1} }\iint \mu(\rmd W_{1,0}) \rmd\hat{Q}_{1,0}e^{-\hat{S}_0[Q,\hat{Q},W]} 
\ee 
Note that r.h.s. of \eqref{norm2-sw} contains the Jacobian from $Q_{1,1}$ to $Z_{0,0}$ which is obtained from \eqref{detinverse}.
We then obtain

\begin{equation}
\label{eq:strict-weak-invariant-measure-rhs}
    \begin{split}
        &RHS= \iiint \mathcal{O}(Z'_f)\mu(\rmd Z_{1,1})\mu(\rmd Z_{1,0} )\mu(\rmd Z_{0,1})e^{-\Tr[\frac{Z_{0,1}}{Z_{1,1}}+\frac{Z_{1,1}}{Z_{1,0}}] } |\Det \frac{Z_{0,1}}{Z_{1,1}}|^{-\alpha \kappa }|\Det\frac{Z_{1,0}}{Z_{1,1}}|^{-\alpha (1-\kappa) }  \\
    \end{split}
\end{equation}

\paragraph{Left hand side LHS.}
From Eqs.~\eqref{eq:average-observable-msr-strictweak} and with analogy to \eqref{eq:MSR-log-gamma-expectation-value}, the LHS is a conditional expectation of the final observables with respect to the initial values

\begin{equation}
\label{eq:strict-weak-invariant-measure-lhs}
\begin{split}
  LHS=&\iiint \mathbb{E}[ \mathcal{O} (Z_f') | Z_i ]
\mu(\rmd Z_{0,0}) \mu(\rmd Z_{1,0})\mu(\rmd Z_{0,1}) e^{-\Tr [\frac{Z_{0,0}}{Z_{1,0}}+\frac{Z_{0,1}}{Z_{0,0}}]} |\Det \frac{Z_{0,0}}{Z_{1,0}}|^{-\alpha \kappa} |\Det  \frac{Z_{0,0}}{Z_{0,1}}|^{-\alpha (1-\kappa)}
\end{split}
\end{equation}

Overall, equating \eqref{eq:strict-weak-invariant-measure-rhs} and \eqref{eq:strict-weak-invariant-measure-lhs}, we obtain our final result

\begin{equation}
\begin{split}
  &\iiint \mathbb{E}[ \mathcal{O} ( Z_f') | Z_i]
\mu(\rmd Z_{0,0}) \mu(\rmd Z_{1,0})\mu(\rmd Z_{0,1}) e^{-\Tr [\frac{Z_{0,0}}{Z_{1,0}}+\frac{Z_{0,1}}{Z_{0,0}}]} |\Det \frac{Z_{0,0}}{Z_{1,0}}|^{-\alpha \kappa} |\Det  \frac{Z_{0,0}}{Z_{0,1}}|^{-\alpha (1-\kappa)} \\
&=\iiint \mathcal{O}(Z'_f)\mu(\rmd Z_{1,1})\mu(\rmd Z_{1,0} )\mu(\rmd Z_{0,1})e^{-\Tr[\frac{Z_{0,1}}{Z_{1,1}}+\frac{Z_{1,1}}{Z_{1,0}}] } |\Det \frac{Z_{0,1}}{Z_{1,1}}|^{-\alpha \kappa }|\Det\frac{Z_{1,0}}{Z_{1,1}}|^{-\alpha (1-\kappa) } 
\end{split}
\end{equation}
which show that the measure on the triplet $Z_{0,0},Z_{1,0},Z_{0,1}$ in the l.h.s 
is transported by the dynamics (i.e., the strict-weak recursion) onto the measure on the triplet $Z_{1,1},Z_{1,0},Z_{0,1}$. 
It is now possible to apply the relations \eqref{AB} to express these measures for the strict-weak polymer more conveniently
on the partition sum ratios, which shows explicitly that these ratios are independent respectively Wishart and inverse Wishart distributed. To this aim we fix the zero mode to be $Z_{0,1}$ for both initial and final sets, and the initial measure reads
\be 
\mu(\rmd Z_{0,1}) \rmd P^{W}_{\alpha(1-\kappa)}[\tilde{r}\left(\frac{Z_{0,1}}{Z_{0,0}} \right)]\rmd P^{iW}_{-\alpha \kappa}[r\left(\frac{Z_{1,0}}{Z_{0,0}} \right)] 
\ee 
while the final measure reads
\be 
\mu(\rmd Z_{0,1}) \rmd P^{W}_{\alpha (1-\kappa)}[\tilde{r}\left(\frac{Z_{1,1}}{Z_{1,0}} \right)] \rmd P^{iW}_{-\alpha \kappa}[r\left(\frac{Z_{1,1}}{Z_{0,1}} \right)]
\ee 
One sees that now these define the invariant measure on the ratios.
\begin{enumerate}
    \item All ratios are sampled independently.
    \item The ratios along the horizontal axis defined with the matrix ordering convention $r$, see \eqref{eq:definition-convention-ratios}, 
    are sampled from the Inverse Wishart distribution with parameter $- \kappa \alpha$.
    \item The ratios along the vertical axis defined with the matrix ordering convention $\tilde{r}$ are sampled from the Wishart distribution with parameter and $(1-\kappa)\alpha$.
    \item The zero mode $Z_{0,1}$ is uniformly distributed over $\mathcal{P}_d$.
    \item We have the constraints $\kappa <-\frac{d-1}{2\alpha}$ and $\alpha > \frac{d-1}{2}$ - the constraints coming from the normalisation condition of the inverse Wishart distribution of the horizontal ratio. 
\end{enumerate}

To obtain the invariant measure over a lattice composed of more than four sites, one can follow a similar procedure as for the log Gamma. The structure of the boundary terms arising from the FD transformation \eqref{eq:FDT-matrix-strict-weak-action} and the associated geometry, see Fig.~\ref{fig:strict-weakpolymer-lightcone-wnt}, are however more complicated in the present case and we leave this derivation for a future work. We anticipate that the invariant measure will have the property that all horizontal ratios will be Wishart distributed, all vertical ratios Inverse Wishart distributed, the zero mode will be uniform and everything will be independent.

\end{widetext}

\end{document}